\documentclass[oneside,11pt]{article}

\usepackage[utf8]{inputenc}
\usepackage[T1]{fontenc}

\usepackage{amsthm}
\usepackage{amsmath}
\usepackage{amsfonts}
\usepackage{amssymb}

\usepackage{graphicx} 
\usepackage{graphbox}
\usepackage{float}
\usepackage{booktabs}
\usepackage{multirow}
\usepackage{rotating}
\usepackage{array}
\usepackage{xcolor}
\usepackage{subcaption}
\usepackage[ruled]{algorithm2e}

\newcolumntype{C}[1]{>{\centering\arraybackslash}p{#1}}
\newcolumntype{R}[1]{>{\raggedleft\let\newline\\\arraybackslash\hspace{0pt}}m{#1}}
\newcolumntype{L}[1]{>{\raggedright\let\newline\\\arraybackslash\hspace{0pt}}m{#1}}

\usepackage{breakcites}

\usepackage{nowidow}

\usepackage{enumitem}

\usepackage[top=1.5cm,bottom=2cm,right=2.5cm,left=2.5cm]{geometry}
\usepackage{titling}

\usepackage{natbib}

\usepackage{ifplatform}

\usepackage{textcomp}
\usepackage{bbm}

\usepackage{comment}
\ifwindows
\fi

\allowdisplaybreaks

\newcommand{\R}{\mathbb{R}}

\newcommand{\rd}{\mathrm{d}}

\newcommand{\bx}{\boldsymbol{x}}
\newcommand{\be}{\boldsymbol{e}}
\newcommand{\bm}{\boldsymbol{m}}

\newcommand{\by}{\boldsymbol{y}}
\newcommand{\bX}{\boldsymbol{X}}
\newcommand{\bY}{\boldsymbol{Y}}

\newcommand{\bR}{\boldsymbol{R}}

\newcommand{\bmu}{\boldsymbol\mu}

\newcommand{\bz}{\boldsymbol{z}}

\newcommand{\bc}{\boldsymbol{c}}

\newcommand{\bbe}{\boldsymbol\beta}

\newcommand{\btheta}{\boldsymbol\theta}

\newcommand{\bSigma}{\boldsymbol\Sigma}

\newcommand{\bA}{\boldsymbol{A}}

\newcommand{\bI}{\boldsymbol{I}}
\newcommand{\bS}{\boldsymbol{S}}

\newcommand{\bT}{\boldsymbol{T}}
\newcommand{\bM}{\boldsymbol{M}}

\newcommand{\xb}{\bx}

\newcommand{\Sd}{\mathbb{S}^{d}}
\newcommand{\Sdm}{\mathbb{S}^{d-1}}

\newcommand{\Hd}{\mathbb{H}^{d}}

\DeclareFontFamily{OT1}{pzc}{}
\DeclareFontShape{OT1}{pzc}{m}{it}{<-> s * [1.10] pzcmi7t}{}
\DeclareMathAlphabet{\mathpzc}{OT1}{pzc}{m}{it}

\newtheorem{definition}{Definition}[section]
\newtheorem{theorem}{Theorem}[section]
\newtheorem{corollary}{Corollary}[section]
\newtheorem{remark}{Remark}[section]
\newtheorem{proposition}{Proposition}[section]
\newtheorem{lemma}{Lemma}[section]

\newcommand{\defin}{:=}

\DeclareMathOperator*{\argmin}{arg\,min}
\DeclareMathOperator*{\argmax}{arg\,max}
\DeclareMathOperator*{\arccosh}{arccosh}
\DeclareMathOperator{\arctantwo}{arctan2}
\newcommand{\convd}{\stackrel{d}{\rightarrow}}
\newcommand{\convp}{\stackrel{\mathsf{P}}{\rightarrow}}
\newcommand{\convn}{\stackrel[n\to\infty]{}{\longrightarrow}}
\newcommand{\Prob}[1]{\mathsf{P}\left\{ #1 \right\}}
\newcommand{\Probbig}[1]{\mathsf{P}\big\{ #1 \big\}}
\newcommand{\Probstar}[1]{\mathsf{P}_*\left\{ #1 \right\}}
\newcommand{\Probbigstar}[1]{\mathsf{P}_*\big\{ #1 \big\}}
\newcommand{\Condprob}[2]{\mathsf{P}\left\{#1 \mid #2\right\}}

\newcommand{\E}[1]{\mathsf{E}( #1 )}
\newcommand{\Elr}[1]{\mathsf{E}\left( #1 \right)}
\newcommand{\Ebig}[1]{\mathsf{E}\big( #1 \big)}

\newcommand{\CondEbig}[2]{\mathsf{E}\big(#1 \mid #2\big)}
\newcommand{\Y}{\mathcal{Y}}
\newcommand{\dY}{d_\mathcal{Y}}

\newcommand{\Roob}{\widehat{R}^{\mathrm{oob}}}
\newcommand{\Var}[1]{\mathsf{Var}( #1 )}

\usepackage[pdftex,final,bookmarksnumbered,bookmarksopen=false,breaklinks,colorlinks]{hyperref}
\hypersetup{
	final,
	bookmarksnumbered=true,
	bookmarksopen=true,
	bookmarksopenlevel=0,
	unicode=false,
	pdftoolbar=true,
	pdfmenubar=true,
	pdffitwindow=false,
	pdftitle={Out-of-bag prediction balls for random forests in metric spaces},
	pdfdisplaydoctitle=true,
	pdftoolbar=true,
	pdfmenubar=true,
	pdflang={English},
	pdfauthor={Diego Serrano, Eduardo Garcia-Portugues},
	pdfsubject={arXiv paper},
	pdfcreator={Eduardo Garcia-Portugues},
	pdfproducer={Diego Serrano},
	pdfkeywords={Confidence regions}{Fréchet mean}{Random objects}{Regression},
	pdfnewwindow=true,
	breaklinks=true,
	hidelinks,       
	linkcolor=black,
	citecolor=black,
	filecolor=magenta,
	urlcolor=cyan
}

\newif\ifmain
\maintrue
\newif\ifsupplement
\supplementtrue

\newif\iffigstabs
\figstabstrue

\begin{document}

\ifmain

\title{Out-of-bag prediction balls for random forests in metric spaces}
\setlength{\droptitle}{-1cm}
\predate{}
\postdate{}
\date{}

\author{Diego Serrano$^{1,2}$ and Eduardo Garc\'ia-Portugu\'es$^{1}$}
\footnotetext[1]{Department of Statistics, Carlos III University of Madrid (Spain).}
\footnotetext[2]{Corresponding author. e-mail: \href{mailto:dieserra@est-econ.uc3m.es}{dieserra@est-econ.uc3m.es}.}
\maketitle

\begin{abstract}
	Statistical methods for metric spaces provide a general and versatile framework for analyzing complex data types. We introduce a novel approach for constructing confidence regions around new predictions from any bagged regression algorithm with metric-space-valued responses. This includes the recent extensions of random forests for metric responses: Fréchet random forests (Capitaine et al., 2024), random forest weighted local constant Fréchet regression (Qiu et al., 2024), and metric random forests (Bulté and Sørensen, 2024). Our prediction regions leverage out-of-bag observations generated during a single forest training, employing the entire data set for both prediction and uncertainty quantification. We establish asymptotic guarantees of out-of-bag prediction balls for four coverage types under certain regularity conditions. Moreover, we demonstrate the superior stability and smaller radius of out-of-bag balls compared to split-conformal methods through extensive numerical experiments where the response lies on the Euclidean space, sphere, hyperboloid, and space of positive definite matrices. A real data application illustrates the potential of the confidence regions for quantifying the uncertainty in the study of solar dynamics and the use of data-driven non-isotropic distances on the sphere.
\end{abstract}
\begin{flushleft}
\end{flushleft}

\section{Introduction}

As the complexity of data structures increases, traditional statistical methods that rely on vector space properties face significant limitations. Specialized methods have been developed for specific data types, including functional data \citep{Li2022}, directional data \citep{Mardia1999}, compositional data \citep{Aitchison1986}, shape data \citep{Dryden2016}, and covariance matrices \citep{Dryden2009}. Random variables from these complex data types are studied by \cite{Marron2021} under the terminology of random objects. A framework that has gained momentum in recent years is to consider random objects as elements in general metric spaces. While exploiting the structure of specific spaces can enhance the flexibility and tractability of statistical modeling, the generality of metric space analysis offers an unparalleled versatile framework that bypasses the need for algebraic structure or local Euclideanity, relying solely on the existence of a distance function.

Key statistical concepts must be adapted to metric spaces. \citet{Frechet1948} generalized the mean through square distance minimization. Extensions of regression models for non-Euclidean data have primarily focused on responses defined in Riemannian manifolds \citep[e.g.,][]{Fletcher2013, Hyunwoo2014}. A foundational nonparametric regression estimator for metric output is the Nadaraya--Watson estimator, which uses local-constant weights determined by a kernel function and a bandwidth parameter. The Nadaraya--Watson estimator was extended by \cite{Hein2009} to the case when both the response and the predictors are random objects using a (generalized) Fréchet mean. Building on the notion of Fréchet mean, \citet{Petersen2019} introduced Fréchet regression, generalizing the standard Euclidean regression to random objects by estimating the conditional Fréchet mean. In the same work, global Fréchet regression was proposed as a parametric model that generalizes linear least squares regression for predictors in $\R^p$ and a metric response, alongside local Fréchet regression as a nonparametric generalization of local linear estimation for random object responses with Euclidean predictors.

Random Forests \citep[RFs;][]{Breiman2001} are a popular nonparametric regression method widely recognized for their flexibility and predictive performance, owing to their ability to capture complex, nonlinear relationships, particularly in high-dimensional settings \citep{Buhlmann2002, Hastie2009, Varian2014}. In recent years, several authors have worked on adaptations of RFs to metric spaces under the framework of Fréchet regression. \citet{Capitaine2024} \citep[first appeared as][]{Capitaine2019} developed Fréchet Random Forests (FRFs), which adapt RFs for metric data by incorporating the Fréchet mean in the splitting criterion and the aggregation of trees. Alternatively, following the ideas of \citet{Meinshausen2006}, \citet{Qiu2024} leveraged the weights generated in the building of each tree to estimate the response as a weighted average of the training responses. \citet{Bulte2024} introduced Fréchet medoids into the splitting criterion of the trees. Fréchet medoids serve as consistent estimators of the Fréchet mean and offer a computationally simpler alternative in settings where no closed-form expression is available and optimization over a manifold is typically employed. 

Despite their predictive power, RFs have historically lacked well-developed tools for statistical inference, particularly for quantifying the uncertainty in a RF prediction. For Euclidean data, \citet{Meinshausen2006} introduced quantile regression forests, which estimate conditional quantiles and naturally lead to prediction intervals. \cite{Barber2021} introduced jackknife+ to construct prediction intervals for regression, based on leave-one-out predictions and residuals. \cite{Kim2020} proposed jackknife+-after-bootstrap, which integrates ensemble learning with the jackknife+. \cite{Wang2023} give sufficient conditions for the robustness of RF predictions against small perturbations in the sample. They obtain stability bounds for the RF prediction, and use them to derive the coverage of RF prediction intervals.

The construction of confidence regions in metric spaces has focused primarily on Split-Conformal (SC) prediction, developed by \cite{Lei2018}. SC prediction is based on ranking non-conformity scores of exchangeable data, and reduces the computational burden of conformal prediction by splitting the data into two subsamples, one for fitting the model and another to rank the conformity scores, at the cost of halving the sample size in each task. Recently, \citet{Lugosi2024} modified the traditional SC ranking of residuals by estimating the quantiles of the errors, while \citet{Zhou2025} introduced Conditional Profile average transport Costs (CPCs) as a method to assess the compatibility of an element of the metric space with the conditional distribution of the response given a predictor value. The SC algorithm can be directly applied to develop confidence sets by using the conditional distribution of CPCs as a conformity score. However, the aforementioned references inherit the inefficiencies of SC inference derived from splitting the data into two subsamples.

In this paper, we propose \textit{OOB prediction balls}, a novel method for constructing confidence regions for random objects in metric spaces using RF Out-Of-Bag (OOB) errors, which are a reliable estimate of predictive performance and are obtained automatically in the forest training process. The OOB balls are constructed around the RF prediction of the response with a radius determined by the $(1-\alpha)$-quantile of the empirical distribution of the OOB errors. For any given value of the response, a prediction ball can be generated with the information obtained when training a single forest, with OOB errors preventing any leakage of information about the response from the training phase. Although we derive OOB balls from RF predictions, they can be applied directly to any bagging regression model in metric spaces. We establish the asymptotic validity of OOB balls under four different coverage types: unconditional, and conditional on combinations of the training sample and prediction point. Our approach is related to SC inference, but is specifically tailored to RFs with metric-space-valued responses. By using OOB observations, OOB balls allow using the full data set for both prediction and uncertainty estimation. We compare OOB and SC balls by simulating from a multivariate Euclidean linear model with different values for the dimensionality of the response, and conclude that the volume of OOB balls is smaller in general than that of SC balls, without compromising coverage. The coverage and radius of OOB balls are assessed using simulated data on the sphere, the hyperboloid, and the space of $2\times2$ symmetric positive definite matrices. Additionally, we study the dynamics of sunspots on the Sun's surface using a real data set. Specifically, we consider the problem of predicting the location of sunspots the last time they were observed, based on the first-ever observed location. The motion of sunspots is complex and difficult to predict, so we use OOB balls to establish regions that contain the true location with a given probability. We use a data-driven distance on the sphere induced by the geodesic distance on the spheroid, illustrating the use of anisotropic distances to adapt the shape of OOB balls to the underlying distribution of the data. 

Our approach builds on \citet{Zhang2020}, who proposed RF prediction intervals using out-of-bag observations and proved their asymptotic validity. The confidence regions of \citet{Zhang2020} have the shape of balls centered at the RF prediction and with radius given by a quantile of the empirical distribution of the OOB errors, which makes their generalization to metric spaces natural. Our method also relates to \citet{Wang2023}, since it is based on the OOB residuals correctly approximating the RF prediction errors; and to \citet{Kim2020}, when the ensemble learner is a bagged regressor. However, \cite{Wang2023} impose conditions on the tails of the square of the response variable, a condition that does not have a natural counterpart in general metric spaces. The approach of \cite{Kim2020} does not extend directly to general metric spaces, since the prediction intervals are built from empirical quantiles of the lower and upper interval endpoints, a step that relies on the order structure of the real line. While \cite{Kim2020} and \cite{Wang2023} provide non-asymptotic lower bounds for the coverage probability, these guarantees concern marginal coverage; our approach provides asymptotic guarantees for four different coverage types, as in \cite{Zhang2020}. However, the absence of vector space structures complicates the proof of the coverage results.

The remainder of this paper is organized as follows. Section \ref{sec:FRF} reviews the literature on Fréchet regression for metric space data, emphasizing adaptations of RFs to metric spaces and the use of OOB errors. In Section \ref{sec:setting}, we define OOB balls based on OOB errors and introduce four distinct coverage types. Concrete examples illustrate the theoretical setting and the conditions under which these coverage types are analyzed. Section \ref{sec:main_results} establishes the asymptotic validity of OOB balls across all coverage types. The performance and coverage of OOB balls are investigated through simulations in Section \ref{sec:experiments} for different metric spaces. Section \ref{sec:oob-vs-sc} delves into the advantages and limitations of OOB balls relative to SC balls. An application to sunspot dynamics is given in Section \ref{sec:real_data}. The article concludes with a discussion in Section \ref{sec:discussion}. The Supplementary Material (SM) contains the proofs omitted from the paper, additional numerical experiments, and a revision of the hyperboloid von Mises--Fisher distribution. The code replicating the results of the paper is available at \url{https://github.com/dieseor/oob_balls}.

\section{Random forests in metric spaces}
\label{sec:FRF}

In this section, we first introduce a framework for regression in metric spaces. We then describe three different methodologies for constructing RFs in this framework.

\subsection{Fréchet regression}

Let $(\Y, \dY)$ denote the metric space where the response $Y$ is defined, with $\dY$ a distance function in ${\Y}$. Denote by $(\mathcal{X}, d_{\mathcal{X}}) \defin (\mathcal{X}_1, d_{\mathcal{X}_1}) \times \dots \times (\mathcal{X}_p, d_{\mathcal{X}_p})$ the metric space of predictors, formed by the product of $p$ metric spaces. Analyzing metric space data requires basic statistical notions like the expectation and variance, which are undefined in the absence of a vector space structure. In order to generalize them for random objects, \citet{Frechet1948} defined the homonymous mean and variance by exploiting the squared distance minimization property of the expectation in the Euclidean space:
\begin{align}\label{frechet_mean_variance}
    y_{\oplus} \defin  \argmin_{y\in (\Y, \dY)} \Elr{\dY(Y, y)^2} , \quad V_{\oplus} \defin  \Elr{ \dY(Y, y_{\oplus})^2 }.
\end{align}
The Fréchet mean may not exist, and in case it does, it may not be unique. If the Fréchet mean exists, then the Fréchet variance exists and is unique. Given a sample $\mathcal{L}_n \defin \{(X_1, Y_1), \ldots, (X_n,Y_n)\}$, an empirical Fréchet mean given $\mathcal{L}_n$~is
$$
\begin{aligned}
\hat{y}_{\oplus}\defin \argmin_{y \in (\Y, \dY)} \frac{1}{n} \sum_{i=1}^n \dY\left(Y_i, y\right)^2
\end{aligned}
$$
and, in this case,
$$
\begin{aligned}
\widehat{V}_{\oplus}\defin \frac{1}{n} \sum_{i=1}^n \dY\left(Y_i, \hat{y}_{\oplus}\right)^2
\end{aligned}
$$
is the empirical Fréchet variance. The Fréchet regression function is defined by conditioning on the predictor in \eqref{frechet_mean_variance}:
\begin{align}\label{frechet+M}
   m(x)\defin \argmin_{y\in (\Y, \dY)} M_\oplus(x,y), \quad 
   M_\oplus(x,y) \defin  \mathsf{E}\left(\dY(Y, y)^{2} \mid X=x\right).
\end{align}
In the case of Euclidean response, then $m(x) = \E{Y \mid X = x}$.

The absence of algebraic structure can hinder the generalization of Euclidean parametric regression models to metric spaces. \citet{Petersen2019} proposed a parametric model, known as global Fréchet regression, as an extension of multiple linear regression. They considered a totally bounded metric space $\Y$ and Euclidean predictors $\bX \in \R^p$. Global Fréchet regression extends the linear regression model for Euclidean response, $m_{\mathrm{L}}(\bx) \defin \beta_0 + \bbe_1^\top(\bx-\bmu)$. Here, let $ \bmu = \E{ \bX }$ and $\bSigma =\Var{\bX}$, and define the linear weight function
$
\omega(\bx_1, \bx_2) \defin 1+(\bx_1-\bmu)^\top \bSigma^{-1}(\bx_2-\bmu),
$
$\bx_1, \, \bx_2 \in \R^p$. \citet{Petersen2019} showed that, when the response is Euclidean, the regression function is
$$
m_{\mathrm{L}}(\bx) = \argmin_{y \in \R} \Elr{\omega(\bx, \bX) d_2(Y, y)^2},
$$
with $d_2$ the standard Euclidean metric. Consequently, they proposed the global Fréchet regression model for $Y \in \Y $ by replacing $d_2$ with the corresponding metric $\dY$:
$$ 
m_{\mathrm{L}}(\bx) = \argmin_{y\in (\Y, \dY)} M(\bx, y), \quad M(\bx, y) \defin \Elr{\omega(\bx, \bX) \dY(Y, y)^2}.
$$
Using an independent and identically distributed (iid) sample $\mathcal{L}_n = \{(\bX_1,Y_1), \ldots, (\bX_n,Y_n)\} $, their proposed estimator of $M(\bx,y)$ is expressed as a weighted Fréchet functional, with weights depending on the value of $\bx$:
\begin{align}\label{fr_weighted}
M_n(\bx, y)\defin \sum_{i=1}^n \omega_{i}(\bx) \dY(Y_i, y)^2.
\end{align}
The empirical weights are defined as
$
n\omega_{i}(\bx)\defin 1+(\bX_i-\bar{\bX})^\top\widehat{\bSigma}^{-1}(\bx-\bar{\bX}),
$
with
$
\bar{\bX}\defin n^{-1} \sum_{i=1}^n \bX_i
$
and
$
\widehat{\bSigma}\defin n^{-1} \sum_{i=1}^n(\bX_i-\bar{\bX})(\bX_i-\bar{\bX})^\top.
$
\citet{Petersen2019} derived pointwise and uniform consistency and convergence rates of the global estimator under several assumptions, which include the existence and uniqueness of $m_\mathrm{L}(\bx)$ and its estimator.

Global Fréchet regression is a parametric model that relies on the assumption of linearity. An alternative nonparametric approach to estimate the conditional Fréchet function $M_\oplus(x,y)$ in \eqref{frechet+M} is through a locally-weighted scheme of the form \eqref{fr_weighted}. Within this framework, \citet{Hein2009} proposed the Nadaraya--Watson estimator, considering weights $\omega_i$ defined through a kernel function $K$ and a bandwidth parameter $h$; that is, $\omega_i(x) = h^{-p}K(d_\mathcal{X}(X_i, x)/h)$, where $p$ is the dimension of the metric space of predictors $(\mathcal{X}, d_\mathcal{X})$. Under certain regularity conditions and when the predictors and the response belong to compact Riemannian manifolds, the kernel estimator is consistent.

\citet{Petersen2019} extended their global estimator to local Fréchet regression, a nonparametric approach of the form \eqref{fr_weighted} that generalizes local linear regression \citep{Fan1996} for random objects and typically outperforms the Nadaraya--Watson estimator, especially near the boundaries of the support of the predictors. \citet{Petersen2019} derived the consistency of local Fréchet regression, and uniform rates of convergence were subsequently provided by \citet{Chen2022}.

\subsection{Random forests in metric spaces}\label{sec:RFs_metric_spaces}

Random forests \citep{Breiman2001} are among the most popular statistical learning techniques for prediction, and are particularly well-suited for extensions to metric spaces. Each decision tree begins with a root node containing all the training observations, which is recursively split into different nodes according to a splitting criterion until a stopping criterion is met. The nodes that are not split further are called terminal nodes. The prediction of the tree for a new observation $x$ is calculated as the average of training responses in the terminal node to which $x$ is assigned. The RF prediction is then computed as the average of the predictions from each tree of $x$. Two main adaptations are needed for RFs to apply to random objects: the splitting criterion and the aggregation of results to obtain predictions. 

For a node $A \subset \mathcal{L}_n$ and a fixed split variable $X^{(j)}$, $j=1,\ldots,p$, \citet{Capitaine2024}~proposed determining the left and right child nodes by the Voronoi partition of $A$ associated~with the points $\left(c_{j, \ell}, c_{j, r}\right) \in \mathcal{X}_j \times \mathcal{X}_j$. This partition divides $A$ based on distances from $c_{j, \ell}$ and~$c_{j, r}$:
\begin{align*}
A_{j, \ell} &\defin \left\{(x, y) \in A: d_{\mathcal{X}_j}(x^{(j)}, c_{j, \ell}) \leq d_{\mathcal{X}_j}(x^{(j)}, c_{j, r})\right\}, \\
A_{j, r} &\defin \left\{(x, y) \in A: d_{\mathcal{X}_j}(x^{(j)}, c_{j, r})<d_{\mathcal{X}_j}(x^{(j)}, c_{j,\ell})\right\}.
\end{align*}
Once the split is performed, the standard way to measure its quality is the CART (Classification And Regression Trees) criterion \citep{Breiman1984}. \citet{Capitaine2024} suggest an adaptation of this criterion for metric spaces, in terms of empirical Fréchet variances:
\begin{align}\label{cart}
    H_j\left(A, c_{j, \ell}, c_{j, r}\right)\defin \widehat{V}_{\oplus}(A)-\frac{\left|A_{j, \ell}\right|}{|A|} \widehat{V}_{\oplus}\left(A_{j, \ell}\right)-\frac{\left|A_{j, r}\right|}{|A|} \widehat{V}_{\oplus}\left(A_{j, r}\right),
\end{align}
where $|S|$ denotes the cardinal of the set $S$ and $\widehat{V}_{\oplus}(S)$ is the empirical Fréchet variance of the responses calculated on the samples $(x,y)$ such that $x \in S \subset \mathcal{X}$. 

For each variable $X^{(j)}$, in order to determine the splits $\left(c_{j, \ell}, c_{j, r}\right) $, a split function is defined beforehand, which assigns a pair $(c_{j, \ell}, c_{j, r}) \in \mathcal{X}_j \times \mathcal{X}_j $ to any sample of $\mathcal{X}_j$. \cite{Capitaine2024} proposed the $2$-means algorithm as a flexible and computationally efficient choice for the split function. Once a split function is defined, the best possible split at node $A$ is the pair $\left(c_{j^*,l}, c_{j^*,r}\right)$ that maximizes the decrease in Fréchet variance of the child nodes with respect to its parent node $A$, i.e., $j^* \defin \argmax _{j=1, \ldots, p} H_j\left(A, c_{j, \ell}, c_{j, r}\right)$.
In this way, observations are grouped into more homogeneous~subgroups.

\begin{remark}
The splits depend on the structure of the metric space of the predictors. For instance, if $\mathcal{X} = \R^p$, one could consider $(\mathcal{X}, d_\mathcal{X}) = (\R, d_2) \times \cdots \times (\R, d_2)$, but also $(\mathcal{X}, d_\mathcal{X}) = (\R^p, d_2)$. The first case connects with standard RFs, since the Fréchet mean and variance coincide with the mean and variance. In the second case, the data is split by a hyperplane $\{ \bx \in \R^p: d_2(\bx, \bc_l) = d_2(\bx,\bc_r)\}$, for a given pair $(\bc_l, \bc_r) \in \R^p \times \R^p$. This is conceptually related to oblique random forests \citep{Li2023}, where the split is given by the hyperplane that maximizes the decrease in variance.
\end{remark}

Trees based on the previous splitting are built in the same fashion as standard trees for Euclidean data. Let $\btheta$ represent the randomization parameter vector that governs the growth of a tree, determining which variables are considered for each split. Once a Fréchet tree $T$ is built with a set of terminal nodes $\widetilde{T}$, the process of obtaining a tree prediction $\widehat{m}_T(x, \btheta)$ for a new observation $x\in\mathcal{X}$ starts with dropping $x$ from the root node, so that it ends up in a certain terminal node $\tau_x \in \widetilde{T}$. Subsequently, $\widehat{m}_T(x, \btheta)$ is predicted as the Fréchet mean of training responses in the terminal node~$\tau_x$:
\begin{align}\label{tree}
\widehat{m}_T(x, \btheta) \defin  \argmin_{y\in (\Y, \dY)} \frac{1}{\left|\tau_x\right|} \sum_{i=1}^n \dY\left(Y_i, y\right)^2 1_{\left\{\left(X_i, Y_i\right) \in \tau_x\right\}}.
\end{align}

A similar idea allows the construction of RFs as an ensemble of decision trees in the metric case. This procedure, proposed by \citet{Capitaine2024}, receives the name of Fréchet Random Forests (FRFs), which consist of a fixed number $B \in \mathbb{N}$ of decision trees, each built with a random resample $\mathcal{L}_n^{* b}$ of $\mathcal{L}_n$, $b=1,\ldots, B$. Commonly, nonparametric bootstrap resamples of size $n$ with replacement are employed. In addition, the best split in each non-terminal node of each tree is determined using a subset of $\ell \leq p$ features selected uniformly at random. The bagging process of subsampling from the training set stabilizes the variance of the trees that constitute the forest. In addition, random feature selection decreases the correlation between trees, generating different trees even if they were trained on the same subsample. Let $\widehat{m}_{T_b}(x, \btheta_b)$ denote the prediction of the $b$-th tree for the point $x\in\mathcal{X}$ as in \eqref{tree}, where $\btheta_b$ is the randomization parameter. The prediction generated by the FRF at the point $x$ is defined as the Fréchet mean of the predictions of each of the trees:
$$
\widehat{m}_{\mathrm{FRF}}(x) \defin \argmin_{y\in (\Y, \dY)} \frac{1}{B} \sum_{b=1}^B \dY\left(\widehat{m}_{T_b}(x, \btheta_b), y\right)^2.
$$
FRFs are applicable when both the predictors and the response are random objects. However, \citet {Capitaine2024} do not provide a theoretical analysis of the consistency of FRFs. 

\citet{Qiu2024} refined and advanced the research conducted by \citet{Petersen2019} by developing a locally adaptive kernel for nonparametric Fréchet regression generated by Fréchet trees. Their approach, called Random Forest Weighted Local Constant Fréchet Regression (RFWLCFR), is equivalent to FRFs when $\Y = \R^q$, $q\in \mathbb{N}$, which can be seen from the property that the weighted sample mean minimizes the weighted sum of squared errors. Following the ideas of \citet{Meinshausen2006}, the tree-building procedure in equation \eqref{tree} naturally generates RF weights that can be used by an estimator of the form \eqref{fr_weighted}. Given a value $\bx$ of the predictors in $\mathcal{X} = [0,1]^p$, a sample $\mathcal{L}_n$, and a bounded metric space $(\Y, \dY)$, the $b$-th tree provides the weights
\begin{align}\label{weights_tree}
\omega^b_i(\bx, \btheta_b) \defin  \frac{1_{\left\{\left(\bX_i, Y_i\right) \in \tau^b_{\bx}\right\}}}{\left|\tau^b_{\bx}\right|},
\end{align}
where $\tau^b_{\bx}$ denotes the terminal node within the $b$-th tree where $\bx$ ends up. The weights corresponding to the individual trees are then averaged to obtain a single weight
\begin{align}\label{weights_rf}
    \omega_i(\bx) \defin  \frac{1}{B} \sum_{b=1}^B \omega_i^b(\bx, \btheta_b)
\end{align}
for each element of $\mathcal{L}_n$. These weights, based on Fréchet trees, are used by RFWLCFR to estimate the conditional Fréchet function $M_\oplus(\bx,y)$ in \eqref{frechet+M} using the weighted scheme in \eqref{fr_weighted}. Furthermore, \citet{Qiu2024} derived pointwise consistency and convergence rates based on the theory of infinite-order $U$-statistics and $U$-processes, under assumptions that include the existence and uniqueness of $m(\bx)$ and its estimator.

\citet{Bulte2024} performed a comprehensive analysis of RFs for their use in Fréchet regression, and proposed the Metric Random Forest (MRF), which applies a new splitting rule that alleviates the computational burden associated with calculating Fréchet means. Using the $2$-means algorithm as split function, the CART criterion \eqref{cart} requires calculating (or estimating) $2\ell$ Fréchet means per split (where $\ell$ is the number of variables considered at each split), yielding $O(n\ell)$ Fréchet mean computations for the whole tree. Since computing a single Fréchet mean can be costly, particularly with non-Euclidean data, \citet{Bulte2024} suggested replacing the empirical Fréchet mean with the Fréchet medoid estimator
\begin{align}\label{medoid}
    \tilde{y}_{\oplus}(S)\defin \argmin_{y \in \Y_n} \sum_{i=1}^n \dY\left(Y_i, y\right)^2 1_{\left\{ \bX_i \in S\right\}},
\end{align}
where $S \subset [0,1]^p$ and $\Y_n = \{Y_1, \ldots, Y_n\}$. In other words, \citet{Bulte2024} simplified the minimization problem needed to calculate the Fréchet mean by searching among the sample points, instead of the whole metric space. For this, obtaining the matrix of pairwise distances of $\mathcal{Y}_n$ is sufficient, a step that can be completed prior to model fitting. The Fréchet mean is replaced with Fréchet medoids within the CART splitting criterion \eqref{cart} in MRFs, building on RFWLCFR. The medoid approximation in \eqref{medoid} was shown to yield a consistent estimation of the Fréchet population mean under conditions related to the consistency of $M$-estimators. These assumptions include the existence and uniqueness of $m(\bx)$ for every $\bx$, and a compact metric space $\mathcal{Y}$.

\subsection{Out-of-bag errors}
\label{subsec:oob}

In the construction of RFs, certain observations are left out of each random resample $\mathcal{L}_n^{*b}$, and hence do not participate in the construction of the $b$-th tree. Such instances of the data receive the name of Out-Of-Bag (OOB) observations. That is, for the resample $\mathcal{L}_n^{*b}$ we say that $(X_i,Y_i)$ is OOB if $(X_i,Y_i) \in \mathcal{L}_n \setminus \mathcal{L}_n^{*b}$. We denote by $\widehat{Y}_{(i)}$ the OOB prediction of $Y_i$, which is based on the forest $\widehat{m}_{(i)}$ that only takes into account the trees in which $(X_i, Y_i)$ is OOB. Note that the forests $\widehat{m}_{(i)}$, $i=1,\ldots, n$ are produced automatically during training, and are conceptually different from the leave-one-out forest $\widehat{m}_{-i}$. This procedure can be applied beyond RFs, since OOB observations arise in any bagged model.

OOB predictions provide a reliable estimate of the predictive power of a bagged regression model, since, by construction, data leakage is prevented. In addition, OOB predictions come at no additional training cost, since the trees are already grown, and it is only required to select the correct trees for each individual in the sample. This is a massive advantage with respect to leave-one-out cross-validation predictions, requiring retraining.

We define the OOB (radial) errors as
\begin{align}\label{oob_errors}
    \Roob_i: = \dY(Y_i, \widehat{Y}_{(i)}).
\end{align}
These OOB errors are intended to estimate the actual prediction errors $\dY(Y_i, \widehat{Y}_i)$. Note that given a sample $\mathcal{L}_n$ of iid observations, the OOB prediction errors $\Roob_1, \ldots, \allowbreak \Roob_n$ are identically distributed but not independent. For any observation $(X,Y)$ independent of $\mathcal{L}_n$, the difference between the distributions of each $\Roob_i$ and $\dY(Y, \widehat{Y})$ lies in the number of observations and trees used to build each forest \citep{Zhang2020}.

We list the procedures to obtain an OOB prediction using the three RFs revised in this subsection. The procedures are simple, but the steps must be followed precisely to avoid any leakage of data that could lead $\widehat{Y}_{(i)}$ to be influenced by $(X_{i},Y_{i})$. Let $\mathcal{B}_i$ denote the set of resamples $\mathcal{L}_n^{*b}$, $b=1,\ldots, B$, for which $(X_{i}, Y_{i})$ is OOB:
\begin{itemize}
    \item In FRFs \citep{Capitaine2024}, the construction of an OOB prediction $\widehat{Y}_{(i)}$ is simple: select the trees built with resamples $\mathcal{L}_n^{*b} \in \mathcal{B}_{i}$ and aggregate their predictions,
    \begin{align*}
    \widehat{Y}_{(i)} = 
    \argmin_{y\in (\Y, \dY)} \sum_{\{b: \, \mathcal{L}_n^{*b} \in \mathcal{B}_i\}} \dY\left(\widehat{m}_{T_b}(X_i, \btheta_b), y\right)^2.
    \end{align*}

    \item In RFWLCFR \citep{Qiu2024}, begin by computing the weights that are generated by the prediction of $Y_{i}$ from each tree. For every $b\notin \mathcal{B}_i$, set $\omega_j^b(\bX_{i}, \btheta_b) = 0$ for $j=1,\ldots,n$ and aggregate the weights as $\omega_j(\bX_{i}) =   \frac{1}{| \mathcal{B}_{i} |} \sum_{b=1}^B \omega_j^b(\bX_{i}, \btheta_b)$. Finally, 
    \begin{align*}
    \widehat{Y}_{(i)} = \argmin_{y\in (\Y, \dY)} \sum_{j=1}^n \omega_{j}(\bX_i) \dY(Y_j, y)^2
    \end{align*}
    gives the OOB prediction of $Y_{i}$.

    \item In MRFs \citep{Bulte2024}, the medoids used in the split criterion \eqref{cart} of the trees  $T_b$, $b \in \mathcal{B}_{i}$
    are computed without $(X_i,Y_i)$ by construction. Thus, computing the weights $\omega_{j}$ as for RFWLCFR, and denoting $\Y_n^{(i)} = \Y_n \setminus \{Y_i\}$, we have that
    \begin{align*}
    \widehat{Y}_{(i)} = \argmin_{y \in \Y_n^{(i)}} \sum_{j=1}^n \omega_{j}(\bX_i) \dY(Y_j, y)^2.
    \end{align*}
\end{itemize}

In the following, we simply refer by RFs to either FRFs, RFWLCFR, or MRFs.

\section{Out-of-bag prediction balls}
\label{sec:setting}

We define OOB prediction balls in this section and introduce four scenarios for the analysis of probability coverage. The assumptions required for the asymptotic analysis of OOB balls are presented, followed by illustrative examples in diverse metric spaces.

\subsection{Definition and coverage types}

Our goal is to obtain estimates of the uncertainty in a RF prediction of the response. In other words, the goal is to develop confidence regions that cover the true value of the target variable with a specified probability. This will be achieved by generalizing the prediction confidence intervals proposed in \citet{Zhang2020} to responses in general metric spaces. 

The distribution of the prediction error $\dY(Y_i, \widehat{Y}_i)$ is fundamental to the construction of confidence regions for the true value of the response $Y_i$ based on the prediction  $\widehat{Y}_i\defin  \widehat{m}(X_i)$ from a RF $\widehat{m}$. To reliably estimate the distribution of $\dY(Y_i, \widehat{Y}_i)$, we consider the OOB errors \eqref{oob_errors}, introduced in Section \ref{subsec:oob}. Although the OOB and RF errors have different distributions, it will be shown through Lemmas~\ref{lem:FRhat} and \ref{lem:Deltan} that, under certain assumptions, the distribution of $\dY(Y, \widehat{Y})$ and the empirical distribution of $\Roob_1, \ldots, \Roob_n$ approach the population distribution function $F_R$ of the radial error $R\defin \dY(Y, m(X))$ as the sample size grows. This motivates quantifying the uncertainty in a RF prediction through the quantile of the empirical distribution of the OOB errors \citep{Zhang2020}. Given the non-Euclidean nature of the data, these confidence regions receive the name of \textit{OOB prediction balls}. 

\begin{definition}[Out-of-bag prediction balls]
The \emph{OOB prediction ball} for predictors $x\in \mathcal{X}$ with significance level $\alpha \in (0,1)$ is defined as
\begin{align}\label{PB}
\mathrm{PB}_{1-\alpha}^{\mathrm{oob}}\left(x, \mathcal{L}_{n}\right) \defin \left\{y \in \Y: \dY (\widehat{m}(x), y) < \widehat{R}_{[1-\alpha, n]} \right\},
\end{align}
where $ \widehat{R}_{[1-\alpha, n]}$ denotes the $(1 - \alpha)$-quantile of $F_{\Roob_1, \ldots, \Roob_n}$, the empirical distribution of $\Roob_1, \ldots, \allowbreak \Roob_n$.
\end{definition}
Note that $F_{\Roob_1, \ldots, \Roob_n}$ is an estimator of $F_R$, but is not the empirical distribution function of $R$, since the OOB errors are identically distributed but not independent, and their distribution differs from $F_R$.

Four scenarios will be analyzed in order to study the finite sample and asymptotic properties of OOB balls. For $\alpha \in (0,1)$, we consider the following probability coverage types:
\begin{itemize}
    \item Type I: $\Prob{Y \in \mathrm{PB}_{1-\alpha}^{\mathrm{oob}}\left(X, \mathcal{L}_{n}\right)}$.
    
    \item Type II: $\Condprob{Y \in \mathrm{PB}_{1-\alpha}^{\mathrm{oob}}\left(X, \mathcal{L}_{n}\right)}{\mathcal{L}_{n}}$.

    \item Type III: $\Condprob{Y \in \mathrm{PB}_{1-\alpha}^{\mathrm{oob}}\left(X, \mathcal{L}_{n}\right)}{X=x}$.

    \item Type IV: $\Condprob{Y \in \mathrm{PB}_{1-\alpha}^{\mathrm{oob}}\left(X, \mathcal{L}_{n} \right)}{\mathcal{L}_n, \, X=x}$.
\end{itemize}
Type I convergence is known as marginal coverage, which is studied by \citet{Lei2018} and \cite{Barber2021}. In this scenario, the data set $\mathcal{L}_n$ and the new observation $(X,Y) \sim \mathbb{G}$ are selected at random. This is interesting in simulation studies where we assume a model $\mathbb{G}$ for the data generation and want to analyze OOB balls across different random samples. In convergence Type II, a given data set is fixed, and the coverage rate for any random pair $(X,Y)$ given the data is under analysis. Type III convergence is mainly interesting for theoretical purposes, particularly when we simulate data from a model $\mathbb{G}$ and our interest is in analyzing the prediction for a specific value of $X=x$ across different samples $\mathcal{L}_n$. This is the convergence type studied in \citet{Zhou2025} and \citet{Meinshausen2006}. Type IV coverage is the case that is commonly found in data analysis: a training set $\mathcal{L}_n$ is given and the coverage rate of the response for a given value of the predictors $X=x$ is under study. 

\subsection{Assumptions}

We assume the following conditions in the study of asymptotic coverage rates of OOB balls:
\begin{enumerate}[label=(\textit{c.\arabic*}), ref=(\textit{c.\arabic*})]
    \item Let $(X,Y) \sim \mathbb{G}$ be a pair of random objects, for a law $\mathbb{G}$ defined on $(\mathcal{X}, d_\mathcal{X}) \times (\Y, \dY)$. The sample $\mathcal{L}_n = \{(X_1,Y_1),\ldots,(X_n,Y_n)\}$ consists of $n$ iid copies of $(X,Y)$, which are also independent of $(X,Y)$.\label{c1}
    \item The law $\mathbb{G}$ is such that:\label{c2}
    \begin{enumerate}[label=(\textit{c.2.\arabic*}), ref=(\textit{c.2.\arabic*})]
        \item For every $x\in \mathcal{X}$ $\mathbb{G}$-a.s, there exists a unique conditional Fréchet mean\label{c2.1}
        \begin{align}\label{frechet_regression}
            m(x)=\argmin_{y\in (\Y, \dY)} \CondEbig{\dY(Y, y)^{2}}{X=x}.
        \end{align}
        \item The error $R = \dY(Y,m(X))$ is independent from the predictor $X$.\label{c2.2}
        \item The cumulative distribution function of $R$, $F_R$, is continuous over $\R$.\label{c2.3}
    \end{enumerate}
    \item The RF and its OOB version are consistent estimators:\label{c3}
    \begin{enumerate}[label=(\textit{c.3.\arabic*}), ref=(\textit{c.3.\arabic*})]
        \item $\dY(m(X), \widehat{m}(X)) \convp 0$ as $n \rightarrow \infty$.\label{c3.1}
        \item $\dY(m\left(X_1\right), \widehat{m}_{(1)}\left(X_1\right))\convp 0$ as $n \rightarrow \infty$.\label{c3.2}
    \end{enumerate}
\end{enumerate}

Assumption \ref{c1} ensures that the OOB errors $\Roob_i$ (see \eqref{oob_errors}), $i=1,\ldots,n$, are identically distributed. This is required to prove the consistency of $\smash{F_{\Roob_1, \ldots, \Roob_n}}$ as an estimator of $F_R$ in Lemma~\ref{lem:FRhat} (notice that the dependence between $\smash{\Roob_i}$'s prevents the application of classical results like the Glivenko--Cantelli Theorem). Existence and uniqueness of the conditional Fréchet mean from assumption \ref{c2.1} is a prerequisite for a well-defined analysis of the consistency of the RF estimators. As stated in Section \ref{sec:FRF}, this assumption is typical in works regarding the consistency of estimators of the Fréchet regression function.

Continuity of $F_R$ in condition \ref{c2.3} is required to build a probabilistic version of Polya's Theorem in Lemma~\ref{lem:polyaprob}, and simplifies the proofs of Lemmas~\ref{lem:FRhat}, \ref{lem:Deltan}, and \ref{lem:Deltan_x} (see Section~\ref{supp_sec:proofs} in the SM). Finally, assumption \ref{c3} provides consistency of the estimators of the conditional Fréchet mean $m$. \cite{Qiu2024} and \cite{Bulte2024} provide sufficient conditions for the consistency of RFWLCFR and MRF, respectively.
Condition \ref{c3} is central in the analysis of the coverage rates, since \ref{c3.1} and \ref{c3.2} entail results of consistency for the RF and OOB residuals, respectively (see the proofs of Lemmas~\ref{lem:FRhat} and \ref{lem:Deltan}). Assumption \ref{c3.2} is related to the notion of stability of the RF, recently studied in \cite{Wang2023}. Due to the identical distributions of $\widehat{m}_{(1)} \left( X_1 \right), \ldots, \widehat{m}_{(n)}\left(X_n\right)$ by condition \ref{c1}, assumption \ref{c3.2} entails consistency of each $\widehat{m}_{(i)} \left( X_i \right)$ for $i=1,\ldots,n$. In fact, the consistency of $\widehat{m}$ is implied by the consistency of $\widehat{m}_{(1)}$, as it comprises a subset of the trees in the larger forest represented by $\widehat{m}$. Thus, condition \ref{c3.1} is included for clarity, although it is implied by \ref{c3.2}.

Assumptions \ref{c2.1} and \ref{c2.2} adapt additive error models for Euclidean data. The traditional regression function in our framework is the conditional Fréchet mean function, and the error $R$ is just implicitly defined through condition \ref{c2.2}. This is important in the construction of data generation processes satisfying the above conditions, since the introduction of random noise must be done ``symmetrically about $m(x)$'' so that the regression function $m(x)$ satisfies \eqref{frechet_regression} for $\mathbb{G}$-a.s. $x\in \mathcal{X}$. Furthermore, independence of $R$ from $X$ by condition \ref{c2.2} can be stronger than the typical assumption of homoscedasticity in Euclidean regression. Still, it should be satisfied by construction in most models if $X$ only affects the location of $Y$ (and not its dispersion around the mean $m(X)$). Condition \ref{c2.2} is needed to derive asymptotic coverage rates of Types III and IV, where the probability is conditional on the value of~$X$, and is inherited from \citet{Zhang2020}. 

\subsection{Illustrative cases}
\label{subsec:illustrative_cases}

We present different scenarios to exemplify conditions \ref{c1}--\ref{c2} and the form of the population OOB balls 
\begin{align}\label{PPB}
    \mathrm{PPB}_{1-\alpha}\left(x, \mathcal{L}_{n}\right) \defin \left\{y \in \Y: \dY (m(x), y) < R_{1-\alpha} \right\},
\end{align}
with $R_{1-\alpha}$ such that $\Prob{\dY(m(X), Y) < R_{1-\alpha}\mid X=x} = 1  - \alpha$. Section \ref{supp_subsec:illustration_oob_balls} in the SM gives illustrations of the population OOB balls in different scenarios.

\subsubsection{Euclidean and Hilbert spaces}

Consider the simple case where $\bY$ is a random vector on $(\R^q,d_2)$. The conditional Fréchet mean exists and is equal to the mean, hence \ref{c2.1} is fulfilled. Consider the additive error model $\bY=\bm_0(X)+\boldsymbol{\varepsilon}$, for some regression function $\bm_0$. The distribution of $\boldsymbol{\varepsilon} \mid X$ is decisive to ensure that $\bm_0$ is the conditional Fréchet mean and \ref{c2.2} holds:
\begin{itemize}
    \item If $\boldsymbol{\varepsilon} \mid X = x \sim \mathcal{N}_q(\boldsymbol{0}, \bSigma)$, then $\bY \mid X=x \sim \mathcal{N}_q(\bm_0(x), \bSigma)$. Hence, the conditional Fréchet mean is $\bm_0$. The population OOB balls are spheres centered at $\bm_0(x)$ with constant radius given by the ($1-\alpha)$-quantile of $\left(\sum_{i=1}^q \lambda_i P_i\right)^{1/2}$, where 
    $P_1, \ldots, P_q$ are mutually independent random variables distributed as $\chi_1^2$ and $\lambda_1, \ldots, \lambda_q$ are the eigenvalues of $\bSigma$. In particular, if $\bSigma=\bI_q$, where $\bI_q$ denotes the $q\times q$ identity matrix, then $R_{1-\alpha} = \chi_{q,1-\alpha}$, the $(1-\alpha)$-quantile of a $\chi_q$ distribution.

    \item In the case of heteroscedasticity, that is, $\boldsymbol{\varepsilon}\mid X=x \sim \mathcal{N}_q(\boldsymbol{0},\bSigma(x))$, the independence of the error $R=\|\boldsymbol{\varepsilon}\|_2$ from $X$ in \ref{c2.2} is not satisfied.

    \item Suppose $\boldsymbol{\varepsilon}\mid X=x \sim \mathcal{N}_q(f(x), \bSigma)$, where $f: \R^q \to \R^q$ is non-identically zero. Then \ref{c2.2} is also not satisfied. In this case $\bY\mid X=x \sim \mathcal{N}_q(f(x)+\bm_0(x), \bSigma)$ and $\bm_0$ is not the conditional Fréchet mean.
\end{itemize}

In a general Hilbert space $\mathcal{H}$, for any random variable $Y$ such that $\E{Y}<\infty$ and any $c \in \mathcal{H}$, $\E{\| Y - \E{Y} \|_\mathcal{H}^2} \le  \E{\| Y - c \|_\mathcal{H}^2}$, with equality holding only if $c=\E{Y}$. Thus, \ref{c2.1} holds. For a given regression function $m_0$, since the Fréchet and standard mean coincide, it is straightforward to construct regression models such that $m_0$ is the Fréchet mean by adding mean-zero error terms $\varepsilon$. For example, consider a function-on-scalar regression model $Y(\cdot) = m_0(X)(\cdot) + \varepsilon(\cdot)$ in $L^2[0,1]$, where $\E{\varepsilon(t)} = 0$ for every $t \in [0,1]$, and $m_0$ could be defined as $m_0(x)(t)=x \beta(t)$, $\beta\in L^2[0,1]$. 

\subsubsection{Sphere and hyperboloid}
\label{subsec:sphere_hyp}

Consider $\left(\mathbb{S}^q, d_{\mathbb{S}^q}\right)$, the unit sphere $\mathbb{S}^q\defin\left\{\bx \in \R^{q+1}: \, \| \bx \|_2 = 1 \right\}$ endowed with the geodesic distance induced by the Riemannian metric $d_{\mathbb{S}^q}(\xb, \by) = \arccos \left(\xb^\top \by \right)$. For probability distributions on $\mathbb{S}^q$, the existence of a Fréchet mean is guaranteed since $\by \mapsto \E{d_{\mathbb{S}^q}(\bY, \by)^2 \allowbreak}$ is continuous and $\mathbb{S}^q$ is compact. Uniqueness depends on the distribution of $\bY$. 

Suppose $\bY \mid X=x$ follows a von Mises--Fisher (vMF) distribution, $\bY \mid X=x \sim  \mathrm{vMF}(\bm_0(x), \kappa)$, where the vMF density is $f_{\mathrm{vMF}}(\by; \bmu, \kappa) \propto \exp{\left\{\kappa \bmu^\top \by \right\}}$, $\by, \bmu \in \mathbb{S}^q$. Then, for $\kappa>0$ and every $x$, there exists a unique Fréchet mean, which coincides with $\bm_0(x)$ \citep[][Section 4.3]{McCormack2023}. Furthermore, the von Mises--Fisher distribution is rotationally symmetric about the location parameter $\bm_0(x)$. Therefore, for every $x$, the distribution of $\bY^\top\bm_0(x)$ does not depend on $\bm_0(x)$, and hence \ref{c2.2} holds. A particular regression model is
\begin{align}\label{eq:great_circle}
\bm_0(\theta) = \left(\cos(\theta), \sin(\theta)\bmu \right)^\top, \quad \theta \in[0,2\pi) \text{ and } \bmu\in\mathbb{S}^{q-1},
\end{align}
which generates a great circle on $\mathbb{S}^q$ through the intersection of $\mathbb{S}^q$ with the two-dimensional subspace spanned by $\{\be_1,(0,\bmu^\top)^\top\}$. Note that when $\kappa=0$ the distribution is uniform on the sphere, and every point is a Fréchet mean. The population OOB balls when $\kappa>0$ are spherical caps centered at $\bm_0(x)$ (see Section \ref{supp_subsec:illustration_oob_balls} in the SM).

Hyperbolic space is the model space for negative curvature in differential geometry \citep{McCormack2023}. Consider the Minkowski pseudo-inner product $(\bx, \by)\defin -x_{1} y_{1}+\sum_{i=2}^{q+1} x_i y_i$. The unit hyperboloid $\mathbb{H}^q\defin\left\{\bx \in \R^{q+1}:(\bx, \bx)=-1,\, x_{1}>0\right\}$ is closely related to $\mathbb{S}^q$. In this surface, distances are calculated as $d_{\mathbb{H}^q}(\bx, \by)=\arccosh \allowbreak\left( -(\bx, \by)\right)$. The analogous distribution to the von Mises--Fisher on the hyperboloid $\mathbb{H}^q$ (denoted by HvMF) has density $f_\mathrm{HvMF}(\by ; \bmu, \kappa) \propto\allowbreak \exp \left\{\kappa(\by, \bmu)\right\}$, for $\by, \bmu \in \mathbb{H}^q$, and $\kappa>0$, with respect to the Lebesgue measure on $\mathbb{H}^q$ \citep{Barndorff-Nielsen1978,Jensen1981}. Section \ref{supp_sec:hvmf} in the SM details the explicit connection of the HvMF with the vMF density. \citet[][Section 4.3]{McCormack2023} showed that the Fréchet mean of the distribution $\mathrm{HvMF}(\bmu, \kappa)$ is $\bmu$. Hence, a regression model similar to that on $\mathbb{S}^q$ is $\bY \mid X = x \sim \mathrm{HvMF}(\bm_0(x), \kappa)$. For example,
\begin{align}\label{eq:great_circle_H2}
\bm_0(\theta) = \left(\cosh(\theta),   \sinh(\theta)\bmu\right)^\top, \quad \theta \in \R \text{ and } \bmu\in\mathbb{S}^{q-1},
\end{align}
which for $q=2$ corresponds to a meridian on the hyperboloid. The population OOB balls are the intersection of $\mathbb{H}^q$ with the half-space $L^{-} = \{ \by \in \R^{q+1}: \by^\top \bz < \cosh(R_{1-\alpha})\}$, where $\bz = (m_1, -m_2, \ldots, -m_{q+1})$ and here we denote $\bm = \bm_0(\theta)$ for simplicity. Observe that the subspace $L = \{ \by \in \R^{q+1}: \by^\top \bz = \cosh(R_{1-\alpha}) \}$ is the tangent hyperplane to $\mathbb{H}^q$ at $\bm$, shifted $\|\bz\|^{-1} |\bm^\top \bz - \cosh(R_{1-\alpha})|$ units in the direction $\bz$. Section \ref{supp_subsec:illustration_oob_balls} in the SM shows a population prediction ball on~$\mathbb{H}^2$.

\subsubsection{Positive definite matrices}
\label{subsec:SPD}

Consider $\mathcal{S}_q^+$, the space of Symmetric Positive Definite (SPD) matrices with real entries of size $q\times q$. The problem of equipping $\mathcal{S}_q^+$ with a proper (and computationally efficient) metric has drawn research interest in recent years \citep{Minh2018}. A standard choice in statistical applications is the Affine-Invariant (AI) distance \citep{Pennec2006}
$$
d_{\mathrm{AI}}(\bS_1, \bS_2)^2 \defin \| \log\big( \bS_1^{-1/2} \bS_2 \bS_1^{-1/2} \big) \|^2_\mathrm{F} = \sum_{i=1}^{q} \log(\lambda_i)^2,
$$
where $\|\cdot\|_\mathrm{F}$ denotes the Frobenius norm, $\log(\cdot)$ stands for the matrix logarithm, and the $\lambda_i$'s are the eigenvalues of $\bS_1^{-1} \bS_2$. This metric endows $\mathcal{S}_q^+$ with a structure of a Riemannian symmetric space, invariant under affine transformations of the variables \citep[see, e.g.,][]{Thanwerdas2019}. Expressing $d_{\mathrm{AI}}$ in terms of the eigenvalues of $\bS_1^{-1} \bS_2$ enables faster computation. \citet{McCormack2023} gave closed formulae for the Fréchet mean of invariant families of distributions on $\mathcal{S}_q^+$. One example is the Wishart distribution. Suppose $\bS \sim \mathrm{Wishart}_q(d, \bSigma)$, for $\bSigma \in \mathcal{S}_q^+$. Then the Fréchet mean is 
\begin{align}\label{eq:ai_frechet_mean}
    \bS_\oplus = c_{d,q}\bSigma, \quad \text{where} \quad c_{d,q}\defin 2 \exp\left\{\frac 1q \sum_{i=1}^q \psi\left(\frac{d-i+1}{2}\right)\right\}
\end{align}
and $\psi$ is the digamma function. 

Another possible metric is the Log-Cholesky (LC) distance, introduced by \citet{Lin2019}. Given two matrices $\bS_1$, $\bS_2 \in \mathcal{S}_q^+$, let $\bR_1$, $\bR_2$ denote their respective Cholesky factors. Let also  $\lfloor\bS\rfloor$ represent the strictly lower triangular matrix of $\bS$, $\mathrm{diag}(\bS)$ the vector with the diagonal of $\bS$, and $\log(\cdot)$ be applied entrywise. The LC distance between $\bS_1$ and $\bS_2$ is
$$
d_{\mathrm{LC}}\left(\bS_1, \bS_2\right)^2
\defin 
\|\lfloor\bR_1\rfloor-\lfloor\bR_2\rfloor\|_\mathrm{F}^2+\|\log \left(\mathrm{diag}(\bR_1)\right)-\log \left(\mathrm{diag}(\bR_2) \right) \|_2^2.
$$
The Fréchet mean can be calculated under this distance by projection to a Hilbert space. For a given matrix $\bS \in \mathcal{S}_q^+$, let $\bR$ denote its Cholesky factor. Define $\phi: \mathcal{S}_q^+ \to \R^{q(q+1)/2}$ as $\phi(\bS) \defin (\bx, \by)$, where $\bx \in \R^{q}$ contains the logarithm of the elements of the diagonal of $\bR$ and  $\by \in \R^{q(q-1)/2}$ are the (strictly) lower triangular entries of $\bR$. Then $\phi$ is an isometry, since for any pair $\bS_1, \bS_2\in\mathcal{S}_q^+$, $d_{\mathrm{LC}} ( \bS_1, \bS_2 ) = d_2(\phi(\bS_1), \phi(\bS_2) )$, with $d_2$ the Euclidean distance in $\R^{q(q+1)/2}$. Thus, it is immediate to estimate the Fréchet mean of a random matrix $\bS \in \mathcal{S}_q^+$ by estimating $\E{\phi(\bS)}= \E{\bX, \bY}$ through the sample mean. Moreover, the Fréchet mean of a Wishart-distributed random matrix can be derived using~$\phi$, as shown~next.
\begin{proposition}\label{prop:frechet_mean_lc}
    Let $\bS \sim \mathrm{Wishart}_q(d, \bSigma)$, for $d \in \mathbb{Z}^+$ and $\bSigma \in \mathcal{S}_q^+$. Let $\boldsymbol{L}$ be the Cholesky factor of $\bSigma = \boldsymbol{L} \boldsymbol{L}^\top$, with elements denoted by $L_{ij}$. The Fréchet mean of $\bS$ under the log-Cholesky distance is given by $\bS_\oplus = \bT \bT^\top$, where the elements $T_{ij}$ of the lower triangular matrix $\bT$ are given by
    $$
    T_{ii} \defin L_{ii} \sqrt{2} \exp\left\{\frac{1}{2} \psi \left( \frac{d-i+1}{2}\right)\right\}, \quad T_{ij} \defin L_{ij} \sqrt{2} \frac{\Gamma((d-j+2)/2)}{\Gamma((d-j+1)/2)}, \quad 1\le j<i\le q.
    $$
\end{proposition}
\begin{proof}
Let $\bR$ be the Cholesky factor of $\bS = \bR \bR^\top$. Bartlett's decomposition for a $\mathrm{Wishart}_q(d, \bSigma)$ distribution entails that 
$$
\E{\log(R_{ii})} = \log(L_{ii}) + \frac{1}{2}\left(\log (2) + \psi\left(\frac{d-i+1}{2}\right)\right), \quad \E{R_{ij}} = L_{ij} \sqrt{2} \frac{\Gamma((d-j+2)/2)}{\Gamma((d-j+1)/2)},
$$
for $j<i$, which follows from the expectations of a $\chi_{d-j+1}$ distribution and of the logarithm of a $\chi^2_{d-i+1}$ distribution \citep{Pav2015}. Defining $\phi: \mathcal{S}_q^+ \to \R^{q(q+1)/2}$ as $\phi(\bS) = (\bx, \by)$, the Fréchet mean under the LC distance is given by
\begin{align}\label{eq:lc_frechet_mean}
    \phi^{-1}(\E{\phi(\bS)}) = \bT \bT^\top,
\end{align}
where $\bT$ is a lower-triangular matrix defined by
$$
T_{ii} \defin \exp\left\{\E{\log(R_{ii})}\right\}, \quad T_{ij} \defin \E{R_{ij}}, \ j<i.
$$
This concludes the result.
\end{proof}

A third distance is the Log-Euclidean (LE) metric
$$
d_{\mathrm{LE}}\left(\bS_1, \bS_2\right) \defin  
\| \log(\bS_1) - \log(\bS_2) \|_\mathrm{F}.
$$
This metric is an approximation of $d_{\mathrm{AI}}(\bS_1, \bS_2)$ when $\bS_1$ and $\bS_2$ are sufficiently close to $\bI_q$ (see, e.g., Theorem~2.11 in Section~2.5 of \cite{Minh2018}). 

A possible regression model can be constructed by a convex interpolation of covariance matrices. Let $\bSigma_1, \bSigma_2 \in \mathcal{S}_q^{+}$, and consider a weight function $w:[0,1]\to[0,1]$. We can write $\bM_0(x) = w(x) \bSigma_1+(1-w(x)) \bSigma_2$ for $x\in [0,1]$, and consider $\bS \mid X=x \sim \mathrm{Wishart}_q(d, c_{d,q}^{-1} \bM_0(x))$, so that $\bM_0$ is the conditional Fréchet mean. For the AI distance, $R=d_{\mathrm{AI}}(\bM_0(X), \bS)$ is independent of $X$. To see this, observe that from the properties of the Wishart distribution, it holds that $\bM_0(X)^{-1/2} \bS \bM_0(X)^{-1/2} \sim \mathrm{Wishart}_q(d, c_{d,q} \bI_q)$. Hence \ref{c2.2} is satisfied. For the LC and LE distances, $\bS \mid X=x$ must follow an alternative distribution to ensure that $\bM_0(x)$ is the conditional Fréchet mean. The task of finding this distribution is less tractable for the LE distance, as the Fréchet mean of a Wishart-distributed matrix remains unknown under this metric. For the LC distance, a specific regression model can be proposed using Proposition~\ref{prop:frechet_mean_lc}. Specifically, consider $\bS \mid X=x \sim \mathrm{Wishart}_q(d, \boldsymbol{L} \boldsymbol{L}^\top)$, where $\boldsymbol{L}$ is a lower-triangular matrix such that 
$$
L_{ii} = \frac{T_{ii}(x) \exp\left\{-\frac{1}{2}\psi \left( \frac{d-i+1}{2}\right)\right\} }{\sqrt{2}} \quad L_{ij} = \frac{T_{ij}(x)\Gamma((d-j+1)/2)}{\sqrt{2}\Gamma((d-j+2)/2)}, \quad 1\le j<i\le q,
$$
where $\bT = (T_{ij}(x))$ is the Cholesky factor of $\bM_0(x)$, which is the conditional Fréchet mean given $x$. However, under this model specification, the error $R=d_{\mathrm{LC}}(\bM_0(X), \bS)$ does depend on the predictor $X$.

Section \ref{supp_subsec:illustration_oob_balls} in the SM illustrates the population OOB balls in $\bS \in \mathcal{S}_q^+$ for different distances. Observe that the AI and LE distances provide sensible OOB balls localized at the shape of the central matrix.

\subsubsection{Probability distributions on the real line}

Consider $\mathcal{W}_2(\R)$, the $2$-Wasserstein metric space of probability distributions over $\R$ that have finite second moment. This space is isomorphic to the subset of $L^2[0,1]$ of quantile functions, which motivates the expression of the $2$-Wasserstein metric $d_{\mathcal{W}_2}$ between two probability distributions $\mathrm{P}$ and $\mathrm{Q}$ of $\mathcal{W}_2(\R)$ through their quantile functions $F_{\mathrm{P}}^{-1}$ and $F_{\mathrm{Q}}^{-1}$,
$$
d_{\mathcal{W}_2}(\mathrm{P}, \mathrm{Q}) \defin \sqrt{\int_0^1\big|F_{\mathrm{P}}^{-1}(u)-F_{\mathrm{Q}}^{-1}(u)\big|^2 \mathrm{~d} u}.
$$
The same idea characterizes the Fréchet mean of $\mathrm{P} \in \mathcal{W}_2(\R)$ as the unique measure $\mathrm{P}_\oplus \in \mathcal{W}_2(\R)$ whose quantile function satisfies $F_{\mathrm{P}_\oplus}^{-1} =\E{F_\mathrm{P}^{-1}} $ \citep[see, e.g., Theorem~3.2.11 of][]{Panaretos2020}, thus ensuring \ref{c2.1}. 

The definition of regression models with response on $W_2(\R)$ is relatively straightforward. For a given probability distribution $\mathrm{Y} \in \mathcal{W}_2(\R)$, consider a slight modification of the regression model in \citet[Section 6.2]{Petersen2019}:
$$
m_0(x)(\cdot)=\E{F_\mathrm{Y}^{-1}(\cdot) \mid X=x}=\gamma_0+ f(x) + \left(\sigma_0+ g(x) \right) F_\mu^{-1}(\cdot),
$$
for some $\mu \in \mathcal{W}_2(\R)$ and some finite real functions $f$ and $g$, under the restriction that $\sigma_0 + g(x)>0$ for all $x$ in the support of $X$, so that $m(x)(\cdot)$ is a valid quantile function in $L^2[0,1]$. In order to generate the response, consider $F_\mathrm{Y}^{-1}(\cdot)=\gamma+ f(X) + (\sigma + g(X)) F_\mu^{-1}(\cdot)$, with $\gamma$ and $\sigma$ random variables independent of $X$ and following distributions of mean $\gamma_0$ and $\sigma_0$, respectively. See \cite{Serrano2025} for a detailed numerical example with this regression model. Thus, this general model satisfies \ref{c2.2}, and the conditional Fréchet mean is $m_0(x)(\cdot)$. The population OOB balls in this setting are formed by the distributions whose quantile functions are within a constant $L_2$-distance on $[0,1]$ equal to $R_{1-\alpha}$. If we assume, for simplicity, that $F_\mu^{-1}(\cdot)$ is the quantile of a distribution with mean zero and unit variance, then $R_{1-\alpha}$ is the $(1-\alpha)$-quantile of $((\gamma-\gamma_0)^2 + (\sigma-\sigma_0)^2 )^{1/2}$.

\section{Asymptotic results}
\label{sec:main_results}

The proofs of the types of asymptotic coverages for OOB balls are organized in the following way. Theorem~\ref{thm:typeII} deals with Type II convergence, which is the case of conditional coverage for a given data set $\mathcal{L}_n$. The asymptotic marginal coverage of Type I is then derived as a corollary of Theorem~\ref{thm:typeII}, by averaging over the distribution of the data sets $\mathcal{L}_n$ in Corollary~\ref{cor:typeI}. Theorem~\ref{thm:typeIV} addresses Type IV convergence, conditional on $\mathcal{L}_n$ and a particular value of the predictor $X=x$. Finally, the asymptotic Type III coverage is derived from Theorem~\ref{thm:typeIV} in Corollary~\ref{cor:typeIII} by averaging over all possible data sets. 

The proofs are based on several lemmas. All the results are proved in Section~\ref{supp_sec:proofs} of the SM.

\begin{theorem}\label{thm:typeII}
    Under conditions \ref{c1}, \ref{c2.1}, \ref{c2.3}, and \ref{c3}, the OOB ball has asymptotically correct conditional coverage rate given $\mathcal{L}_n$ for any significance level $\alpha\in(0,1)$; that is, as $n$ diverges to infinity
    $$
   \Condprob{Y\in \mathrm{PB}_{1-\alpha}^{\mathrm{oob}}(X,\mathcal{L}_n)}{\mathcal{L}_n} \convp 1-\alpha.
    $$
\end{theorem}

\begin{corollary}\label{cor:typeI}
    Under the same conditions established for Theorem~\ref{thm:typeII},
\begin{align*}
\Prob{Y \in \mathrm{PB}_{1-\alpha}^{\mathrm{oob}}\left(X, \mathcal{L}_{n}\right)}\to 1-\alpha,    
\end{align*}
as $n$ diverges to infinity.
\end{corollary}

\begin{theorem}\label{thm:typeIV}
Assume conditions \ref{c1}--\ref{c3} are satisfied and let $x \in \mathcal{X}$ be a fixed value of the predictor such that $\widehat{m}(x) \convp m(x)$ as $n \rightarrow \infty$. Then, for any $\alpha \in (0,1)$, it holds that
$$
\Condprob{Y \in \mathrm{PB}_{1-\alpha}^{\mathrm{oob}}\left(X, \mathcal{L}_n\right)}{\mathcal{L}_n,\, X=x} \convp 1-\alpha
$$
as $n$ diverges to infinity.
\end{theorem}

\begin{corollary}\label{cor:typeIII}
Under the same conditions of Theorem~\ref{thm:typeIV}, for every $\alpha \in (0,1)$, we have
$$
\Condprob{Y \in \mathrm{PB}_{1-\alpha}^{\mathrm{oob}}\left(X, \mathcal{L}_n\right)}{X=x} \to 1-\alpha
$$  
as $n$ diverges to infinity.
\end{corollary}

\section{Numerical experiments}
\label{sec:experiments}

In this section, we assess the performance of OOB balls through simulations with responses in the Euclidean space $\R^q$, on the unit sphere $\mathbb{S}^2$ and unit hyperboloid $\mathbb{H}^2$, and in the space of SPD matrices $\mathcal{S}_2^+$. In $\mathbb{R}^q$, OOB balls are compared with SC balls for metric data \citep{Lugosi2024} in terms of coverage (Types I--IV), radius, and computational time.

To assess coverage for Types I--IV, each probability is estimated with $M=1000$ Monte Carlo samples. In the case of Type I, each probability is estimated as follows:
\begin{align}\label{type_I_estimation}
    p \defin \Prob{Y \in \mathrm{PB}_{1-\alpha}^{\mathrm{oob}}(X, \mathcal{L}_n) }  \approx \frac 1M \sum_{j=1}^M 1_{\left\{Y_j \in \mathrm{PB}_{1-\alpha}^{\mathrm{oob}} \big( X_j, \mathcal{L}_n^{(j)} \big)\right\}}
\end{align}
for $\big((X_j, Y_j), \mathcal{L}_n^{(j)}\big)$ such that, for every $j=1,\ldots,M$, $(X_j, Y_j)$ is independent of $\mathcal{L}_n^{(j)}$ and $(X_j, Y_j) \sim \mathbb{G}$, and $\mathcal{L}_n^{(j)} \sim \mathbb{G}^n$. For Type IV, the probabilities are estimated as
\begin{align}\label{type_IV_estimation}
    p_{x,\mathcal{L}_n}\defin\Probbig{Y \in \mathrm{PB}_{1-\alpha}^{\mathrm{oob}}(X, \mathcal{L}_n) \mid X = x, \mathcal{L}_n }  \approx \frac 1M \sum_{j=1}^M 1_{\left\{Y_j \in \mathrm{PB}_{1-\alpha}^{\mathrm{oob}} (x, \mathcal{L}_n)\right\}}
\end{align}
for $Y_j\mid X=x$ induced by $\mathbb{G}$ and a fixed $x \in \mathcal{X}$. The estimation of probabilities for Types II and III is analogous. 
Observe that for Type I (similarly for Type III), each probability estimation using \eqref{type_I_estimation} requires training $M$ RFs, one on each sample $\mathcal{L}_n^{(j)}$, $j=1,\ldots,M$. We tried to measure the variability of the estimated Type I probabilities, without training more RFs beyond the initial $M$ fits needed to obtain the estimation of Type I probability using \eqref{type_I_estimation} (analogously for Type III). To achieve this, a bootstrap procedure was employed to obtain $K=500$ probability estimations $\{\hat{p}_1, \ldots, \hat{p}_K\}$ (Type I). The standard deviation of the estimated probabilities was calculated as follows:
\begin{align}\label{sd_boot}
  \widehat{\sigma}^2 = \frac{1}{K - 1} \sum_{k=1}^K \left(\hat{p}_k - \overline{p} \right)^2, \quad \text{where } \overline{p} = \frac{1}{K} \sum_{k=1}^K \hat{p}_{k}.  
\end{align}
To estimate each probability $\hat{p}_k$ we used \eqref{type_I_estimation}, resampling $M$ training samples $\mathcal{L}_n^{(j)}$ and $M$ test pairs $(X_j, Y_j)$ independently from $\{ \mathcal{L}_n^{(1)}, \ldots, \mathcal{L}_n^{(M)} \}$ and $\{(X_1, Y_1), \ldots, (X_M, Y_M)\}$, respectively. The standard deviation for Type III was estimated analogously to Type I.

In the simulations, we explored combinations of sample size $n \in \{50, 100, \allowbreak 200, 500\}$ and dispersion specification, generating $N=1000$ data sets for each combination. For Type II (analogously for Type IV), on each data set, one probability can be estimated using \eqref{type_IV_estimation} at the cost of training only one RF, yielding $N$ estimates $\{\hat{p}_{\mathcal{L}^{(1)}_n},\ldots,\hat{p}_{\mathcal{L}^{(N)}_n}\}$ (Type II) and $\{\hat{p}_{x,\mathcal{L}^{(1)}_n},\ldots,\hat{p}_{x,\mathcal{L}^{(N)}_n}\}$ (Type IV).

Each tree was trained with a size $n$ subsample generated with nonparametric bootstrap with replacement. We used $B=200$ trees, so that the number of trees in each OOB prediction is closer to the default parameter $B=100$ in \texttt{scikit-learn} \citep{Pedregosa2011} and works related to RFs in metric spaces \citep{Bulte2024, Qiu2024}, without significantly increasing the computational cost of fitting and predicting. We investigated the effect of increasing $B$ on Type I coverage results, for OOB and SC balls, and found no significant improvement in coverage when the number of trees exceeded $B=200$. 

The CART criterion was selected as impurity measure, and the split method used was the $2$-means algorithm. The minimal split size, which determines the minimal number of samples allowed in the leaf nodes, was tuned from $\{1, 5, 10\}$ using grid-search $5$-fold cross-validation with the MSE as error measure. Using the same cross-validation procedure, the number of random candidate variables to consider at each split was tuned from $\{1,2, \ldots, p\}$. This parameter is denoted by \texttt{mtry} in the \texttt{pyfréchet} package and in the most popular R programming language implementations \citep{Liaw2002, Wright2017}. Three different values were considered for the significance level $\alpha$: $0.01, 0.05$, and $0.10$. 

\subsection{Euclidean space}
\label{subsec:numerical_euclidean}

We analyzed a multiple linear model using the Euclidean distance to compare the performance of OOB and SC balls. Specifically, we considered
\begin{align} \label{eq:mult_linear_model}
    Y = X_1 - X_2 + X_3 +\varepsilon, \quad \text{with } X_{i} = 2\sqrt{5}(W_{i} - 1/2),
\end{align}
where for $i=1,2,3$ the random variables $W_{i}$ are independent, $W_i \sim \mathrm{Beta}(2,2)$. The error term $\varepsilon$ follows a $\mathcal{N}(0, \sigma^2)$ distribution, where $\sigma = \frac{\sqrt{3}}{2}$, $\sqrt{3}$ so that the population coefficient~of determination $R^2$ was equal to $0.8$ and $0.5$, respectively. Each RF was trained using the RF regressor implemented in \texttt{scikit-learn} with default parameters, except for \texttt{mtry} and~the minimal split size, which were tuned under the CV procedure explained previously in this section. The structure of the metric space of predictors was $(\R,d_2) \times (\R,d_2) \times (\R,d_2)$~($p=3$).

For each scenario and prediction region, we estimated the probability of Type I using \eqref{type_I_estimation} and the standard deviation of the estimated probabilities using \eqref{sd_boot}, and gathered the results in Table \ref{tab:euc_typeI_PB_SC}. OOB balls presented mean coverages closer to the nominal level, and reduced standard deviations. The superior performance of OOB balls is primarily explained by the higher predictive capability of the RFs that generate them, compared to SC balls. This is highlighted in the MSE analysis presented in Table \ref{tab:mse_PB_SC}, where it is shown that OOB balls require half the sample size that SC balls need to achieve a given MSE.

\begin{table}[H]
\setlength{\tabcolsep}{3pt}
\centering 
\begin{tabular}{llcccc}
\toprule
$\sigma$ & Method & $n=50$ & $n=100$ & $n=200$ & $n=500$\\
\midrule
\multirow[c]{2}{*}{$\frac{\sqrt{3}}{2}$} & Out-of-bag & 1.36 (0.17) & 1.15 (0.11) & 1.02 (0.07) & 0.92 (0.05)
\\
& Split-conformal & 1.73 (0.34) & 1.38 (0.18) & 1.15 (0.10) & 1.00 (0.06) \\
\multirow[c]{2}{*}{$\sqrt{3}$}
 & Out-of-bag & 4.02 (0.36) & 3.69 (0.25) & 3.48 (0.19) & 3.29 (0.16) 
\\
& Split-conformal & 4.52 (0.63) & 4.01 (0.39) & 3.69 (0.24) & 3.43 (0.18) \\
\bottomrule
\end{tabular}
\caption{\small For each sample $\mathcal{L}_n^{(j)}$, $j=1,\ldots,N$, the MSE was computed using RF prediction errors from $1000$ new observations drawn from the multiple linear model \eqref{eq:mult_linear_model}. The reported values are the sample mean and sample standard deviation of the $N$ computed MSEs.}
\label{tab:mse_PB_SC}
\end{table}

\begin{table}[H]
\setlength{\tabcolsep}{5pt}
\centering
\begin{tabular}{llccc|ccc}
\toprule
 &  & \multicolumn{3}{c}{Out-of-bag} & \multicolumn{3}{c}{Split-conformal} \\
 \cmidrule(lr){3-5} \cmidrule(lr){6-8}
  $\sigma$ & $n$ & $\alpha=0.01$ & $\alpha=0.05$ & $\alpha=0.10$ & $\alpha=0.01$ & $\alpha=0.05$ & $\alpha=0.10$ \\
\midrule
\multirow[c]{4}{*}{$\frac{\sqrt{3}}{2}$} & 50 & 98.3 (0.39) & 94.7 (0.71) & 89.4 (1.02) & \underline{96.1} (0.58) & 93.6 (0.80) & 90.0 (0.98)
\\
 & 100 & \underline{98.1} (0.43) & 94.0 (0.76) & 89.8 (0.97) & \underline{97.7} (0.45) & 93.8 (0.73) & 88.7 (0.97) \\
 & 200 & 98.2 (0.40) & \underline{93.1} (0.78) & \underline{87.5} (1.04) & \underline{97.7} (0.46) & 93.6 (0.77) & 88.4 (1.07) \\
 & 500 & 98.8 (0.34) & 95.0 (0.65) & 90.4 (0.92) & 98.7 (0.35) & 95.4 (0.70) & 89.4 (0.98) \\
\cline{1-8}
\multirow[c]{4}{*}{$\sqrt{3}$} & 50 & \underline{97.4} (0.47) & \underline{93.3} (0.79) & \underline{87.6} (1.02) & \underline{96.0} (0.60) & \underline{92.5} (0.85) & 89.3 (1.03) \\
 & 100 & 98.2 (0.43) & 94.3 (0.75) & 88.2 (1.00) & \underline{97.9} (0.41) & 95.5 (0.72) & 88.8 (1.01) \\
 & 200 & 98.6 (0.37) & 94.5 (0.68) & 90.1 (0.93) & 98.4 (0.41) & 94.6 (0.71) & 88.8 (0.95) \\
 & 500 & 99.1 (0.37) & 95.0 (0.74) & 90.1 (0.90) & 98.7 (0.37) & 94.8 (0.73) & 89.7 (0.95) \\
\bottomrule
\end{tabular}
\caption{\small Comparison of the estimated Type I probability (in $\%$) for OOB and SC balls using \eqref{type_I_estimation}, and the standard deviation of $K$ probability estimations calculated with \eqref{sd_boot}. The significance levels considered were $\alpha=0.01$, $0.05$, and $0.10$. The values that fall outside the corresponding Wilson $95\%$ confidence interval are underlined.}
\label{tab:euc_typeI_PB_SC}
\end{table}

Larger prediction errors lead to larger radii of the confidence regions, and the impact of an increase in the radius on the volume of the confidence regions becomes more pronounced as the dimensionality of the response increases. To investigate this situation, we considered three different response dimensions: $q\in \{1, 5, 10\}$, and generated $N$ samples for each combination of $q$ and $n$, according to the following multivariate linear model:
\begin{align}\label{eq:multidim_euc_model}
    \mathbf{Y} = \mathbf{X} \mathbf{B} + \boldsymbol{\varepsilon}.
\end{align}
Let $1 \le i\le n$, $1 \le j \le 3$ and $1 \le k \le q$. Each predictor $X_{ij}$ was generated as $X_{ij} = 2\sqrt{5}(W_{ij} - 1/2)$, where $W_{ij}$ are mutually independent and follow a $\mathrm{Beta}(2,2)$ distribution. The coefficients matrix $\mathbf{B}$ has elements $\beta_{jk} = \sqrt{j} \sin\big(\frac{k \pi}{q+1}\big)$. The error matrix $\boldsymbol{\varepsilon}$ is an $n \times q$ matrix, with each row distributed according to a $\mathcal{N}_q(\boldsymbol{0}, \bSigma)$ distribution, where $\bSigma$ is an AR(1) covariance matrix with first row $(1, 0.75, \ldots, 0.75^q)$. For each value of $q$, we summarized the relative errors in radius and volume of SC with respect to OOB balls in Figure \ref{fig:radius_volume_dimension}. The results show positive relative errors in the radii, especially for small sample sizes. This is a consequence of the larger prediction errors in the SC method caused by halving the sample size. The relative error of the radii causes an exponential increase in the relative error of the volume as the dimensionality of the response grows.

\begin{figure}[H]
    \centering
    \begin{minipage}{0.48\textwidth}
        \centering
        \includegraphics[height=5cm]{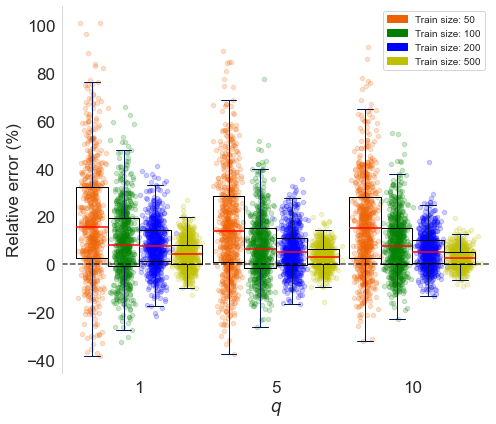}  
        \subcaption{Radius}\label{fig:radius_dimension}
    \end{minipage}
    \begin{minipage}{0.48\textwidth}
        \centering
        \includegraphics[height=5cm]{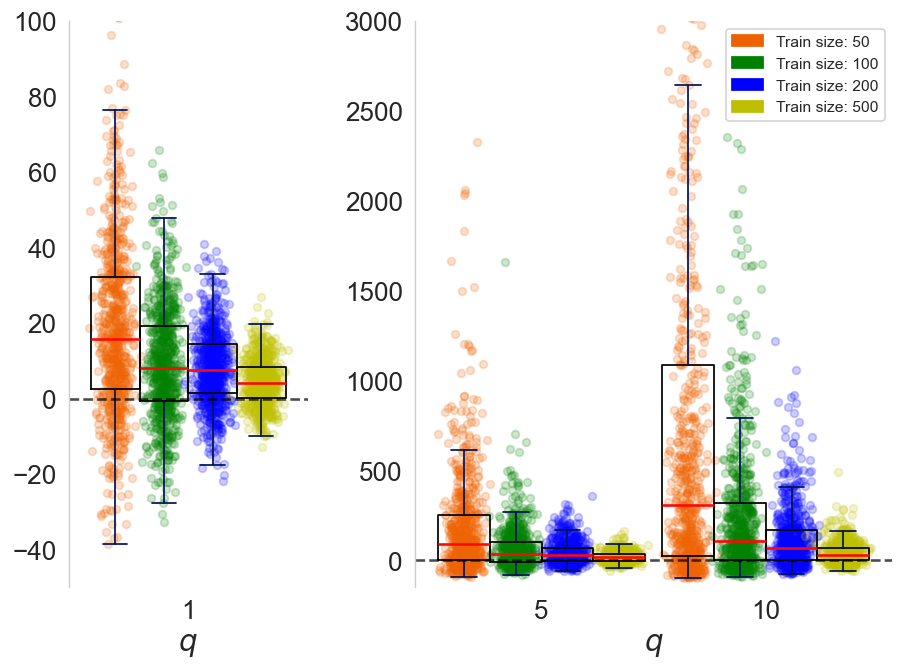}  
        \subcaption{Volume}\label{fig:volume_dimension}
    \end{minipage}
    \caption{\small Each dot is the relative error (\%) of the SC ball with respect to the OOB ball~given a sample $\mathcal{L}_n^{(j)}$, $j=1, \ldots, N$, for the radii and volumes in the simulations of model \eqref{eq:multidim_euc_model}. 
    }
    \label{fig:radius_volume_dimension}
\end{figure}

Section \ref{supp_subsec:euc} in the SM contains the results for Type III, as well as for Types II and IV with parameters $\alpha = 0.10$ and $\sigma = \frac{\sqrt{3}}{2}$. For Types III and IV, we considered a fixed value of the predictors $\bX = (X_{0.25}, X_{0.25}, X_{0.25})^\top$, where $X_{0.25}$ is the $0.25$-quantile of the distribution of $X_i$, $i=1,2,3$. Overall, OOB balls outperformed SC balls across Types I, II, and IV in terms of a mean coverage closer to the nominal level, and the standard deviations were reduced in coverage Types I and II. The computation times of each method are reported in Section \ref{supp_subsec:euc}. The analogous results with $\alpha = 0.05$ and $\alpha = 0.01$, and for Types II and IV with $\sigma = \sqrt{3}$, can be found in Section \ref{supp_subsec:github} of the SM.

\subsection{Sphere}

In the subsequent simulation scenarios, the focus is on the performance of OOB balls (radius and coverage types). We used RFWLCFR to train RFs using the implementation from \cite{Bulte2024} in the \texttt{pyfréchet} package for Python. For the unit sphere $\mathbb{S}^2$, we considered the regression function $\bm_0$ (great circle) given in \eqref{eq:great_circle}, with $\bmu = (1/\sqrt{2}, 1/\sqrt{2})^\top$. The predictor variable $\Theta$ was generated as $\Theta = \arctantwo(X_2,X_1)$, where $\bX = (X_1, X_2) \sim \mathrm{vMF}((1,0)^\top, 1)$. For the response, we considered $\bY \mid \Theta = \theta \sim \mathrm{vMF}(\bm_0(\theta), \kappa)$ for $\kappa = 50$, $200$. The data generation process is illustrated in Figure \ref{fig:data_gen_process} for different values of~$\Theta$.

The coverage results of Types I and III are collected in Table \ref{tab:typeIandIIIcoveragesphere}. 
For Type III, we considered the fixed value of the predictor $\theta = \Theta_{0.25}$, where $\Theta_{0.25}$ is the $0.25$-quantile of the distribution of $\Theta$. Increasing $n$ improved the mean and standard deviations of the OOB ball radius and the coverage of Types II and IV (also for $\theta = \Theta_{0.25}$). As $\kappa$ grows, the data becomes increasingly concentrated around the Fréchet mean, leading to narrower OOB balls. However, despite this reduction in dispersion and ball radius, the coverage of Types III and IV at $\kappa=200$ deviated further from the nominal levels (exhibiting over-coverage) than at $\kappa=50$.~This likely occurs because larger errors arise in regions of low data density, inflating the radius of the OOB balls. To support this claim, we repeated the experiments with $\Theta \sim U(0,1)$ (constant density), and found no significant deviation from the nominal levels for $\kappa=200$. Section \ref{supp_subsec:github} in the SM contains the results of Types II and IV for $\alpha=0.05$, $0.01$. The~conclusions are similar to those for $\alpha=0.10$, albeit with a less pronounced effect of increasing~$n$.

\begin{figure}[H]
    \centering
    \begin{subfigure}{0.5\textwidth}
        \centering
        \includegraphics[height=5cm]{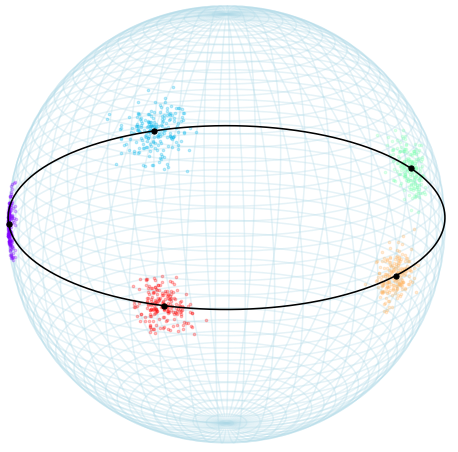}
        \subcaption{Regression on the unit sphere $\mathbb{S}^2$} 
    \end{subfigure}\hfill
    \begin{subfigure}{0.5\textwidth}
        \centering
        \includegraphics[height=5cm]{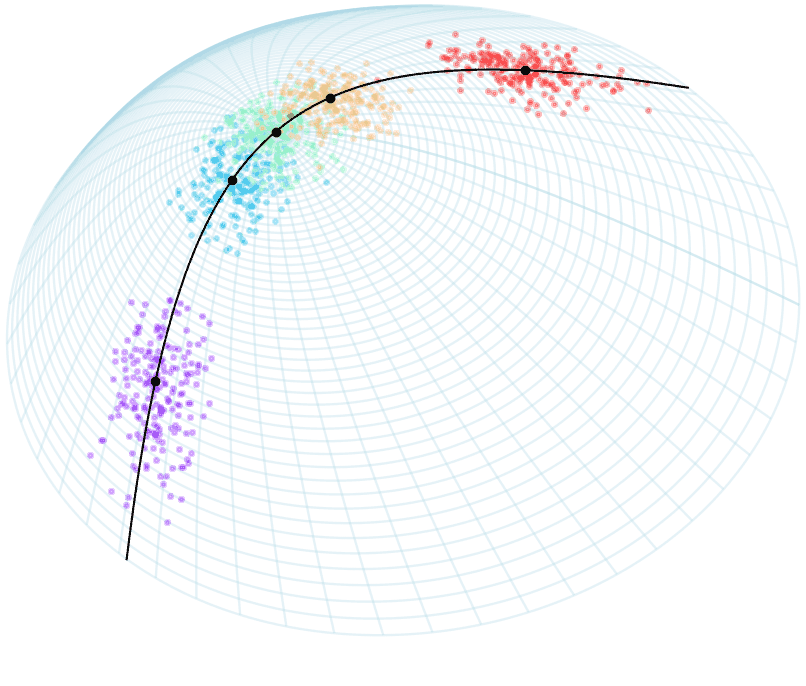}
        \subcaption{Regression on the unit hyperboloid $\mathbb{H}^2$} 
    \end{subfigure}    
    \begin{subfigure}{\textwidth}
        \centering
        \includegraphics[width=\linewidth]{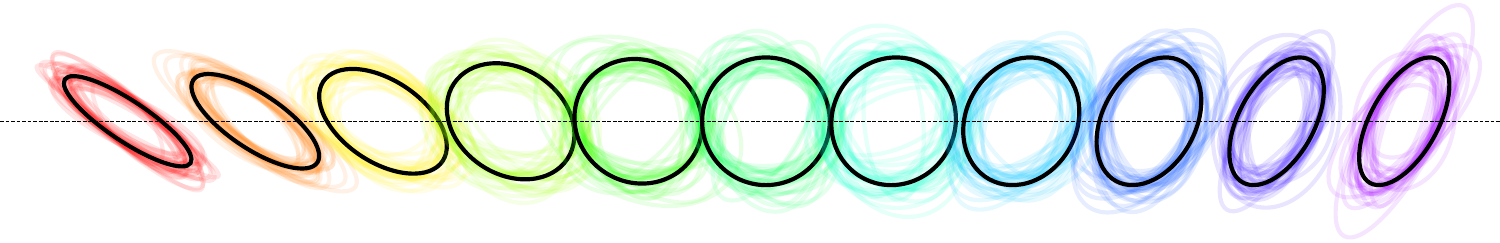} 
        \subcaption{Regression on the space of SPD matrices $\mathcal{S}_2^+$} 
    \end{subfigure}
    \caption{\small Illustration of the data generation processes in $\mathbb{S}^2$, $\mathbb{H}^2$, and $\mathcal{S}_2^+$. On $\mathbb{S}^2$, we highlighted five foci on the great circle corresponding to $\theta \in \{0, \frac{2\pi}{5}, \frac{4\pi}{5}, \frac{6\pi}{5}, \frac{8\pi}{5}\}$, and generated $200$ points from $\bY \mid \Theta = \theta \sim \mathrm{vMF}(\bm_0(\theta), \kappa)$ for each specification of $\theta$. For $\mathbb{H}^2$, the foci correspond to the $0.01$, $0.25$, $0.5$, $0.75$, and $0.99$ quantiles of $\Theta \sim \mathcal{N}(0, \frac{1}{4})$, and $\bY \mid \Theta = \theta \sim \mathrm{HvMF}(\bm_0(\theta), \kappa)$. SPD matrices ($q=2$ and $d=15$) are illustrated through their eigenvectors and eigenvalues, which are plotted as ellipses on the plane. The central ellipses (thicker outline, higher opacity) represent $\bM_0(x)$ for $x\in \{0, 0.1, \ldots, 1\}$. The regression model starts with covariance matrices with negative correlation, goes through a perfect circle (zero correlation), and ends with matrices with positive correlations, as the change in the eccentricity in the ellipses shows. Around each central ellipse, $30$ additional ellipses (thinner outline, lower opacity) are plotted, generated from $\bS \mid X=x \sim \mathrm{Wishart}_2(d, c_{15,2}^{-1} \bM_0(x))$.}\label{fig:data_gen_process} 
\end{figure}

\begin{table}[H]
\setlength{\tabcolsep}{5pt}
\centering
\begin{tabular}{llccc|ccc}
\toprule
 &  & \multicolumn{3}{c}{Type I} & \multicolumn{3}{c}{Type III} \\
 \cmidrule(lr){3-5} \cmidrule(lr){6-8}
 $\kappa$ & $n$ & $\alpha=0.01$ & $\alpha=0.05$ & $\alpha=0.10$ & $\alpha=0.01$ & $\alpha=0.05$ & $\alpha=0.10$ \\
\midrule
\multirow{4}{*}{$50$} & $50$ & \underline{97.9} (0.40) & 94.1 (0.73) & 88.4 (1.00)  & \underline{100.0} (0.08) & \underline{98.6} (0.33) & \underline{94.4} (0.70) \\
      & $100$ & 98.5 (0.41) & 94.4 (0.74) & 89.5 (0.97) & 99.3 (0.28) & 95.6 (0.66) & 90.5 (0.89)  \\
      & $200$ & 98.5 (0.37) & 93.8 (0.78) & 88.3 (0.98) & 99.3 (0.31) & 95.0 (0.67) & 90.0 (0.91) \\
      & $500$ & 98.5 (0.37) & 93.9 (0.72) & 88.9 (0.99) & 99.3 (0.30) & \underline{96.4} (0.63) & \underline{92.1} (0.88) \\
\cline{1-8}
\multirow{4}{*}{$200$} & $50$ & 98.8 (0.42) & 95.3 (0.73) & 89.4 (1.03) & \underline{100.0} (0.02) & \underline{100.0} (0.16) & \underline{98.2} (0.45)\\
      & $100$ & 98.2 (0.42) & 93.6 (0.78) & 89.0 (1.02) & \underline{100.0} (0.10) & \underline{98.8} (0.44) & \underline{94.6} (0.71) \\
      & $200$ & 98.7 (0.39) & 94.0 (0.79) & 88.3 (1.07) & \underline{99.7} (0.20) & \underline{96.9} (0.58) & \underline{92.7} (0.83) \\
      & $500$ & 98.4 (0.38) & 95.2 (0.73) & 90.7 (1.01) & 99.3 (0.33) & 94.9 (0.70) & 91.1 (0.96) \\
\bottomrule
\end{tabular}
\caption{\small Comparison of the estimated Type I probability (in $\%$) using \eqref{type_I_estimation}, and the standard deviation of $K$ probability estimations calculated with \eqref{sd_boot}, for OOB balls on $\mathbb{S}^2$. The same procedure was applied for Type III, with predictors value $\Theta=\Theta_{0.25}$. The values that fall outside the corresponding Wilson $95\%$ confidence interval are underlined.}
\label{tab:typeIandIIIcoveragesphere}
\end{table}

\begin{figure}[H]
    \centering
    \begin{minipage}{0.33\textwidth}
        \centering
        \includegraphics[height=5cm]{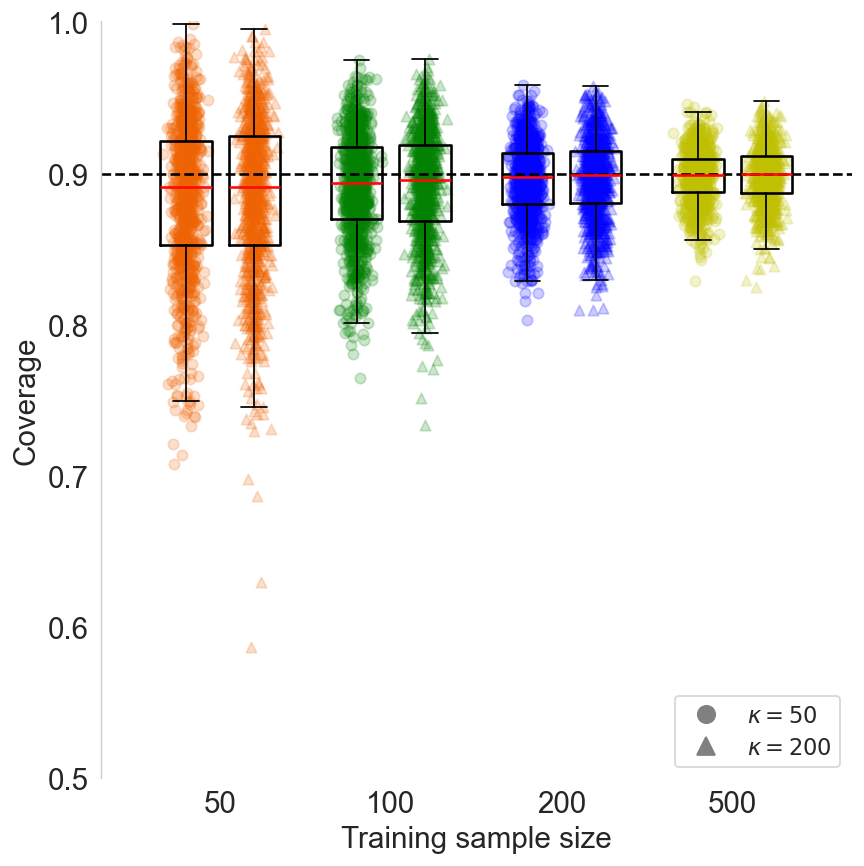}  
        \subcaption{Type II}\label{fig:sphere_coverage_radius_1_a}
    \end{minipage}\hfill
    \begin{minipage}{0.33\textwidth}
        \centering
        \includegraphics[height=5cm]{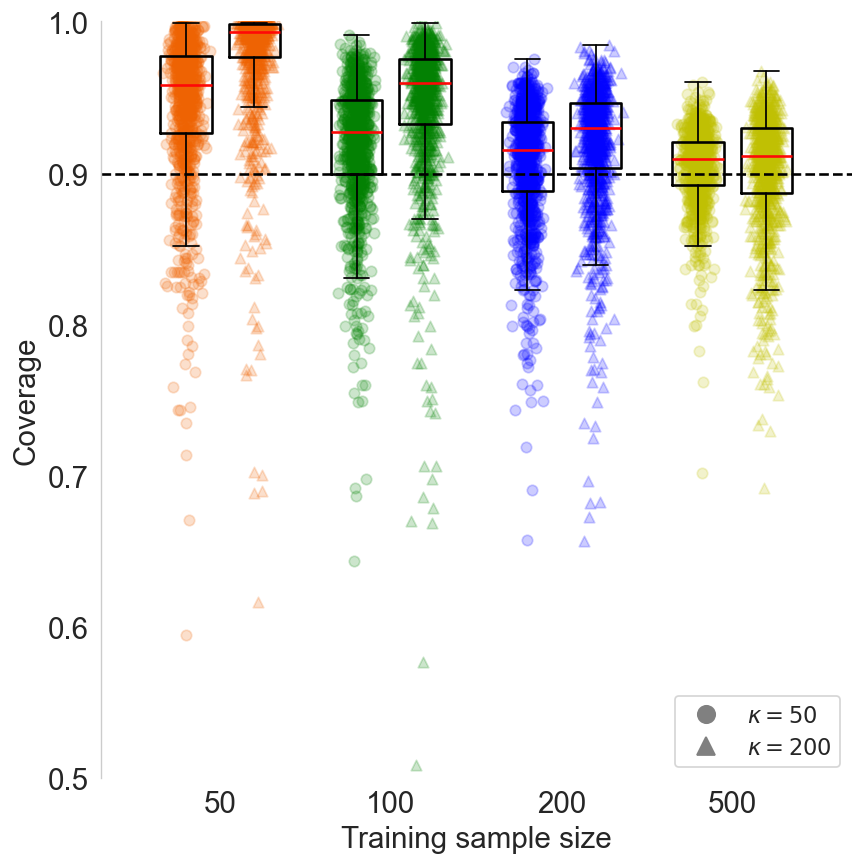}  
        \subcaption{Type IV ($\Theta = \Theta_{0.25}$)}\label{fig:sphere_coverage_radius_1_b}
    \end{minipage}
    \begin{minipage}{0.33\textwidth}
        \centering
        \includegraphics[height=5cm]{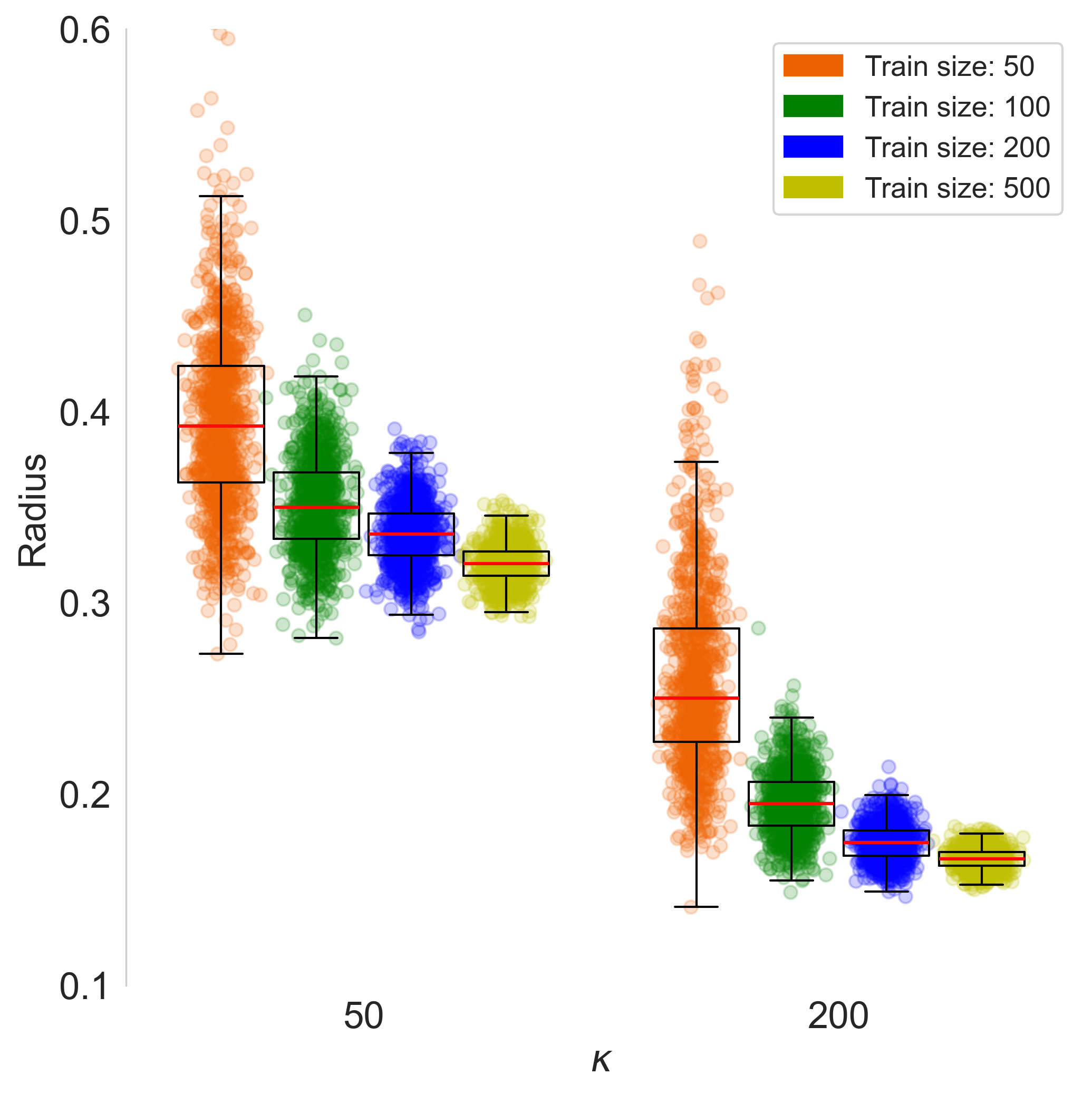}  
        \subcaption{Ball radius}\label{fig:sphere_coverage_radius_1_c}
    \end{minipage}
    \caption{\small Each dot in the boxplots is associated with a data set $\mathcal{L}_n^{(j)}$, $j=1,\ldots, N$. In Figure \ref{fig:sphere_coverage_radius_1_a}, the vertical position represents the Type II reported coverage for the given sample on $\mathbb{S}^2$. In Figure \ref{fig:sphere_coverage_radius_1_b}, we considered $\Theta=\Theta_{0.25}$ (Type IV). For Figure \ref{fig:sphere_coverage_radius_1_c}, the vertical position represents the estimated ball radius given the data set. All three plots use $\alpha=0.10$.}\label{fig:sphere_coverage_radius_1}
\end{figure}
    
\subsection{Hyperboloid}

We considered the regression function $\bm_0$ on $\mathbb{H}^2$ given in \eqref{eq:great_circle_H2} with $\bmu = (1/\sqrt{2}, 1/\sqrt{2})^\top$, which generates the analog of a great circle on the sphere. The predictor variable $\Theta$ was generated as $\Theta \sim \mathcal{N}(0, 1/4)$ and $\bY \mid \Theta = \theta \sim \mathrm{HvMF}(\bm_0(\theta), \kappa)$, for $\kappa = 50$, $200$. The data generation process is illustrated in Figure \ref{fig:data_gen_process}.

\begin{table}[H]
\setlength{\tabcolsep}{5pt}
\centering
\begin{tabular}{llccc|ccc}
\toprule
 &  & \multicolumn{3}{c}{Type I} & \multicolumn{3}{c}{Type III} \\
 \cmidrule(lr){3-5} \cmidrule(lr){6-8}
 $\kappa$ & $n$ & $\alpha=0.01$ & $\alpha=0.05$ & $\alpha=0.10$ & $\alpha=0.01$ & $\alpha=0.05$ & $\alpha=0.10$ \\
\midrule
\multirow{4}{*}{$50$} & $50$ & \underline{97.0} (0.48) & \underline{91.7} (0.80) & \underline{85.1} (1.10) & 98.6 (0.33) & \underline{96.3} (0.63) & 91.7 (0.90)\\
      & $100$ & \underline{98.0} (0.48) & 93.8 (0.74) & 88.0 (0.96) & \underline{98.1} (0.43) & 94.6 (0.70) & 89.3 (0.99) \\
      & $200$ & 98.4 (0.38) & 94.2 (0.76) & 88.7 (0.98) & \underline{98.1} (0.44) & 94.9 (0.73) & 90.1 (0.96) \\
      & $500$ & 98.7 (0.38) & 94.0 (0.71) & 90.6 (0.90) & 98.5 (0.37) & 94.3 (0.71) & 88.7 (0.98) \\
\cline{1-8}
\multirow{4}{*}{$200$} & $50$ & \underline{97.9} (0.47) & 94.4 (0.75) & 90.1 (1.03) & \underline{99.7} (0.26) & \underline{97.2} (0.55) & \underline{92.1} (0.87) \\
      & $100$ & \underline{97.8} (0.46) & \underline{92.7} (0.74) & \underline{87.9} (0.98) & 98.5 (0.35) & 94.4 (0.69) & 88.5 (0.99) \\
      & $200$ & 98.8 (0.38) & \underline{92.6} (0.72) & \underline{87.9} (1.03) & 99.2 (0.28) & 95.1 (0.66) & 90.0 (0.93) \\
      & $500$ & 98.8 (0.35) & 95.8 (0.67) & 91.1 (0.92) & 99.2 (0.28) & 94.3 (0.72) & 89.2 (0.99) \\
\bottomrule
\end{tabular}
\caption{\small Same description as Table \ref{tab:typeIandIIIcoveragesphere}, for $\mathbb{H}^2$. For Type III, we considered $\Theta = \frac{1}{4}\Phi^{-1}(0.25)$.}
\label{tab:typeIandIIIcoverageH2}
\end{table}

\vspace{-0.32cm}

\begin{figure}[H]
    \centering
    \begin{minipage}{0.33\textwidth}
        \centering
        \includegraphics[height=5cm]{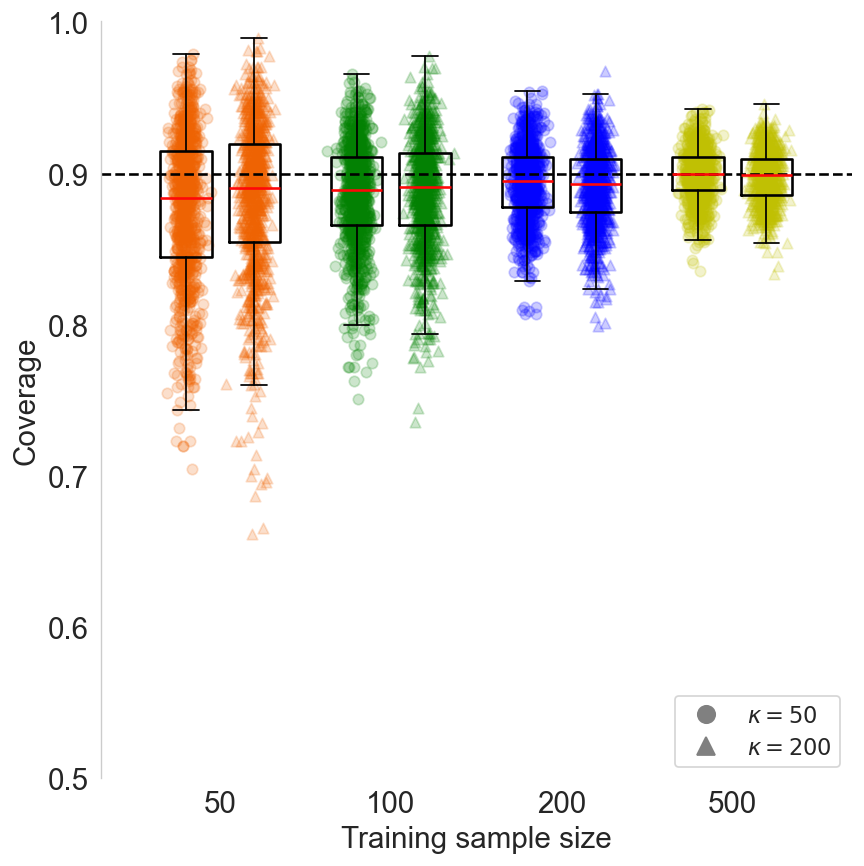}  
        \subcaption{Type II}\label{fig:H2_coverage_radius_1_a}
    \end{minipage}\hfill
    \begin{minipage}{0.33\textwidth}
        \centering
        \includegraphics[height=5cm]{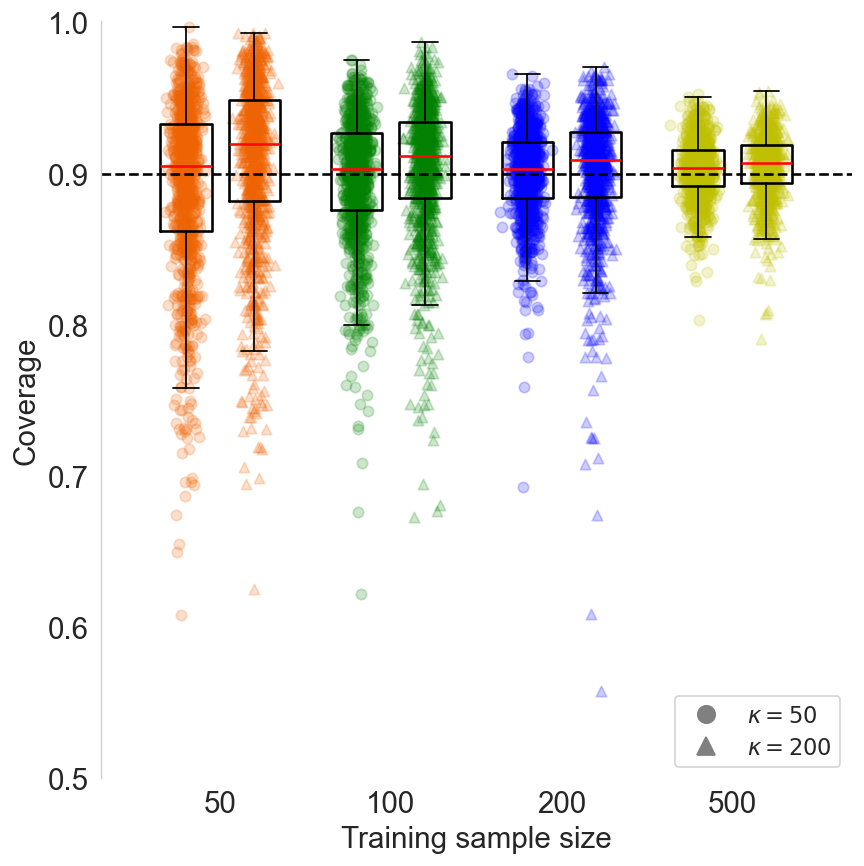}  
        \subcaption{Type IV ($\theta = \frac{1}{4}\Phi^{-1}(0.25)$)}\label{fig:H2_coverage_radius_1_b}
    \end{minipage}
    \begin{minipage}{0.33\textwidth}
        \centering
        \includegraphics[height=5cm]{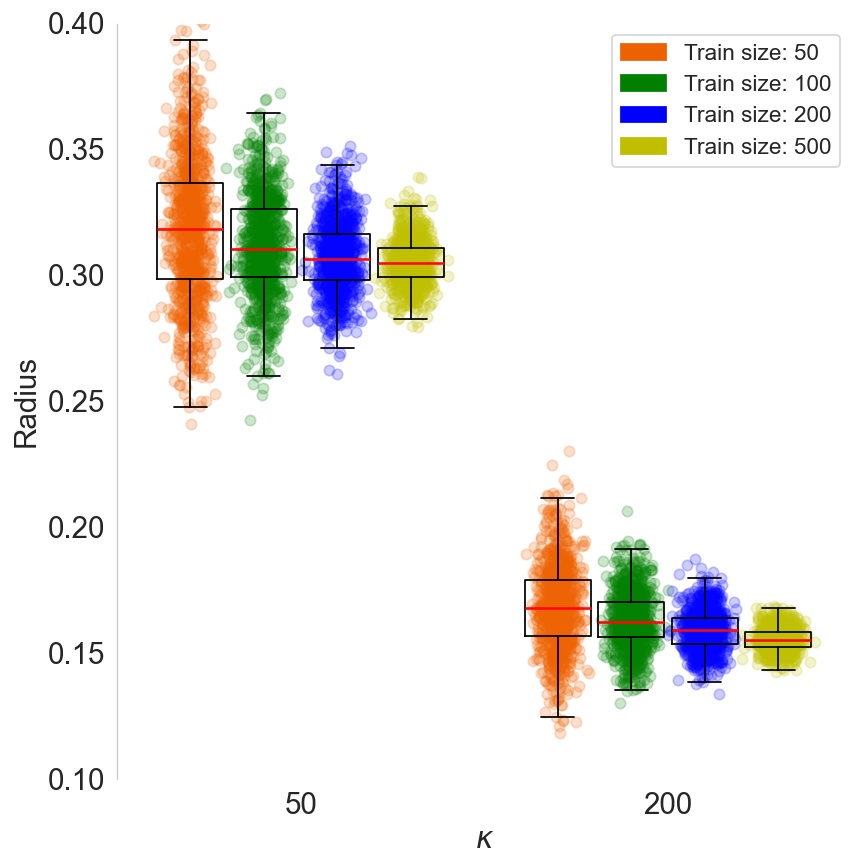}
        \subcaption{Ball radius}\label{fig:H2_coverage_radius_1_c}
    \end{minipage}
    \caption{\small Same description as Figure \ref{fig:sphere_coverage_radius_1}, for $\mathbb{H}^2$. In Figure \ref{fig:H2_coverage_radius_1_b}, we considered $\Theta = \frac{1}{4}\Phi^{-1}(0.25)$.}\label{fig:H2_coverage_radius_1}
\end{figure}

Table \ref{tab:typeIandIIIcoverageH2} contains the coverage results of Types I and III, and Figure \ref{fig:H2_coverage_radius_1} shows the results for Types II and IV as well as the radii of the OOB balls. For Types III and IV, we considered the fixed predictor value $\Theta = \frac{1}{4}\Phi^{-1}(0.25)$, where $\Phi$ is the CDF of the standard normal distribution. Increasing $n$ had a positive impact on the mean coverage for Types II and III, and for Type IV with $\kappa=200$. The reported standard deviation decreased as $n$ grew for Types II, III, and IV. The mean and standard deviation of the radius of OOB balls were also reduced with $n$. See Section~\ref{supp_subsec:github} in the SM for $\alpha = 0.05$, $0.01$.

\subsection{Symmetric positive definite matrices}

We consider the space $\mathcal{S}_q^+$ of SPD $q\times q$ matrices, and study the impact of metric selection on the performance of OOB balls. We compared the coverage and radius of OOB balls in $\mathcal{S}_2^+$ using the three metrics. The regression model is constructed based on the idea of a convex interpolation between SPD matrices introduced in Section \ref{subsec:SPD}. In particular, we considered the interpolation function 
$$
\bM_0(x) = 
\begin{cases} 
\cos^2(\pi x) \bSigma_1 + \sin^2(\pi x) \bSigma_2 \ \text{ if } \lfloor x + \frac{1}{2} \rfloor \text{ is even}, \\
\sin^2(\pi x) \bSigma_2 + \cos^2(\pi x) \bSigma_3 \ \text{ if } \lfloor x + \frac{1}{2} \rfloor \text{ is odd},
\end{cases}
$$
with matrices
$$
\bSigma_1 = 
\begin{pmatrix}
1 & -0.6 \\
-0.6 & 0.5
\end{pmatrix}, \quad
\bSigma_2 =
\begin{pmatrix}
1 & 0 \\
0 & 1
\end{pmatrix}, \quad
\bSigma_3 = 
\begin{pmatrix}
0.5 & 0.4 \\
0.4 & 1
\end{pmatrix}.
$$
We considered a single predictor $X = 2\sqrt{5}(W - 1/2)$, $W \sim \mathrm{Beta}(2,2)$, and generated the response $\bS$ according to a $\mathrm{Wishart}_2(d, c_{d,2}^{-1}\bM_0(X))$, for degrees of freedom $d \in \{5,15\}$. See Figure \ref{fig:data_gen_process} for a visualization of the data generation process with $d=15$.

The coverages of Types I and III from the experiments are collected in Table \ref{tab:typeIandIIIcoverageAI} for the AI distance. For the significance level $\alpha = 0.10$, Types II and IV coverage results are shown in Figure \ref{fig:AI_SPD_coverage_radius_1}. For Types III and IV, the fixed value of the predictor $X=X_{0.25}$ was considered, where $X_{0.25}$ denotes the $0.25$ quantile of the distribution of $X$. Increasing the sample size had a positive impact on coverage Types I, II, and IV, and on the radius of the OOB balls. Since the data is generated according to a $\mathrm{Wishart}_2(d, c_{d,2}^{-1}\bM_0(X))$, the variance decreases as $d$ grows. Thus, $d$ acts as a concentration parameter, similarly to the role of $\kappa$ in vMF and HvMF distributions. This results in a smaller ball radius for larger values of $d$.

\begin{table}[H]
\setlength{\tabcolsep}{5pt}
\centering
\begin{tabular}{llccc|ccc}
\toprule
 &  & \multicolumn{3}{c}{Type I} & \multicolumn{3}{c}{Type III} \\
 \cmidrule(lr){3-5} \cmidrule(lr){6-8}
 $d$ & $n$ & $\alpha=0.01$ & $\alpha=0.05$ & $\alpha=0.10$ & $\alpha=0.01$ & $\alpha=0.05$ & $\alpha=0.10$ \\
 \midrule
\multirow{4}{*}{$5$} & $50$ & \underline{97.2} (0.51) & \underline{92.8} (0.81) & \underline{86.1} (1.10) & 98.2 (0.38) & 94.6 (0.71) & 89.5 (0.99) \\
 & $100$ & \underline{97.8} (0.48) & \underline{93.2} (0.76) & 89.1 (1.01) & 98.6 (0.42) & 94.6 (0.72) & 90.1 (0.93) \\
 & $200$ & 99.3 (0.36) & 94.1 (0.69) & 88.5 (0.95) & \underline{97.8} (0.42) & 94.7 (0.83) & 88.6 (1.03) \\
 & $500$ & \underline{97.9} (0.40) & 93.7 (0.75) & 88.3 (1.01) & 99.2 (0.30) & 96.2 (0.64) & 91.7 (0.93) \\
\cline{1-8}
\multirow{4}{*}{$15$} & $50$ & 98.9 (0.39) & 96.2 (0.75) & 90.2 (0.97) & 99.3 (0.26) & \underline{96.8} (0.56) & \underline{92.7} (0.83) \\
 & $100$ & 98.4 (0.40) & 94.9 (0.68) & 91.1 (0.92) & 98.4 (0.33) & 95.9 (0.61) & \underline{91.9} (0.85) \\
 & $200$ & 98.8 (0.38) & 95.0 (0.70) & \underline{92.1} (0.94) & 98.9 (0.29) & 96.1 (0.63) & \underline{92.0} (0.86) \\
 & $500$ & 98.8 (0.29) & 95.0 (0.67) & 90.0 (0.89) & 99.2 (0.25) & 96.1 (0.62) & 91.7 (0.88) \\
\bottomrule
\end{tabular}
\caption{\small Same description as Table \ref{tab:typeIandIIIcoveragesphere}, for the space of SPD matrices $\mathcal{S}_2^+$ endowed with the AI distance. For Type III, the value of the predictors $X=X_{0.25}$ was considered.}
\label{tab:typeIandIIIcoverageAI}
\end{table}

\begin{figure}[H]
    \centering
    \begin{minipage}{0.33\textwidth}
        \centering
        \includegraphics[height=5cm]{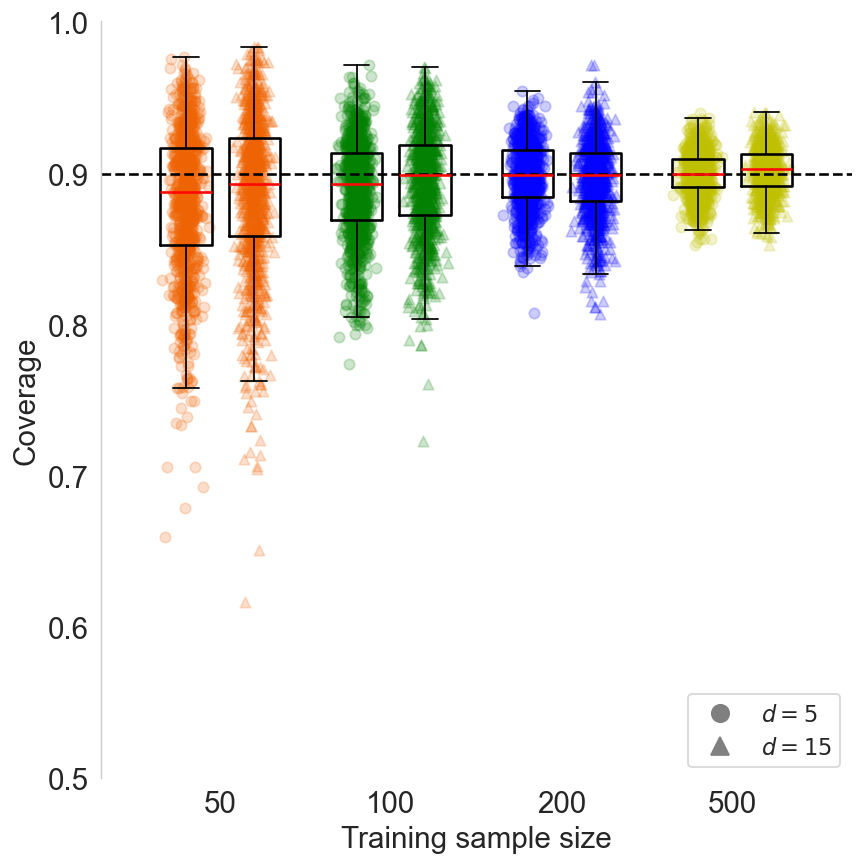}  
        \subcaption{Type II}\label{fig:AI_SPD_coverage_radius_1_a}
    \end{minipage}\hfill
    \begin{minipage}{0.33\textwidth}
        \centering
        \includegraphics[height=5cm]{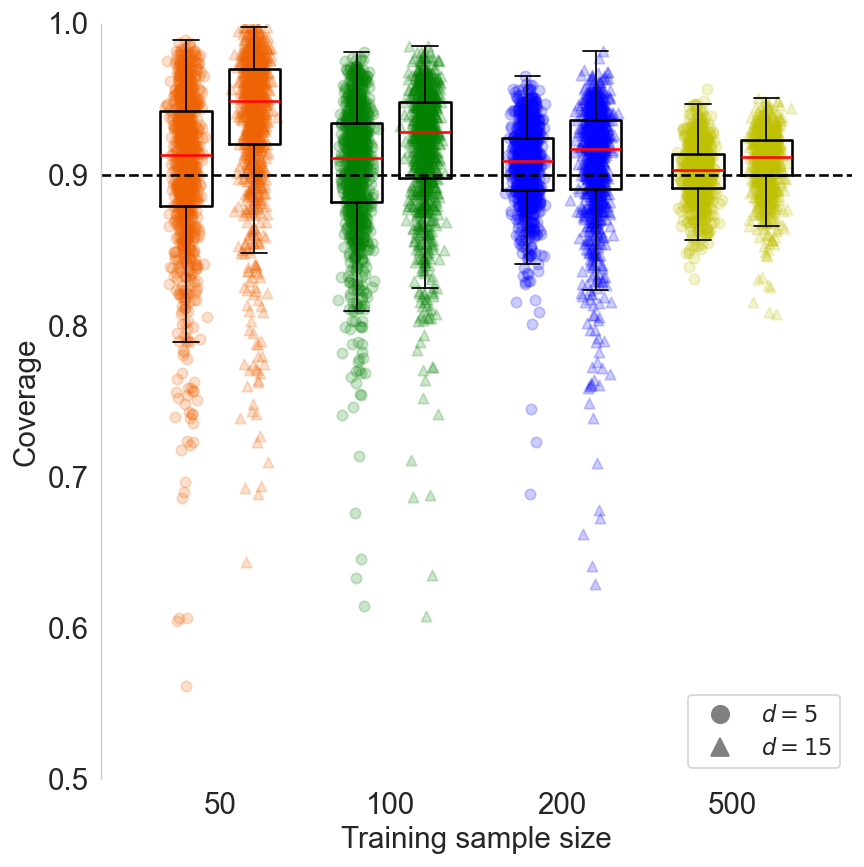}  
        \subcaption{Type IV ($X=X_{0.25}$)}\label{fig:AI_SPD_coverage_radius_1_b}
    \end{minipage}
    \begin{minipage}{0.33\textwidth}
        \centering
        \includegraphics[height=5cm]{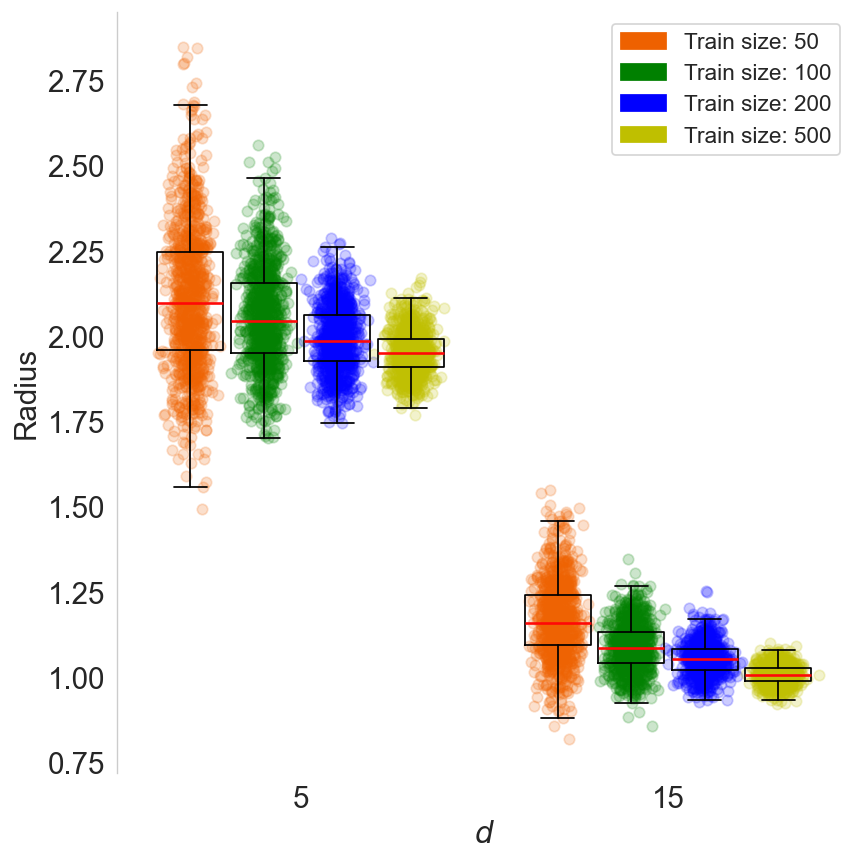}  
        \subcaption{Ball radius}\label{fig:AI_SPD_coverage_radius_1_c}
    \end{minipage}
    \caption{\small Same description as Figure \ref{fig:sphere_coverage_radius_1}, for the space of SPD matrices $\mathcal{S}_2^+$ endowed with the AI distance. In Figure \ref{fig:AI_SPD_coverage_radius_1_b}, we considered $X=X_{0.25}$ (Type IV).}\label{fig:AI_SPD_coverage_radius_1}
\end{figure}

Section~\ref{supp_subsec:github} in the SM collects the results with $\alpha=0.05$, $0.01$, and for the LC and LE distances. The results show that the performance of OOB balls is consistent across the three different metrics. For the LC and LE metrics, even though $\bM_0(x)$ is not the conditional Fréchet mean \ref{c2.1}, the empirical coverages were satisfactory. The comparable results for the AI and LE distances align with their theoretical connection, as the LE metric approximates the AI distance. This is also reflected in the nearly identical geometries of their corresponding population OOB balls in Section \ref{supp_subsec:illustration_oob_balls} of the~SM.

\section{Comparison of out-of-bag and split-conformal balls}
\label{sec:oob-vs-sc}

SC methods split the sample into two subsamples, one to fit the forest (training set) and the other to estimate the ball radius (test set). Below, we highlight the main differences between SC and OOB balls, drawing on the results obtained in Section \ref{subsec:numerical_euclidean}.

\begin{itemize}
    \item In general, a RF is trained using the full sample to increase the accuracy and reliability for a specific task (e.g., prediction or variable importance analysis). Once the forest is fitted, one may be interested in quantifying the uncertainty in the predictions of the given forest. This can not be done with SC balls, since it would require training a different forest (trained on half of the sample). OOB balls can estimate the uncertainty in the predictions of the given forest, and at no additional training cost, since they only require selecting the trees for which the target observation is OOB. 
    
    \item For OOB balls, every RF is trained using the complete sample. Table \ref{tab:mse_PB_SC} compares the Mean Squared Error (MSE) of the forests that generate the OOB and SC balls in the multiple Euclidean linear model \eqref{eq:mult_linear_model}. SC balls from a sample of size $n$ showed means and standard deviations of the MSEs very close to those of OOB balls with size $n/2$.
    
    \item The radius of SC balls is calculated using the prediction errors from the test set. Thus, the number of prediction error estimates used to calculate the radius is also halved, compared to OOB balls. In practice, this advantage, along with the smaller MSE, results in a smaller and less variable radius for OOB balls. This is illustrated in Figure \ref{fig:radius_dimension} for the multivariate Euclidean linear model \eqref{eq:multidim_euc_model}, across various dimensionalities of the response. Note that when the response is high-dimensional, a small deviation in radius can lead to a significant increase in the volume of the ball, as Figure \ref{fig:volume_dimension} shows.
    
    \item Consider a forest comprised of $B$ trees, each trained with a full-size bootstrap resample of $\mathcal{L}_n$, and an observation $(x, y) \in \mathcal{L}_n$. The probability that $(x, y)$ is not selected in a specific bootstrap resample is $(1 - \frac{1}{n})^n$. Thus, for large $n$, approximately $B e^{-1}$ trees are trained without $(x,y)$, and can be used to produce an OOB prediction of $(x,y)$, which is ultimately employed to calculate the ball radius. For SC balls, a test set (independent of the training set) is used to compute the prediction errors. Thus, all the trees in the forest are employed for radius estimation.
    
    \item Since the RF is trained on a reduced sample, SC balls present lower computational costs than OOB balls, as illustrated in Section \ref{supp_subsec:euc} of the SM.
\end{itemize}

\section{Out-of-bag balls to model sunspot dynamics}
\label{sec:real_data}

Due to its plasma composition, the Sun's equatorial regions rotate faster than the polar regions (differential rotation). This phenomenon distorts the magnetic field, stretching the Sun's initial poloidal magnetic field into toroidal configurations, as plasma drags field lines along their rotational trajectories. Each solar cycle lasts around $11$ years and culminates when these twisted fields undergo a global polarity reversal \citep[see, e.g.,][]{Babcock1961}. The entanglement of field lines creates localized regions of intensified magnetic pressure that suppress convective heat transfer, manifesting as cooler, darker sunspots. Sunspots have a variable lifetime and serve as direct observational tracers of solar activity, which is linked to space weather events that impact spacecraft, astronauts, and Earth's infrastructure.

We analyzed data based on a merged catalog combining the Greenwich Photoheliographic Results (GPR) and the Debrecen Photoheliographic Data (DPD) sunspot records \citep{Baranyi2016, Gyoeri2016}. The GPR spans the years 1872–1976, while the DPD extends the record from 1974 onward, covering 14 solar cycles to date. The R package \texttt{rotasym} \citep{Garcia-Portugues2020e} provides the data sets \texttt{sunspots\_births} and \texttt{sunspots\_deaths}, containing, respectively, the location on the Sun of the first and last record of each sunspot group (referred to as ``birth'' and ``death'' henceforth). We merged these data sets and excluded the first ($11$th) and last ($24$th) cycles due to the low reliability of their observations. Sunspots that were born and died on the same day were also excluded.

Figure \ref{fig:sunspots_data_balls_a} reveals that longitudinal displacements of sunspots from birth to death are mainly counterclockwise near the equator and clockwise near the poles. This pattern is explained because heliographic coordinates (used in the data set) reference positions to a frame rotating at approximately the Sun's mean rate. Differential rotation causes equatorial sunspots to advance counterclockwise and polar sunspots to retreat clockwise relative to this frame. Apart from this mild tendency, the distribution of death locations of the sunspots in the data based on location at birth is chaotic, with no clear trend or direction in motion, thus making predictions of death locations challenging. For this reason, instead of focusing on the accuracy of death location predictions, the goal of our analysis is to generate prediction regions on the Sun's surface that contain the true death location with a given probability.

For each solar cycle, we predicted death locations from birth locations using RFWLCFR, treating both response and predictors as points on the unit sphere $\mathbb{S}^2$ with geodesic distance $d_{\mathbb{S}^2}$. Although \cite{Qiu2024} consider only real predictors, their method can be applied straightforwardly when the predictors belong to any metric space. The data from each cycle were split into training ($75\%$) and test ($25\%$) sets, and the RFs were tuned with the same parameters as in Section \ref{sec:experiments}, using the implementation in \texttt{pyfréchet}. Although the RF captured the latitude-dependent directional pattern (Figure \ref{fig:sunspots_data_balls_b}), the predicted longitudinal displacements are systematically smaller than observed in the training data. This shrinkage effect likely arises from the model's averaging of opposing trajectories present at all latitudes, a consequence of the chaotic motion of sunspots.

\begin{figure}[H]
    \centering
    \begin{subfigure}{0.33\textwidth}
        \centering
        \includegraphics[height=5cm]{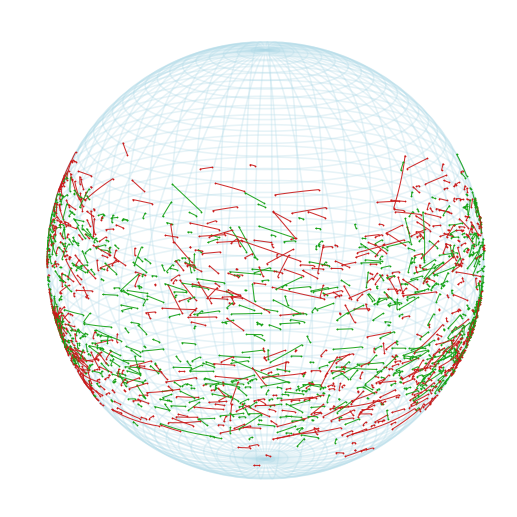}  
        \subcaption{Sunspots trajectories}\label{fig:sunspots_data_balls_a}
    \end{subfigure}\hfill
    \begin{subfigure}{0.33\textwidth}
        \centering
        \includegraphics[height=5cm]{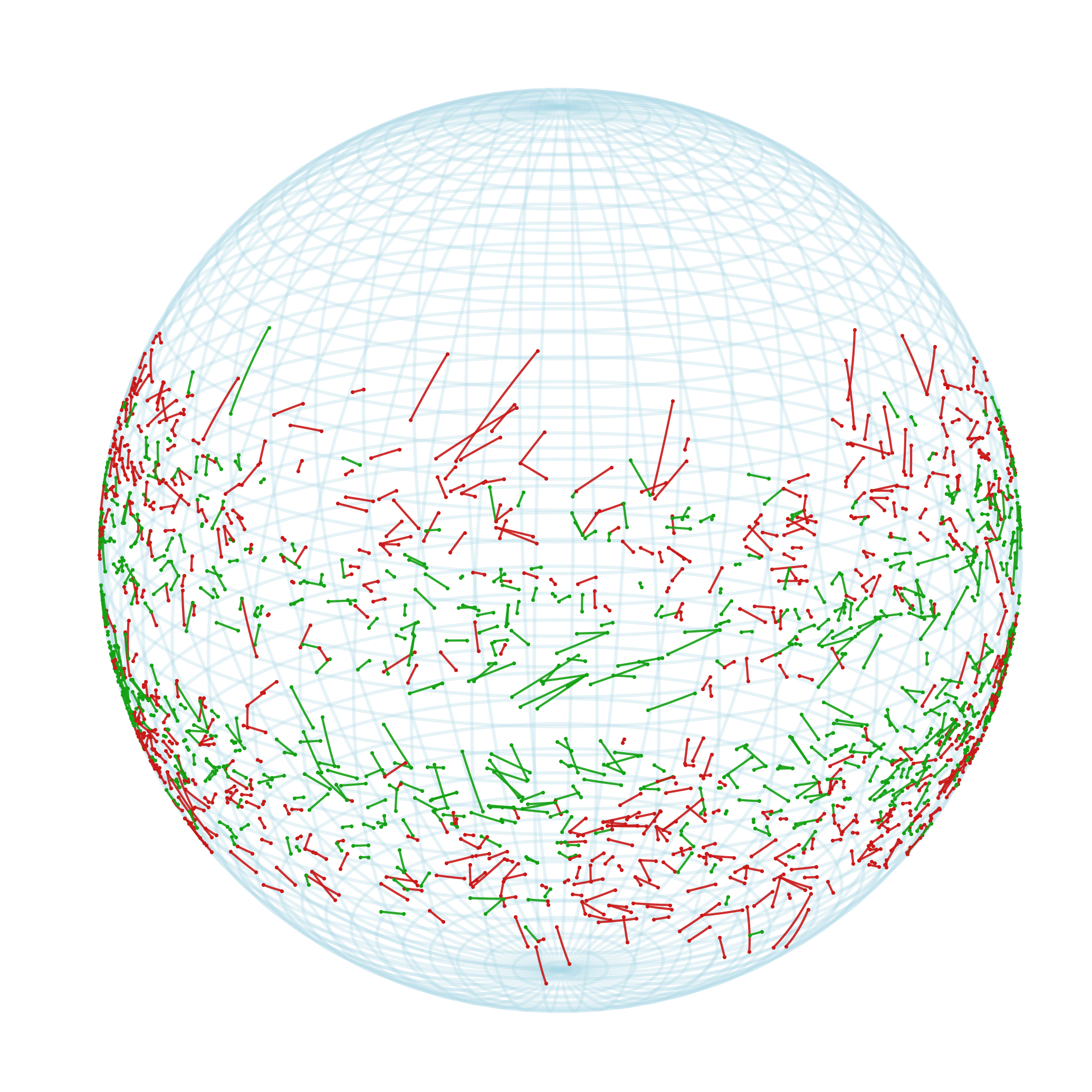}  
        \subcaption{Predicted trajectories}\label{fig:sunspots_data_balls_b}
    \end{subfigure}
    \begin{subfigure}{0.33\textwidth}
        \centering
        \includegraphics[height=5cm]{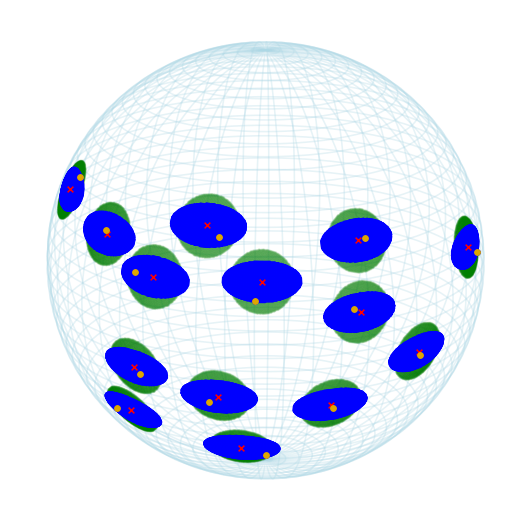}  
        \subcaption{OOB balls}\label{fig:sunspots_data_balls_c}
    \end{subfigure}
    \caption{\small Figure \ref{fig:sunspots_data_balls_a} represents the displacements of sunspot groups for the test sets of cycles 21--23, with paths connecting birth to death locations. Green lines represent positive longitudinal increments (the sunspot moves forward with respect to solar rotation), and red lines for negative increments in longitude (backward). Figure \ref{fig:sunspots_data_balls_b} is the analog plot for predicted death locations. The $90\%$ OOB balls are illustrated in Figure \ref{fig:sunspots_data_balls_c} for some selected points in the test set of cycle $23$. Green balls correspond to $\mathbb{S}^2$ as response space, while blue balls are induced by $S_{0.5,1}^2$. Red crosses and yellow dots denote predicted and true death locations, respectively. In all figures, only points in the frontal hemisphere are shown.}\label{fig:sunspots_data_balls}
\end{figure}

Figure \ref{fig:sunspots_data_balls_a} shows that the displacement of sunspots is larger in the longitudinal direction (along parallels) than latitudinally (along meridians). OOB balls generated using the geodesic distance on $\mathbb{S}^2$ are isotropic (green balls in Figure \ref{fig:sunspots_data_balls_c}), and thus do not capture the anisotropic nature of sunspot motion. To address this limitation, we induced an anisotropic distance on $\mathbb{S}^2$ as follows: first, the data is mapped onto a (prolate) spheroid $S_{a,c}^2$ with semi axes $a$ and $c$, via the mapping $\phi: \mathbb{S}^2 \to S_{a,c}^2$ defined by $\phi(\bx) = (a x_1, a x_2, c x_3)$, where $\bx = (x_1,x_2,x_3) \in \mathbb{S}^2$. Second, a RF is trained on $S_{a,c}^2$ using the intrinsic geodesic distance $d_{S_{a,c}^2}$, and an OOB ball is computed on $S_{a,c}^2$ using $d_{S_{a,c}^2}$. This induces an anisotropic distance on $\mathbb{S}^2$ defined as $(\bx,\by)\in \mathbb{S}^2\times\mathbb{S}^2 \mapsto d_{S_{a,c}^2}(\phi(\bx),\phi(\by))$. Finally, the OOB ball is mapped back to $\mathbb{S}^2$ using the inverse mapping of $\phi$. Figure \ref{fig:sunspots_data_balls_c} (blue balls) shows the OOB balls generated with this procedure. See Figure \ref{fig:sphere_spheroid_mapping} for an illustration of the mapping of the balls from $S_{a,c}^2$ to $\mathbb{S}^2$ for different values of $a$ and $c$.

\begin{figure}[H]
    \centering
    \begin{subfigure}[b]{0.16\textwidth}
        \centering
        \includegraphics[width=\linewidth]{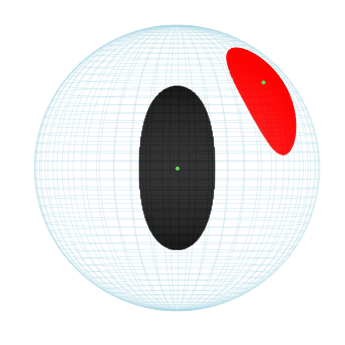}
    \end{subfigure}
    \begin{subfigure}[b]{0.16\textwidth}
        \centering
        \includegraphics[width=\linewidth]{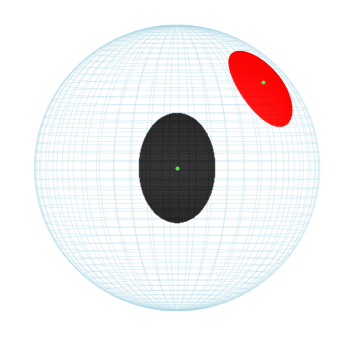}
    \end{subfigure}
    \begin{subfigure}[b]{0.16\textwidth}
        \centering
        \includegraphics[width=\linewidth]{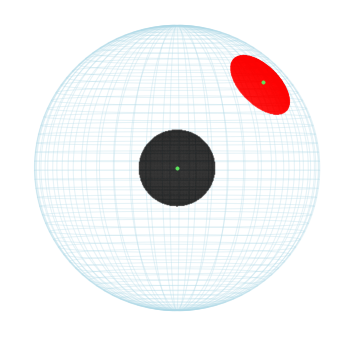}
    \end{subfigure}
    \begin{subfigure}[b]{0.16\textwidth}
        \centering
        \includegraphics[width=\linewidth]{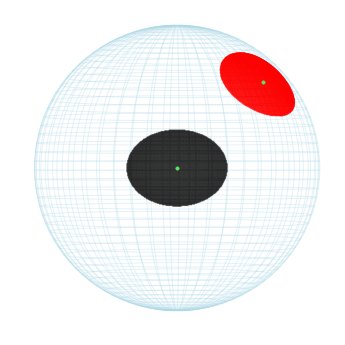}
    \end{subfigure}
    \begin{subfigure}[b]{0.16\textwidth}
        \centering
        \includegraphics[width=\linewidth]{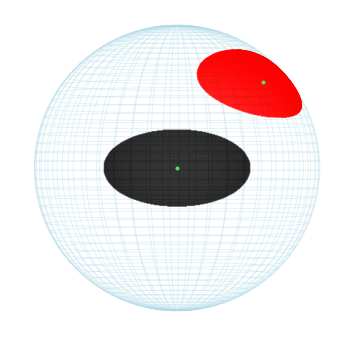}
    \end{subfigure}
    \begin{subfigure}[b]{0.16\textwidth}
        \centering
        \includegraphics[width=\linewidth]{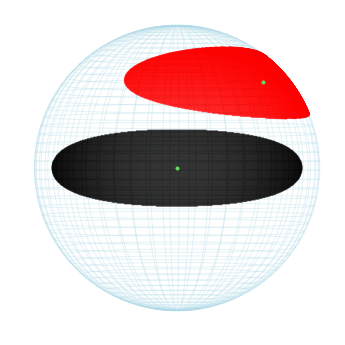}
    \end{subfigure}

    \vspace{1em}

    \begin{subfigure}[b]{0.16\textwidth}
        \centering
        \includegraphics[width=\linewidth]{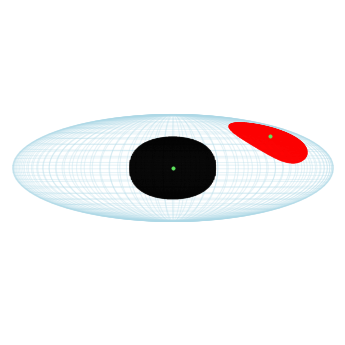}
    \end{subfigure}
    \begin{subfigure}[b]{0.16\textwidth}
        \centering
        \includegraphics[width=\linewidth]{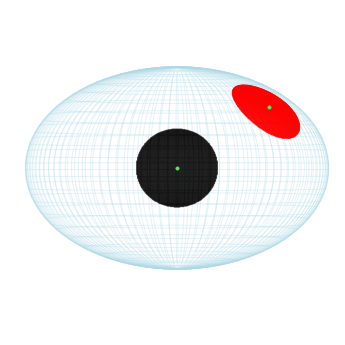}
    \end{subfigure}
    \begin{subfigure}[b]{0.16\textwidth}
        \centering
        \includegraphics[width=\linewidth]{img/sphere_a1.00_c1.00.png}
    \end{subfigure}
    \begin{subfigure}[b]{0.16\textwidth}
        \centering
        \includegraphics[width=\linewidth]{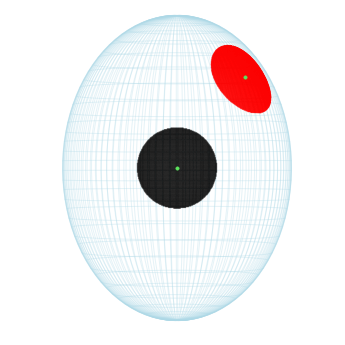}
    \end{subfigure}
    \begin{subfigure}[b]{0.16\textwidth}
        \centering
        \includegraphics[width=\linewidth]{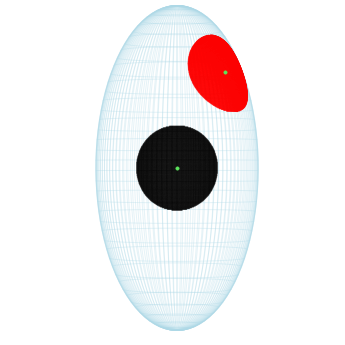}
    \end{subfigure}
    \begin{subfigure}[b]{0.16\textwidth}
        \centering
        \includegraphics[width=\linewidth]{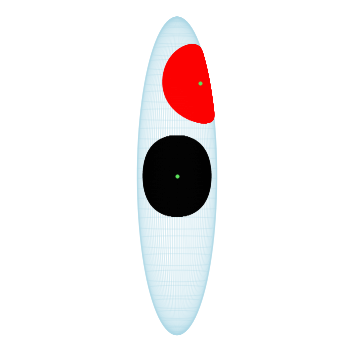}
    \end{subfigure}
    \caption{\small Visualization of OOB balls on $S_{a,c}^2$ (bottom row) and the induced OOB balls on $\mathbb{S}^2$ generated by mapping the balls from the spheroid (top row), for $(a,c) \in \{ (1, \frac 13), \allowbreak (1, \frac 23), \allowbreak (1, 1), (\frac 34, 1),(\frac{1}{2}, 1),(\frac 14, 1)\}$ (from left to right). A green point indicates the center of the balls.}
    \label{fig:sphere_spheroid_mapping}
\end{figure}

The predictors remained on $\mathbb{S}^2$ with the geodesic distance $d_{\mathbb{S}^2}$, while the responses were modeled on $S_{a,c}^2$, equipped with its intrinsic geodesic distance. \cite{Karney2011} provides an algorithm to compute $d_{S_{a,c}^2}$ which can be made arbitrarily accurate and always converges. Although $d_{S_{a,c}^2}$ defines a Riemannian manifold structure on $S_{a,c}^2$, to simplify the implementation, we employed Fréchet medoids instead of Fréchet means in the CART criterion \eqref{cart} and in the aggregation of results (i.e., we minimized \eqref{fr_weighted} over the sample elements). On the sphere ($a=c=1$), medoids were also used for the sake of homogeneity in the comparison.

In Table~\ref{tab:spheroid_metrics}, the area of the OOB balls generated with $d_{S_{a,1}^2}$ is compared with the isotropic OOB balls, for different values of $a$. On the one hand, the area reduction was significant for all prolate configurations $S_{a, 1}^2$ with $0.3\leq a \leq 1$. On the other hand, the area increased for all oblate spheroid configurations considered ($a>1$). This is explained by the predominantly longitudinal motion of sunspots that prolate configurations capture more effectively (see Figure \ref{fig:sphere_spheroid_mapping}). No evidence was found of a significant increase in MSE when $0.4\leq a \leq 1$. Type II coverage probability was computed as the proportion of test responses contained within their OOB balls. The nominal coverage was honored across all spheroidal configurations.

\begin{table}[hpbt]
\setlength{\tabcolsep}{2.5pt}
\centering
\begin{tabular}{lccccccccc|c|ccc}
\toprule
\multicolumn{1}{c}{$a$} & $0.1$ & $0.2$ & $0.3$ & $0.4$ & $\mathbf{0.5}$ & $0.6$ & $0.7$ & $0.8$ & $0.9$ & $1.0$ & $1.1$ & $1.25$ & $1.5$ \\
\midrule
$\Delta_{\mathrm{MSE}}$ & $27.2$ & $11.2$ & $4.0$ & $2.3$ & $\mathbf{1.1}$ & $0.7$ & $0.4$ & $0.2$ & $0.2$ & $0.0$ & $0.2$ & $0.6$ & $0.8$ \\
$\Delta_{\mathrm{area}}$ & $109.6$ & $18.9$ & $-6.5$ & $-15.3$ & $\mathbf{-19.1}$ & $-17.7$ & $-15.1$ & $-10.7$ & $-5.1$ & $0.0$ & $4.1$ & $12.2$ & $24.6$ \\
Coverage & $90.3$ & $90.2$ & $90.1$ & $90.1$ & $\mathbf{89.6}$ & $90.6$ & $90.4$ & $90.2$ & $90.2$ & $90.3$ & $90.2$ & $90.3$ & $90.0$ \\
$p_{\mathrm{MSE}}$ & $0.00$ & $0.00$ & $0.01$ & $0.13$ & $\mathbf{0.47}$ & $0.78$ & $0.91$ & $1.00$ & $1.00$ & --- & $0.91$ & $0.40$ & $0.40$ \\
$p_{\mathrm{area}}$ & $1.00$ & $1.00$ & $0.00$ & $0.00$ & $\mathbf{0.00}$ & $0.00$ & $0.00$ & $0.00$ & $0.00$ & --- & $1.00$ & $1.00$ & $1.00$ \\
\bottomrule
\end{tabular}
\caption{\small For each specification of $a$ (and $c=1$), the errors, areas, and coverages were calculated on the test points of cycles $21$--$23$, and the reported values are weighted means across the three cycles, with weights proportional to the number of test points on each cycle. The areas correspond to OOB balls centered on the predicted value of the test points. The relative MSE difference $\Delta_{\mathrm{MSE}}$ and the relative mean area difference $\Delta_{\mathrm{area}}$ (both in $\%$) were calculated with respect to the balls on the sphere. The reported coverage corresponds to Type II (in \%) for $\alpha=0.10$. The $p$-value of the one-sided paired $t$-test with alternative hypothesis $H_1: \mathrm{MSE}_{\mathbb{S}^2} < \mathrm{MSE}_{S_{a, 1}}$ is $p_{\mathrm{MSE}}$. For $p_{\mathrm{area}}$, we considered a one-sided paired $t$-test with alternative $H_1: \mathrm{area}_{\mathbb{S}^2} > \mathrm{area}_{S_{a, 1}}$, to test the equality of mean areas. False discovery rate correction \cite{Benjamini2001} is applied to the $p$-values at level $\alpha = 0.01$.}\label{tab:spheroid_metrics}
\end{table}

\begin{figure}[H]
    \centering
    \begin{subfigure}{0.45\textwidth}
        \centering
        \includegraphics[height=5cm]{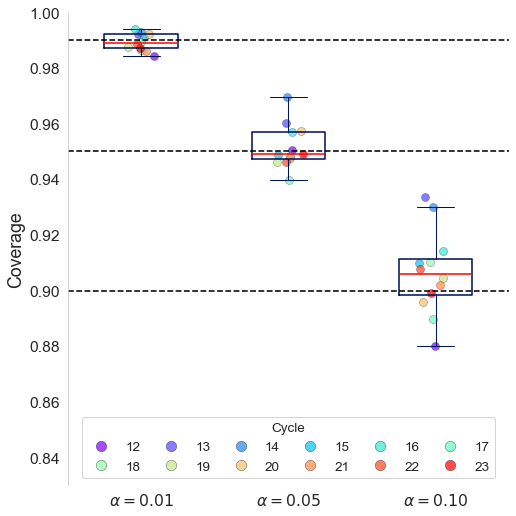}  
        \subcaption{Type II coverage}\label{fig:sunspots_coverage_area_a}
    \end{subfigure}\hfill
    \begin{subfigure}{0.45\textwidth}
        \centering
        \includegraphics[height=5cm]{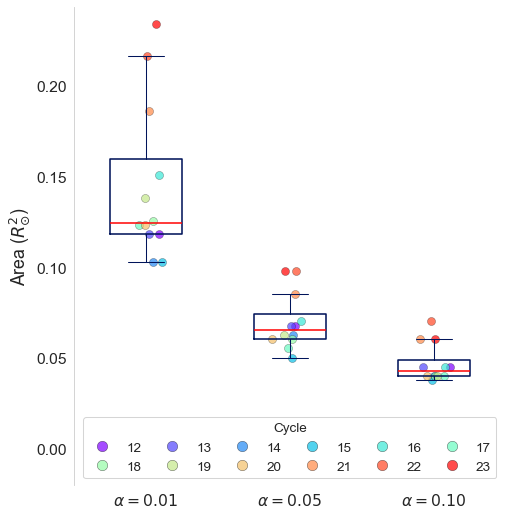}  
        \subcaption{OOB balls area}\label{fig:sunspots_coverage_area_b}
    \end{subfigure}
    \caption{\small In Figure \ref{fig:sunspots_coverage_area_a}, Type II coverage probabilities of OOB balls for each solar cycle. Figure \ref{fig:sunspots_coverage_area_b} shows the area of the OOB balls, measured in units of solar radius squared ($R_\odot^2$).}
    \label{fig:sunspots_coverage_area}
\end{figure}

In order to analyze OOB balls across cycles $12$--$23$, we compared the area and Type II coverage of OOB balls considering $\mathbb{S}^2$ as response space. Figure \ref{fig:sunspots_coverage_area_a} shows the coverage results of Type II for each cycle at significance levels $\alpha = 0.10$, $0.05$, $0.01$. The nominal coverages were honored across the different cycles and significance levels. Figure \ref{fig:sunspots_coverage_area_b} reports the area of the OOB balls, which tends to increase over successive cycles. The growth in area is a result of the larger displacements between the birth and death positions of sunspots observed in more recent cycles. This trend does not reflect heightened solar activity in recent years, but may arise from improved observational coverage in newer cycles. The fraction of days with recorded observations increased from $20\%$ in cycle $11$ to $60\%$ in cycle $23$, reducing time gaps and enabling finer trajectory tracking.

\section{Discussion}
\label{sec:discussion}

OOB prediction balls provide a principled approach to uncertainty quantification for RF predictions in metric data. The nature of a RF organically provides the OOB balls, with a main advantage over existing SC methods: the entire sample is used to train the RFs that generate the OOB balls, and to estimate the ball radius. Thus, OOB balls need half the sample size that standard SC methods require to achieve the same prediction and inference accuracy, as corroborated in simulations. Although OOB balls reduce the effective number of trees per observation by a factor of $e^{-1}$, this can be compensated by increasing the total number of trees.

The extension of the Euclidean prediction intervals \citep{Zhang2020} to the metric setting introduces theoretical and computational challenges. From the theoretical viewpoint, the standard additive-error models are unavailable in the absence of algebraic structure, so they are adapted through conditions~\ref{c2.1} and~\ref{c2.2}. The choice of metric space affects the performance of the RFs and the geometry of the OOB balls. In Section~\ref{sec:real_data}, we exploit this flexibility by constructing a data-driven distance. From the computational viewpoint, the additional cost does not stem from the construction of OOB balls itself, but from extending RFs to metric spaces (see Section \ref{sec:RFs_metric_spaces}). The estimation of Fréchet means is the main computational overhead, and depends strongly on the geometry of the metric space and on the availability of algorithms to compute them. Fréchet medoids provide a practical alternative, as evidenced in Section \ref{sec:real_data}, which also represents, to the best of our knowledge, the first application of RFWLCFR with metric predictors.

OOB balls also offer a natural path toward in-sample inference. Suppose one wants to analyze the confidence regions at the same points where the RF was trained. Rank-one-out procedures \citep{Lei2018} reduce the cost of in-sample SC inference, but still inherit the inefficiency of SC prediction due to halving the sample size, and the computational cost is doubled since one RF is trained on each half sample. A possible alternative is to apply OOB balls for in-sample inference. For each sample element $(X_i,Y_i) \in \mathcal{L}_n$, begin by replacing the RF prediction of $X_i$ with its OOB prediction as the center of the ball. For each $(X_j,Y_j) \in \mathcal{L}_n \setminus \{(X_i,Y_i)\}$, compute the prediction based on the RF comprised of the trees in which both $(X_i,Y_i)$ and $(X_j,Y_j)$ are OOB, and use the predictions to obtain the ball radius. Although the effective number of trees is reduced, only one forest is trained on the complete sample, which is an advantage over rank-one-out procedures for SC inference, especially for small sample sizes.

Our theoretical analysis assumes independence between prediction errors and predictors. Extensions to heteroscedastic settings that preserve asymptotic coverage guarantees are a natural direction for future work. Additionally, the present work focuses on standard regression trees within RFs. Adapting richer tree-based models to metric spaces, such as linear RFs \citep{Raymaekers2024} via global Fréchet regression, is another promising avenue. Future work could also focus on deriving finite-sample guarantees for OOB balls.

\section*{Supplementary materials}

Supplementary materials provide proofs omitted from the paper, additional numerical experiments, and a revision of the hyperboloid von Mises--Fisher distribution.

\section*{Acknowledgments}

The first author acknowledges funding by ``PIPF Programa Inteligencia Artificial'' by Universidad Carlos III of Madrid, and partial support from the INFLUENTIA-CM-UC3M project, funded by Comunidad de Madrid under the Multiannual Agreement with UC3M. The second author also acknowledges support from INFLUENTIA-CM-UC3M and ``Convocatoria de la Universidad Carlos III de Madrid de Ayudas para la recualificación del sistema universitario español para 2021--2023'', funded by Spain's Ministerio de Ciencia, Innovación y Universidades. The authors acknowledge access to computational resources provided by the Centro de Supercomputación de Galicia (CESGA) through the EuroCC Spain testbed program. The authors would also like to acknowledge Jesús Carretero Rodríguez for his contribution to initial stages of this work. The second author acknowledges the input by Prof. Antonio Cuevas regarding Lemma~\ref{lem:polyaprob}.

\bibliographystyle{apalike}

\fi

\ifsupplement

\newpage
\title{Supplementary materials for ``Out-of-bag prediction balls for random forests in metric spaces''}
\setlength{\droptitle}{-1cm}
\predate{}
\postdate{}
\date{}

\author{Diego Serrano$^{1,2}$ and Eduardo Garc\'ia-Portugu\'es$^{1}$}
\footnotetext[1]{Department of Statistics, Carlos III University of Madrid (Spain).}
\footnotetext[2]{Corresponding author. e-mail: \href{mailto:dieserra@est-econ.uc3m.es}{dieserra@est-econ.uc3m.es}.}
\maketitle

\begin{abstract}
In Section~\ref{supp_sec:proofs}, Theorems \ref{thm:typeII} and \ref{thm:typeIV} and Corollaries \ref{cor:typeI} and \ref{cor:typeIII} are proved, relying on four technical lemmas. Section~\ref{supp_sec:num_exp_frechet_mean} shows numerical experiments to validate the Fréchet means of Wishart-distributed random variables under the AI and LC distances, given in \eqref{eq:ai_frechet_mean} and \eqref{eq:lc_frechet_mean}, respectively. Section~\ref{supp_sec:other_sim_results} contains a comprehensive list of figures and tables from Section \ref{sec:experiments}. In Section~\ref{supp_sec:hvmf}, we reapproach the derivations by \citet[pp. 196--197]{Jensen1981|SM} to clarify the connection of the HvMF and vMF distributions, and to detail the derivation and computation of the normalizing constant. 
\end{abstract}
\begin{flushleft}
	\small\textbf{Keywords:} Confidence regions; Fréchet mean; Random objects; Regression. 
\end{flushleft}

\appendix

\section{Proofs of the main results}\label{supp_sec:proofs}

We provide the proof of Theorems \ref{thm:typeII} and \ref{thm:typeIV} and Corollaries \ref{cor:typeI} and \ref{cor:typeIII} of the main paper. The proofs rely on four technical lemmas. Lemma~\ref{lem:FRhat} states uniform convergence in probability of $F_{\Roob_1, \ldots, \Roob_n}$ to $F_R$, which is required to prove Theorems~\ref{thm:typeII} and \ref{thm:typeIV}. Lemma~\ref{lem:Deltan} provides convergence in probability to zero of a remainder term in the proof of Theorem~\ref{thm:typeII}, and Lemma~\ref{lem:Deltan_x} is the analogous result for Theorem~\ref{thm:typeIV}. Finally, Lemma~\ref{lem:polyaprob}, a probabilistic version of Polya's Theorem, guarantees uniform convergence in probability to a continuous distribution $F$ from simple convergence in probability. This result is necessary to prove Lemmas \ref{lem:FRhat}, \ref{lem:Deltan}, and \ref{lem:Deltan_x}.

First, the lemmas are stated and proved.
\begin{lemma} \label{lem:FRhat}
Under conditions \ref{c1}, \ref{c2.3}, and \ref{c3.2}, as $n\to \infty$,
\begin{align}\label{eq:sup0}
\sup_{t \in \R}\left|F_{\Roob_1, \ldots, \Roob_n}(t)-F_{R}(t)\right| \convp 0. 
\end{align}
In particular, $F_{R}\big(\widehat{R}_{[1-\alpha, n]}\big) \convp 1-\alpha$.
\end{lemma}
\begin{proof}[Proof of Lemma~\ref{lem:FRhat}] \!
Since $F_R$ is continuous by condition \ref{c2.3}, applying the probabilistic version of Polya's Theorem from Lemma~\ref{lem:polyaprob}, it is sufficient to show that, for every $t \in \R$,
\begin{align}\label{eq:pointwise}
F_{\Roob_1, \ldots, \Roob_n}(t) - F_R(t) \convp 0  
\end{align}
as $n$ diverges to infinity to obtain \eqref{eq:sup0}.

To arrive at \eqref{eq:pointwise}, first apply the triangle and reverse triangle inequalities, to obtain
\begin{align*}
U_i \defin &\; \left| \dY\left( Y_i, m(X_i)\right) - \dY\left( m(X_i), \widehat{m}_{(i)}(X_i) \right) \right|
\le \Roob_i 
\\
\le &\; 
L_i\defin  \dY\left( Y_i, m(X_i)\right) + \dY\left( m(X_i), \widehat{m}_{(i)}(X_i) \right),
\end{align*}
for every $i=1,\ldots,n$. Hence, for every $t_1, t_2 \in \R$,
$$
\{L_1  \le t_1, L_2 \le t_2\}  \subset \{\Roob_1 \le t_1,\Roob_2 \le t_2\}\subset\{U_1 \le t_1, U_2 \le t_2 \}.
$$
Therefore, it is verified
\begin{align*}
P_1(t_1,t_2) \defin&\;
\Prob{
L_1  \le t_1,
L_2 \le t_2}
\le
\Probbig{
\Roob_1 \le t_1,
\Roob_2 \le t_2
}
\\
\le
&\; \Prob{
U_1 \le t_1,
U_2 \le t_2
} 
=:P_2(t_1,t_2).
\end{align*}
Observe that since by \ref{c3.2}, $\dY(m(X_{1}),\widehat{m}_{(1)}(X_{1})) \convp 0$ and $\dY(m(X_{2}),  \allowbreak \widehat{m}_{(2)} (X_{2})) \allowbreak \convp 0$, Slutsky's Theorem provides both 
$P_1 (t_1,t_2) \to \Probbig{\{ \dY(Y_1, m(X_1))  \le t_1\}, \allowbreak 
\{ \dY( Y_2, m(X_2))) \le t_2 \} }$ 
and 
$P_2(t_1,t_2) \to \Probbig{\{ \dY( Y_1, m(X_1))  \le t_1\}, 
\{ \dY( Y_2, m(X_2))) \le t_2 \} }$ for every $t_1, t_2 \in \R$. Hence, we conclude that 
\begin{align}\label{bivariate_convergence_lem:FRhat}
    \big(\Roob_1, \Roob_2\big) \convd \big( \dY\left(Y_1, m(X_1)\right), \dY\left(Y_2, m(X_2)\right)\big) \text{ as } n \to \infty.
\end{align}
By condition \ref{c1}, both components of the limit vector $(\dY\left(Y_1, m(X_1)\right), \dY\left(Y_2, m(X_2)\right))$ are iid with CDF $F_R$.

Even though \eqref{bivariate_convergence_lem:FRhat} provides convergence in distribution of the OOB errors to $R$,  $F_{\Roob_1, \ldots, \Roob_n}$ is not the distribution of the OOB errors. Thus, to prove \eqref{eq:pointwise}, we first verify that by \eqref{bivariate_convergence_lem:FRhat}, for every $t\in \R$,
$$
\Elr{F_{\Roob_1, \ldots, \Roob_n}(t)} = \frac{1}{n} \sum_{i=1}^{n}\Elr{1_{\{\Roob_i \le t\}}} = \Prob{ \Roob_1 \le t} \convn F_R(t).
$$
Notice that $\Roob_1, \ldots, \Roob_n$ are not necessarily independent, but they are identically distributed. Finally, observe that
$$
\begin{aligned}
\mathsf{Var}\left[F_{\Roob_1, \ldots, \Roob_n}(t)\right] =&\; \frac{1}{n} \mathsf{Var}\left[1_{\{\Roob_1 \le t\}} \right] + \frac{n(n-1)}{n^2} \mathsf{Cov}\left[1_{\{\Roob_1 \le t\}}, 1_{\{\Roob_2\le t\}}\right]
\\
\le&\; \frac{1}{n} + \Prob{\Roob_1 \le t, \Roob_2 \le t} - \Prob{ \Roob_1 \le t}^2
\convn 0,
\end{aligned}
$$
as $n\to \infty$, where convergence to $0$ is guaranteed thanks to \eqref{bivariate_convergence_lem:FRhat}, since $ \Roob_1$ and  $\Roob_2$ are not independent in general.
In conclusion, as $n\to \infty$, $F_{\Roob_1, \ldots, \Roob_n}(t) - F_R(t) \convp 0$ for every $t \in \R$, and hence \eqref{eq:pointwise} holds.

The second claim follows in a similar manner to Lemma~$1$ in \cite{Zhang2020|SM}. We want to prove that $F_R\big(\widehat{R}_{[n, \gamma]}\big) \convp \gamma$ as $n\to \infty$ for any $\gamma \in(0,1)$. We define
$$
\begin{aligned}
a\defin \inf \left\{t \in \R: F_R(t) \geq \gamma\right\}, \quad b\defin \sup \left\{t \in \R: F_{R}(t) \leq \gamma\right\}.
\end{aligned}
$$
We observe that $a\le b$ and $F_R(a-\varepsilon) < \gamma < F_R(b+\varepsilon)$ for every $\varepsilon>0$. By the first claim of Lemma~\ref{lem:FRhat}, we have that $\Probbig{ F_{\Roob_1, \ldots, \Roob_n}(a-\varepsilon) < \gamma < F_{\Roob_1, \ldots, \Roob_n} (b+\varepsilon)  } \to 1$ for every $\varepsilon>0.$ The event $ F_{\Roob_1, \ldots, \Roob_n}(a-\varepsilon) < \gamma < F_{\Roob_1, \ldots, \Roob_n} (b+\varepsilon)  $ implies $\widehat{R}_{ [n, \gamma] } \in [a-\varepsilon , b + \varepsilon]$,~so 
$$
\Big|F_R\big(\widehat{R}_{[n, \gamma]}\big)-\gamma \Big| \leq \Lambda(\varepsilon) \defin F_R(b+\varepsilon) - F_R(a-\varepsilon),
$$
due to the monotonicity of $F_R$. Since 
$$
\lim _{n \rightarrow \infty}\Prob{\left|F_R\big(\widehat{R}_{[n, \gamma]}\big)-\gamma \right| \leq \Lambda(\varepsilon)}=1 \text{ for every }  \varepsilon>0
$$
and $\lim _{\varepsilon \to 0} \Lambda(\varepsilon)=0$ because of \ref{c2.3} (which implies $F_R(a)=F_R(b)=\gamma$), we then conclude that $F_R\big(\widehat{R}_{[n, \gamma]}\big) \convp\gamma$.
\end{proof}

\begin{lemma}\label{lem:Deltan}
Under conditions \ref{c1}, \ref{c2.1}, \ref{c2.3}, and \ref{c3.1}, as $n\to \infty$,
\begin{align}\label{eq_lem:Deltan}
\Delta_{n} \defin \sup_{t \in \R} \big|\Probstar{\dY\left(Y, \widehat{m}(X) \right)<t}-F_{R}(t) \big| \convp 0, 
\end{align}
where $\mathsf{P}_*(\cdot) = \mathsf{P}(\cdot | \mathcal{L}_n).$
\end{lemma}

\begin{proof}[Proof of Lemma~\ref{lem:Deltan}] \!
Denote $f(t^-) \defin  \lim_{s\to t^-} f(s)$ and $f(t^+) \defin  \lim_{s\to t^+} f(s)$ for any given function $f$. First, observe that the equality $\Delta_n = \sup_{t \in \R} \big| \Probstar{\dY\left(Y, \widehat{m}(X) \right) \le t}-F_{R}(t) \big|$ holds by applying $ \Probstar{ \dY\left(Y, \widehat{m}(X) \right) \le t^-} = \allowbreak \Probstar{ \dY\left(Y, \widehat{m}(X) \right) < t}$ and $\Probbig{\dY (Y, \allowbreak \widehat{m}(X)) \le t}  = \Probstar{ \dY\left(Y, \widehat{m}(X) \right) < t^+}$, together with the continuity of $F_R$ by condition \ref{c2.3}. Now, we prove $\sup_{t \in \R} \big| \Probstar{\dY\left(Y, \widehat{m}(X) \right) \le t}-F_{R}(t) \big| \convp 0$. By an application of Lemma~\ref{lem:polyaprob}, it is only required to verify that for every $t \in \R,$
\begin{align}\label{no_uniforme_prob}
    \Probstar{\dY\left(Y, \widehat{m}(X) \right) \le t}-F_{R}(t) \convp 0 \ \text{ as } n\to \infty.
\end{align}
Notice that the hypotheses of Lemma~\ref{lem:polyaprob} are verified since, by \ref{c2.3}, $F_R$ is continuous. The result is trivial for $t<0$ due to the nonnegativity of distance functions, so if we prove \eqref{no_uniforme_prob} for $t>0$, then the continuity of $F_R$ by condition \ref{c2.3} entails \eqref{no_uniforme_prob} for every $t\in \R$. To see this, take $t=0$ and suppose that \eqref{no_uniforme_prob} holds for positive $t$; then, for all $\varepsilon >0$, 
$$ 
0\le
\Probstar{\dY\left( Y, \widehat{m}(X)\right) = 0 }
\le
\Probstar{\dY\left( Y, \widehat{m}(X)\right) \le \varepsilon },
$$
and hence, by \ref{c3.1}, it holds that $0 \le \mathrm{plim}_{n\to\infty}\Probstar{\dY\left( Y, \widehat{m}(X)\right)=0}\leq F_R(\varepsilon).$
Due to the continuity of $F_R$ by \ref{c2.3}, $F_R(0^+) = 0$ and hence $\Probstar{\dY\left( Y, \widehat{m}(X)\right)=0}\convp0$.

It is only left to prove \eqref{no_uniforme_prob} for positive $t$. First, observe that for every $t>0$, 
\begin{align}\label{consistency_P_*_Deltan}
\Probstar{\dY(m(X), \widehat{m}(X)) > t } \convp 0 \ \text{ as } n \to \infty.
\end{align}
This result follows from consistency in probability of the estimator $\widehat{m}(X)$ as stated in condition \ref{c3.1}, since for every $t>0$,
$$
\Elr{ \Probstar{\dY\left( m(X),\widehat{m}(X)\right) > t}} 
 = 
 \Prob{ \dY\left( m(X), \widehat{m}(X) 
\right) > t
} 
\rightarrow 0 \text{ as } n \to \infty.
$$ 
Thus, since convergence in $1$-mean implies convergence in probability, \eqref{consistency_P_*_Deltan} holds.

Now, we will use the triangle and reverse triangle inequalities to provide (lower and upper) bounds for $\Probstar{\dY\left( Y, \widehat{m}(X)\right) > t}$, and show that they converge in probability to $1-F_R(t)$.

We will use the following notation:
$$
    A_n \defin  \{ \dY\left(m(X), \widehat{m}(X)\right) > t - \dY\left( Y, m(X)\right) \}, \quad 
    B \defin  \{ t - \dY\left(Y, m(X) \right) > 0 \}.
$$
Notice that $\mathsf{P}_*(A_n \mid B) \convp 0 $ by \eqref{consistency_P_*_Deltan}, and $\mathsf{P}_*\big(A_n \mid \overline{B}\big)  = 1$. Note also that $\mathsf{P}_*(B) = \mathsf{P}\{\dY\left(Y, m(X)\right) \allowbreak < t\} \allowbreak = F_R(t)$ due to the independence of $R = \dY\left(Y, m(X)\right)$ from $\mathcal{L}_n$ by condition \ref{c1}. Using the triangle inequality, for $t>0$ we have that
$$
\{ \dY\left( Y, \widehat{m}(X)\right) > t \}
\subset
\{
\dY\left( Y, m(X)\right)
+
\dY\left(m(X), \widehat{m}(X)\right) > t
\}
= A_n
$$
and, therefore,
\begin{align*}
\mathsf{P}_*\left(\dY\left( Y, \widehat{m}(X)\right) > t \right)
\le 
\mathsf{P}_*\left(A_n \mid B\right) \mathsf{P}_*\left(B\right) + \mathsf{P}_*\left(A_n \mid \overline{B} \right) \mathsf{P}_*\left( \overline{B} \right)
\defin U_n(t)
\convp 1 - F_R(t)
\end{align*}
as $n \to \infty$, which provides the upper bound.

Second, to obtain the lower bound, let
$$
\begin{aligned}
C_n \defin&\; \left\{\dY\left(m(X), \widehat{m}(X)\right) > \dY\left(Y, m(X) \right) + t  \right\}, 
\\
D_n \defin &\; \left\{\dY\left(m(X), \widehat{m}(X)\right) < \dY\left(Y, m(X) \right) - t\right\}, 
\\
L_n(t) \defin&\; \mathsf{P}_*(C_n) + \mathsf{P}_*\big(D_n  \mid \overline{B} \big)\mathsf{P}_*\big(\overline{B}\big).
\end{aligned}
$$
Notice that $\dY\left(Y, m(X) \right) + t >0$ and hence $\mathsf{P}_*(C_n) \convp 0$ by \eqref{consistency_P_*_Deltan}. Similarly, \eqref{consistency_P_*_Deltan} implies $\mathsf{P}_*\big(D_n \mid \overline{B}\big) \convp 1$, and hence $L_n(t) \convp \mathsf{P}_*\big(\overline{B}\big) = 1-F_R(t)$ for all $t>0$. Also, $\mathsf{P}_*(D_n \mid B) = 0$.

Applying the reverse triangle inequality, for every $t > 0$
$$
\{
|\dY\left( Y, m(X)\right)
-
\dY\left(m(X), \widehat{m}(X)\right)| > t
\}
\subset
\{ \dY\left( Y, \widehat{m}(X)\right) > t \},
$$
and therefore
$$
\begin{aligned}
\Probstar{\dY\left( Y, \widehat{m}(X)\right) > t } \ge&\;
\Probstar{ \left| \dY\left( Y, m(X)\right) - \dY\left(m(X), \widehat{m}(X)\right) \right| > t
} 
=
\mathsf{P}_*( C_n )
+ 
\mathsf{P}_* ( D_n )
\\
=&\;
\mathsf{P}_*(C_n)
+ \mathsf{P}_*(D_n  \mid B )\mathsf{P}_*(B)
+ \mathsf{P}_*(D_n  \mid \overline{B} )\mathsf{P}_*(\overline{B})
L_n(t)
\convp 1-F_R(t)
\end{aligned}
$$
as $n\to \infty$. Finally, applying the squeeze theorem in probability (Remark~\ref{rem:squeeze}), since $L_n\le$ $\Probstar{\dY\left( Y, \allowbreak \widehat{m}(X)\right) > \allowbreak t }$ $\le U_n$ for every $n > 0$, then \eqref{no_uniforme_prob} is proved.
\end{proof}

\begin{remark} \label{rem:squeeze}
To verify that the squeeze theorem holds for convergence in probability, notice that if $Y_n \le X_n \le Z_n$ and $Y_n \convp X$, $Z_n \convp X$ for some random variable $X$, then for every $\varepsilon>0$, it holds that $\left\{|X_n - X | >\varepsilon  \right\} \subset \left\{|Y_n - X|>\varepsilon\right\} \cup\left\{|Z_n-X|>\varepsilon\right\}$. Hence, $X_n \convp X$. 
\end{remark}

\begin{lemma}\label{lem:Deltan_x}
Let $\mathsf{P}_*(\cdot) = \mathsf{P}(\cdot | \mathcal{L}_n, X=x).$ Under assumptions \ref{c1}, \ref{c2} and \ref{c3.1}, for every $x \in \mathcal{X}$ such that $\dY(\widehat{m}(x), m(x)) \convp 0$ as $n\to \infty$, we have that 
\begin{align}\label{eq_lem:Deltan_x}
\Delta_{n}^\prime \defin \sup_{t \in \R} \left| \Probstar{\dY\left(Y, \widehat{m}(X) \right)<t}-F_{R}(t) \right| 
\convp 0,
\end{align}
as $n$ diverges to infinity.
\end{lemma}
\begin{proof}[Proof of Lemma~\ref{lem:Deltan_x}] \!
The proof follows in the same way as that of Lemma~\ref{lem:Deltan}, with the new definition of conditional probability $\mathsf{P}_*(\cdot) = \mathsf{P}(\cdot \mid \mathcal{L}_n, X=x)$. The only difference is that $\dY(\widehat{m}(x), m(x)) \convp 0$ is used instead of condition \ref{c3.1} to prove \eqref{consistency_P_*_Deltan}, and $\mathsf{P}_*(B) = \Prob{\dY\left(Y, m(X)\right) < t} = F_R(t)$ due to the independence of $R = \dY\left(Y, m(X)\right)$ from $\mathcal{L}_n$ by condition \ref{c1} and independence from $X$ by condition \ref{c2.2}.
\end{proof}

\begin{lemma}
\label{lem:polyaprob}
Let $F:{\mathbb R}\rightarrow [0,1]$ be a continuous CDF.
Let $F_n:{\mathbb R}\rightarrow [0,1]$ be
 a random sequence of CDFs such that, for all $t$,
 \begin{align}\label{hip}
 	F_n(t)\convp F(t) \mbox{ as } n\to\infty.
 \end{align}
Then,
\begin{align}\label{tesis}
	\sup_{t\in\R}|F_n(t)- F(t)| \convp 0
\end{align}
as $n$ diverges to infinity.
\end{lemma}

\begin{proof}[Proof of Lemma~\ref{lem:polyaprob}] \!
Take $\varepsilon>0$. We want to prove 
$$
Q_n(\varepsilon)\defin {\mathsf{P}}\bigg\{\sup_{t\in \R}|F_n(t)-F(t)|> \varepsilon\bigg\}\to 0 \mbox{ as } n\to\infty.
$$
The sequence $\{Q_n(\varepsilon)\}$ converges to zero if, given any subsequence $\{Q_m(\varepsilon)\}$, there is a further subsequence $\{Q_k(\varepsilon)\}$ converging to zero as $k\to\infty$. We prove this.

Since $F$ is a continuous probability CDF, there exist points $x_0<\cdots<x_N$ such that 
$F(x_0)<\varepsilon/2$, $F(x_N)>1-\varepsilon/2$, and $F(x_i)-F(x_{i-1})<\varepsilon/2$, for $i=1,\ldots,N$ and $N \ge 1$.

From the characterization of convergence in probability in terms of subsequences, we have that, given the subsequence $\{F_{m}\}$, there is 
a further subsequence, denoted $\{F_k\}$, such that $F_{k}(x_i)\to F(x_i)$ as $k\to\infty$, almost surely for all $i=0,\ldots,N$. From the monotonicity of $F_k$ and the assumption \eqref{hip}, we have
\begin{align}
    &\limsup_{k \ge 1}{\mathsf{P}}\bigg\{\sup_{t\leq x_0}|F_k(t)-F(t)|> \varepsilon\bigg\}\leq {\mathsf{P}}\bigg\{\limsup_{k \ge 1}\Big\{\sup_{t\leq x_0} |F_k(t)-F(t)|>\varepsilon\Big\}\bigg\}=0, \label{eq:Fx0}
\end{align}
where the inequality holds by (Reverse) Fatou's Lemma. To see the equality in \eqref{eq:Fx0}, observe that, by the monotonicity of $F_k$ and $F$,
\begin{align*}
    \sup_{t\leq x_0}|F_k(t)-F(t)|\leq \sup_{t\leq x_0}\{F_k(t)+F(t)\} \leq F_k(x_0) + F(x_0)< F_k(x_0)+\varepsilon/2,
\end{align*}
and hence $ \{F_k(x_0)+\varepsilon/2\leq \varepsilon\}\subset \{\sup_{t\leq x_0} |F_k(t)-F(t)|\leq \varepsilon\}$. Therefore, 
$$
{\mathsf{P}}\bigg\{\limsup_{k \ge 1}\Big\{\sup_{t\leq x_0} |F_k(t)-F(t)|\leq\varepsilon\Big\}\bigg\}\geq {\mathsf{P}}\bigg\{\limsup_{k \ge 1}\{ F_k(x_0)\leq\varepsilon/2\}\bigg\}=1
$$
since $F_k(x_0)\to F(x_0)<\varepsilon/2$ a.s. Similarly,
\begin{align*}
	&\;\limsup_{k \ge 1}{\mathsf{P}}\bigg\{\sup_{t\geq x_N}|F_k(t)-F(t)|> \varepsilon\bigg\}\leq {\mathsf{P}}\bigg\{\limsup_{k \ge 1}\Big\{\sup_{t\geq x_N} |F_k(t)-F(t)|>\varepsilon\Big\}\bigg\}=0.
\end{align*}
Now, for $t\in I_i\defin [x_{i-1},x_i]$, with $i=1,\ldots,N$,
\begin{align*}
    \limsup_{k \ge 1}{\mathsf{P}}\bigg\{\sup_{t\in I_i}|F_k(t)-F(t)|> \varepsilon\bigg\} \leq&\; {\mathsf{P}}\bigg\{\limsup_{k \ge 1}\Big\{\sup_{t\in I_i} |F_k(t)-F(t)|>\varepsilon\Big\}\bigg\}
    \\
    \leq&\; {\mathsf{P}}\bigg\{\limsup_{k \ge 1} \Big\{\max\big\{ |F_k(x_{i-1})-F(x_{i})|,
    \\
    & \qquad   |F_k(x_{i})-F(x_{i-1})|\big\}>\varepsilon\Big\}\bigg\}
    = 0,
\end{align*}
because $F(x_{i})-F(x_{i-1})<\varepsilon/2$. To see the second inequality, note that
$$
F_k(x_{i-1})-F(x_{i})\leq F_k(x_{i-1})-F(t)\leq F_k(t)-F(t)\leq F_k(x_i)-F(x_{i-1}).
$$
We have proved that there is a subsequence $\{Q_k(\varepsilon)\}$ of the subsequence $\{Q_m(\varepsilon)\}$ such that $Q_k(\varepsilon)\to 0$, as $k\to\infty$.
\end{proof}

\begin{remark} \label{rem:convergence_x}
    The assumption $\dY(m(x), \widehat{m}(x)) \convp 0$ as $n\to \infty$ for every $x \in \mathcal{X}$ is implied (a.s.) by condition \ref{c3.1}. To see this, let $A$ be a subset of the support of $X$ verifying that for every $x\in A$, there exists $\delta(x)>0$ such that $\lim_{n\to\infty} \Prob{\dY\left(m(x), \widehat{m}(x)\right) > \delta(x)} > 0$ (i.e., $\dY(m(x), \widehat{m}(x)) \not\convp 0$ as $n\to\infty$). Define $A_{\delta} \defin \{x\in A: \, \lim_{n\to \infty} \Prob{\dY\left(m(x), \widehat{m}(x) \right) >  \delta} > 0 \}$ for $\delta>0$. Notice that $\{ A_\delta\}_{\{\delta>0\}}$ is a nested sequence of subsets of $A$ that grows to $A$ as $\delta \to 0$. Now, suppose $\mathsf{P}(A)>0$. We have that $\mathsf{P}(A) = \mathsf{P}(\cup_{\delta>0} A_\delta) = \mathsf{P}(\cup_{n=1}^\infty A_{1/n}) \le \sum_{n=1}^\infty \mathsf{P}(A_{1/n}) $. Since $\mathsf{P}(A) > 0$, then necessarily $\mathsf{P}(A_{1/n_0}) > 0$ for some $n_0 \in \mathbb{N}$.
    Hence, we have that
    \begin{align*}
    \lim_{n\to \infty} \Prob{\dY\left(m(X), \widehat{m}(X)\right) > n_0^{-1}} = \lim_{n\to \infty} \Prob{\dY\left(m(X), \widehat{m}(X)\right) > n_0^{-1} \ \Big\lvert \big. \ A_{1/n_0}} \Prob{A_{1/n_0}}
    > 0,
    \end{align*}
    in contradiction with $\dY(m(X), \widehat{m}(X) )\convp 0$ in \ref{c3.1}. We thus conclude that $\mathsf{P}(A)=0$, i.e., $\dY(m(x), \widehat{m}(x)) \allowbreak \convp 0$ as $n\to \infty$ for all $x\in \mathcal{X}$ $\mathbb{G}$-a.s.
\end{remark}

\begin{remark}
The assumption $\dY(m(x), \widehat{m}(x)) \convp 0$ as $n\to \infty$, for every $x\in\mathcal{X}$, clearly implies condition \ref{c3.1}. Indeed,
$$
\Prob{\dY(m(X), \widehat{m}(X))>\varepsilon}
=\int_{\mathcal{X}} \Prob{\dY(m(x), \widehat{m}(x))>\varepsilon}\,\rd \mathbb{G}_X(x) \to 0
$$
by the dominated convergence theorem applied to the bounded function $\Prob{\dY(m(x), \widehat{m}(x))>\varepsilon} \allowbreak \to 0$, where $\mathbb{G}_X$ is the law of $X$ on $\mathcal{X}$.
\end{remark}

Using the above lemmas, Theorems~\ref{thm:typeII} and \ref{thm:typeIV}, and Corollaries~\ref{cor:typeI} and \ref{cor:typeIII} can be proved.
\begin{proof}[Proof of Theorem~\ref{thm:typeII}] \!
    Denoting $\mathsf{P}_*(\cdot) \defin  \mathsf{P}(\cdot | \mathcal{L}_n)$, we have that
    \begin{align}\label{eq:thm1_decomposition}
   \Prob{ Y\in \mathrm{PB}_{1-\alpha}^{\mathrm{oob}}(X,\mathcal{L}_n) \mid \mathcal{L}_n}=\Probstar{
    \dY\left(Y,\widehat{m}(X)\right) < \widehat{R}_{[1-\alpha, n]}
    } = 
    F_{R} (\widehat{R}_{[1-\alpha, n]})
    +
    S_n,
    \end{align}
    where $S_n \defin  \mathsf{P}_*\big\{ \dY\left(Y,\widehat{m}(X)\right) < \widehat{R}_{[1-\alpha, n]}\big\} - F_{R} \big(\widehat{R}_{[1-\alpha, n]}\big)$. On the one hand, by Lemma~\ref{lem:FRhat}, $F_{R} \big(\widehat{R}_{[1-\alpha, n]}\big) \convp 1-\alpha$. On the other hand, $S_n\convp 0$ because by Lemma~\ref{lem:Deltan}, $ | S_n | \le \Delta_{n} \convp 0 $ with $\Delta_n$ as defined in \eqref{eq_lem:Deltan}.
\end{proof}

\begin{proof}[Proof of Corollary~\ref{cor:typeI}] \!
Since $\Probbig{Y \in \mathrm{PB}_{1-\alpha}^{\mathrm{oob}}\left(X, \mathcal{L}_{n}\right)}=\Ebig{\Probbigstar{Y \in \mathrm{PB}_{1-\alpha}^{\mathrm{oob}}\left(X, \mathcal{L}_{n}\right)}}$, then the result follows if $\Probbigstar{Y \in \mathrm{PB}_{1-\alpha}^{\mathrm{oob}}\left(X, \mathcal{L}_{n}\right)}$ converges in $1$-mean to $1-\alpha$. This follows since $\Probbigstar{Y \in \mathrm{PB}_{1-\alpha}^{\mathrm{oob}}\left(X, \mathcal{L}_{n}\right)}\convp 1-\alpha$ by Theorem~\ref{thm:typeII} and the sequence is uniformly bounded by one.
\end{proof}

\begin{proof}[Proof of Theorem~\ref{thm:typeIV}] \!
Redefining the conditional probability $\mathsf{P}_*$ in Theorem~\ref{thm:typeII} as
$
\mathsf{P}_*(\cdot)\defin \mathsf{P}(\cdot \mid \mathcal{L}_n,\allowbreak \, \allowbreak X\allowbreak=x),
$
the proof of Theorem~\ref{thm:typeII} can be directly adapted by applying \eqref{eq:thm1_decomposition} and defining $S_n$ with the new formulation for $\mathsf{P}_*$. Lemma~\ref{lem:FRhat} still applies, so $F_{R} \big(\widehat{R}_{[1-\alpha, n]}\big) \convp 1-\alpha$, and Lemma~\ref{lem:Deltan_x} provides $S_n\convp0$ because $ | S_n | \le \Delta'_{n} \convp 0 $ with $\Delta'_n$ as defined in \eqref{eq_lem:Deltan_x}.
\end{proof}

\begin{proof}[Proof of Corollary~\ref{cor:typeIII}] \!
Denoting 
$\mathsf{P}_*(\cdot)\defin \mathsf{P}\left(\cdot \mid \mathcal{L}_n,\, X=x\right)$,
this result can be proved from Theorem~\ref{thm:typeIV} in the same way as Corollary~\ref{cor:typeI} was proved from Theorem~\ref{thm:typeII}, since $\Probbig{Y \in \mathrm{PB}_{1-\alpha}^{\mathrm{oob}}\left(X, \mathcal{L}_{n}\right) \mid X=x}=\Ebig{\Probbigstar{Y \in \mathrm{PB}_{1-\alpha}^{\mathrm{oob}}\left(X, \mathcal{L}_{n}\right)}}$ and $\Probbigstar{Y \in \mathrm{PB}_{1-\alpha}^{\mathrm{oob}}\left(X, \mathcal{L}_{n}\right)}\convp 1-\alpha$ by Theorem~\ref{thm:typeIV}.
\end{proof}

\section{\texorpdfstring{Numerical experiments for the Fréchet means in $\mathcal{S}_q^+$}{Numerical experiments for the Fréchet means in Sq+}}\label{supp_sec:num_exp_frechet_mean}

We present numerical experiments to validate the Fréchet means of Wishart-distributed random variables under the AI and LC distances, given in \eqref{eq:ai_frechet_mean} and \eqref{eq:lc_frechet_mean}, respectively.
We considered SPD random matrices $\bS \sim \mathrm{Wishart}_q(d, \bSigma)$, with different choices of $q$, $d$, and $\bSigma$. The candidates for the true Fréchet mean of $\bS$ are defined through an interpolation between the extrinsic mean $\bM_{\mathrm{Ext}}$ (i.e., the Euclidean mean of the matrices in the ambient space), the Fréchet mean for the AI distance $\bM_{\mathrm{AI}}$, and the analog for the LC distance $\bM_{\mathrm{LC}}$:
$$
\bM(t) \defin
\begin{cases}
(1+t) \bM_{\mathrm{AI}} - t \bM_{\mathrm{Ext}}, & t \in [-1,0], \\
(1-t) \bM_{\mathrm{AI}} + t \bM_{\mathrm{LC}}, & t \in (0,1], \\
(2-t) \bM_{\mathrm{LC}} + (t-1) \bM_{\mathrm{Ext}}, & t \in (1,2].
\end{cases}
$$
The values of $t$ are selected from an equispaced grid of $600$ values in $[-1,2]$. To locate the true Fréchet mean empirically for the AI distance (analogously for the LC distance), the normalized Fréchet loss function is estimated at each candidate matrix $\bM(t)$
\begin{equation}\label{normalized_frechet_loss}
    \mathcal{L}_{\mathrm{AI}}(\bM(t)) \defin \frac{\mathbb{E}[d^2_{\mathrm{AI}}(\bM(t),\bS)] - \mathbb{E}[d^2_{\mathrm{AI}}(\bM_\mathrm{AI}, \bS)]}{\mathbb{E}[d^2_{\mathrm{AI}}(\bM_\mathrm{AI}, \bS)]}.
\end{equation}
Each expectation in \eqref{normalized_frechet_loss} is estimated by sampling $M=25000$ values of $\bS \sim \mathrm{Wishart}_q(d, \bSigma)$.

Figure~\ref{fig:frechet_loss} compares the normalized Fréchet loss function for the AI and LC metrics, under three different scenarios for $\bS \sim \mathrm{Wishart}_q(d, \bSigma)$, depending on the scale matrix $\bSigma$, the matrix dimension $q$, and the degrees of freedom $d$. The results show that the loss function is minimized at the Fréchet mean \eqref{eq:ai_frechet_mean} or \eqref{eq:lc_frechet_mean}, respectively for the AI and LC distances.

\begin{figure}[H]
    \centering
    \begin{subfigure}{0.32\textwidth}
        \includegraphics[height=5cm]{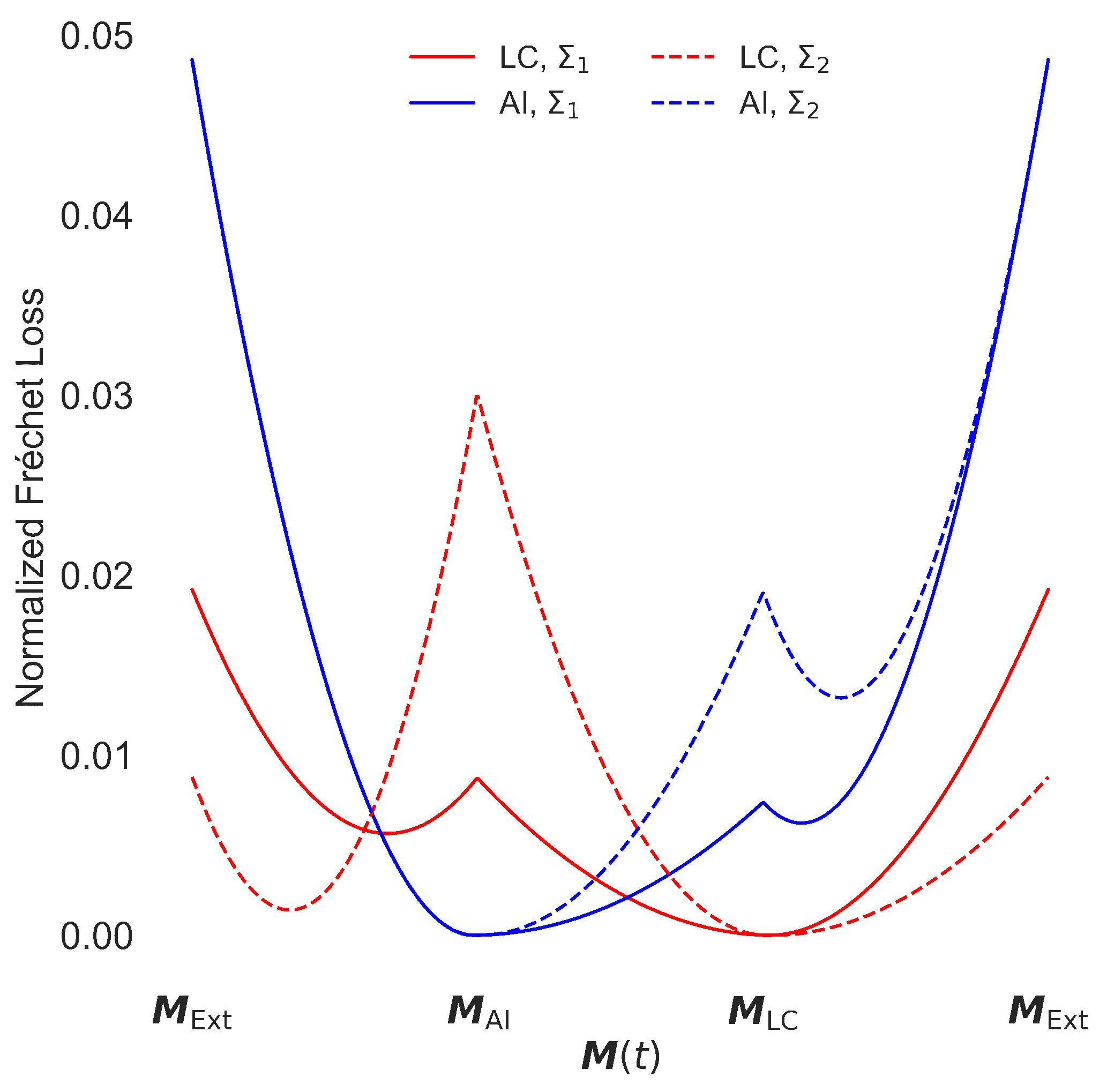}
        \caption{\small $q=2$, $d=15$.}
    \end{subfigure}
    \begin{subfigure}{0.32\textwidth}
        \includegraphics[height=5cm]{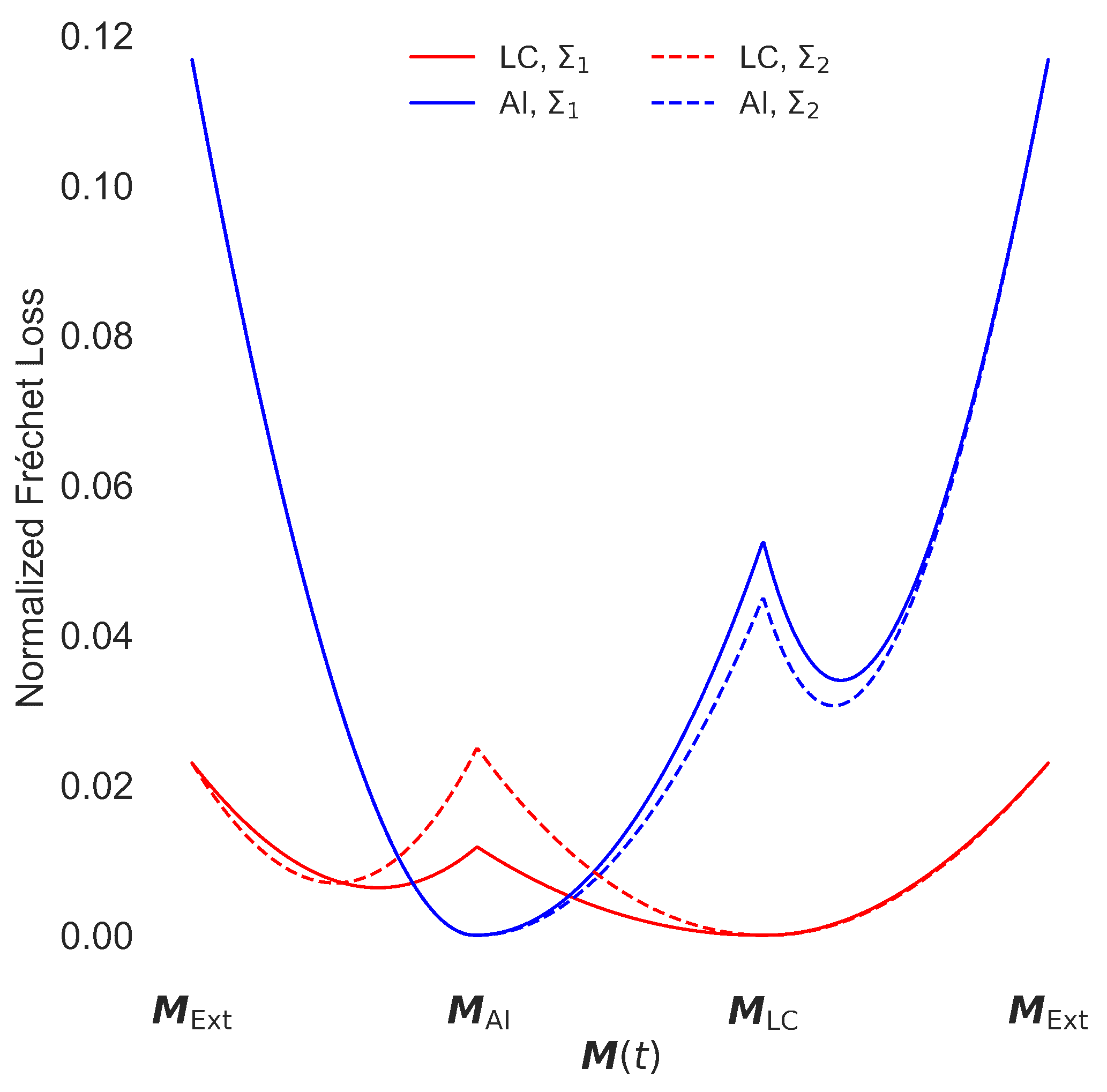}
        \caption{\small $q=6$, $d=15$.}
    \end{subfigure}
    \begin{subfigure}{0.32\textwidth}
        \includegraphics[height=5cm]{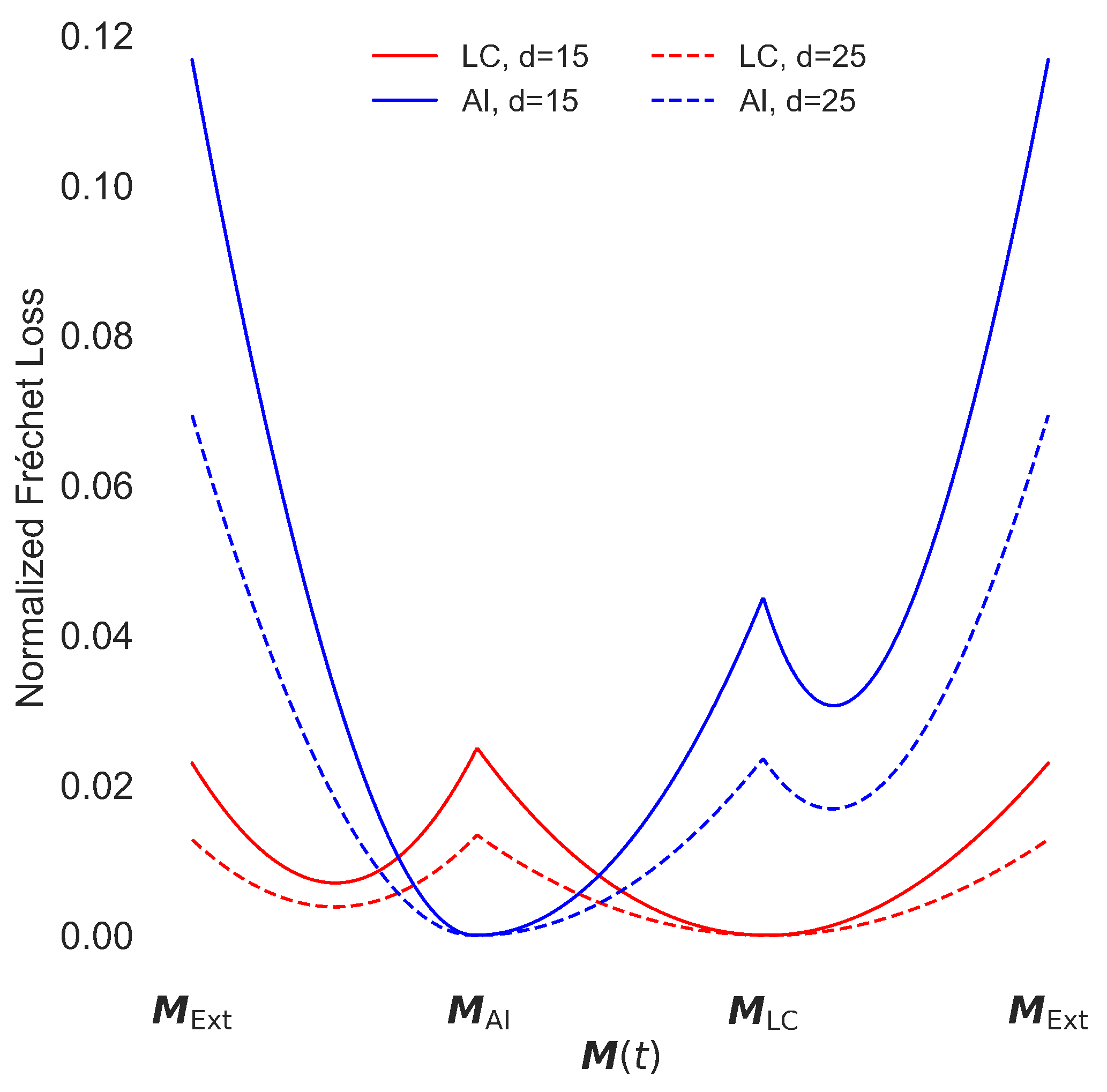}
        \caption{\small $q=6$, $d=15$, $25$.}
    \end{subfigure}
    \caption{\small Normalized Fréchet loss functions $t \mapsto \mathcal{L}_{\mathrm{AI}}\big(\bM(t)\big)$ and $t \mapsto \mathcal{L}_{\mathrm{LC}}\big(\bM(t)\big)$ under different scenarios for $\bS \sim \mathrm{Wishart}_q(d, \bSigma)$.}
    \label{fig:frechet_loss}
\end{figure}

\section{Other simulation results}\label{supp_sec:other_sim_results}

\subsection{Euclidean space}\label{supp_subsec:euc}

Table \ref{tab:euc_typeIII_PB_SC} reports the results of Type III for OOB and SC balls, and Figure \ref{fig:euc_coverage_sigma_09_1} deals with Types II and IV ($\bX = \allowbreak (X_{0.25}, X_{0.25}, X_{0.25})^\top$). The computation times of SC and OOB balls in the simulations of model \eqref{eq:mult_linear_model} are collected in Table \ref{tab:computation_times}.

\begin{table}[H]
\setlength{\tabcolsep}{5pt}
\centering
\begin{tabular}{llccc|ccc}
\toprule
 &  & \multicolumn{3}{c}{Out-of-bag} & \multicolumn{3}{c}{Split-conformal} \\
\cmidrule(lr){3-5} \cmidrule(lr){6-8}
$\sigma$ & $n$ & $\alpha=0.01$ & $\alpha=0.05$ & $\alpha=0.10$ & $\alpha=0.01$ & $\alpha=0.05$ & $\alpha=0.10$ \\
\midrule
\multirow[c]{4}{*}{$\frac{\sqrt{3}}{2}$} & 50 & 99.1 (0.31) & 96.2 (0.61) & \underline{91.8} (0.92) & 98.3 (0.40) & 95.1 (0.64) & \underline{92.8} (0.80)
\\
 & 100 & 98.8 (0.33) & 95.0 (0.61) & 91.0 (0.86) & 99.0 (0.30) & 96.2 (0.56) & \underline{91.8} (0.82) \\
 & 200 & 98.6 (0.33) & 95.0 (0.69) & 91.0 (0.91) & 99.0 (0.36) & 95.3 (0.65) & 90.7 (0.90) \\
 & 500 & 99.4 (0.29) & 95.4 (0.64) & 91.6 (0.95) & 99.3 (0.31) & 95.6 (0.64) & 90.3 (0.93) \\
\cline{1-8}
\multirow[c]{4}{*}{$\sqrt{3}$} & 50 & 98.7 (0.39) & 94.9 (0.69) & 89.1 (0.97) & \underline{97.0} (0.52) & 93.7 (0.79) & 91.0 (0.98) \\
 & 100 & 98.4 (0.42) & 94.6 (0.69) & 89.8 (0.94) & 98.5 (0.39) & 95.4 (0.66) & 89.6 (0.97) \\
 & 200 & 99.2 (0.30) & \underline{96.7} (0.63) & 90.5 (0.95) & 99.0 (0.34) & 95.5 (0.67) & 90.8 (0.94) \\
 & 500 & 98.8 (0.37) & 94.3 (0.71) & 90.5 (0.93) & 99.0 (0.35) & 94.7 (0.67) & 91.1 (0.90) \\
\bottomrule
\end{tabular}
\caption{\small Same description as Table \ref{tab:euc_typeI_PB_SC}, but for Type III coverage. The fixed value $\bX = \allowbreak (X_{0.25}, X_{0.25}, X_{0.25})^\top$ was considered for the predictors.}
\label{tab:euc_typeIII_PB_SC}
\end{table}

\begin{table}[hpbt]
\setlength{\tabcolsep}{5pt}
\centering 
\begin{tabular}{lcccc}
\toprule
Method & $n=50$ & $n=100$ & $n=200$ & $n=500$ \\
\midrule
Out-of-bag & 9.37 (0.70) & 9.64 (0.66) & 10.06 (0.67) & 10.75 (0.70) \\
Split-conformal & 9.12 (0.82) & 9.44 (0.71) & 9.69 (0.69) & 10.26 (0.64) \\
\bottomrule
\end{tabular}
\caption{\small For each sample $\mathcal{L}_n^{(j)}$, $j=1, \ldots, N$, the elapsed time for fitting the forest and calculating the ball radius was measured. The reported values are the sample mean and sample standard deviation of the $N$ computation times (in seconds), in the simulations of model \eqref{eq:mult_linear_model}.}
\label{tab:computation_times}
\end{table}

\begin{figure}[H]
    \centering
    \begin{minipage}{0.45\textwidth}
        \centering
        \includegraphics[height=5cm]{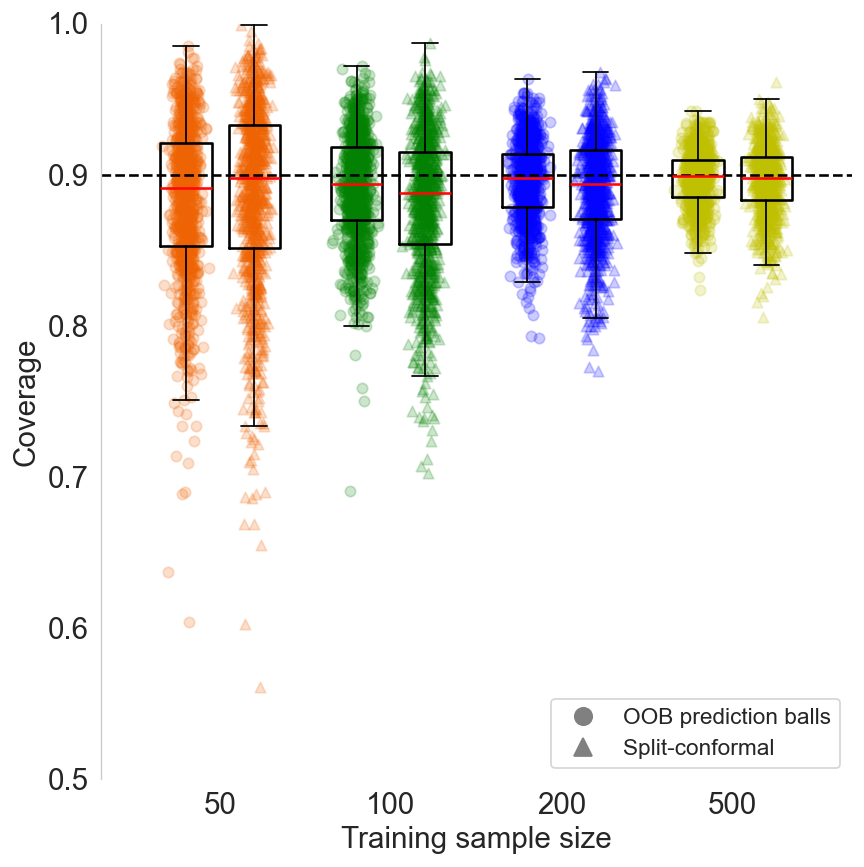}
        \subcaption{Type II}
    \end{minipage}\hfill
    \begin{minipage}{0.45\textwidth}
        \centering
        \includegraphics[height=5cm]{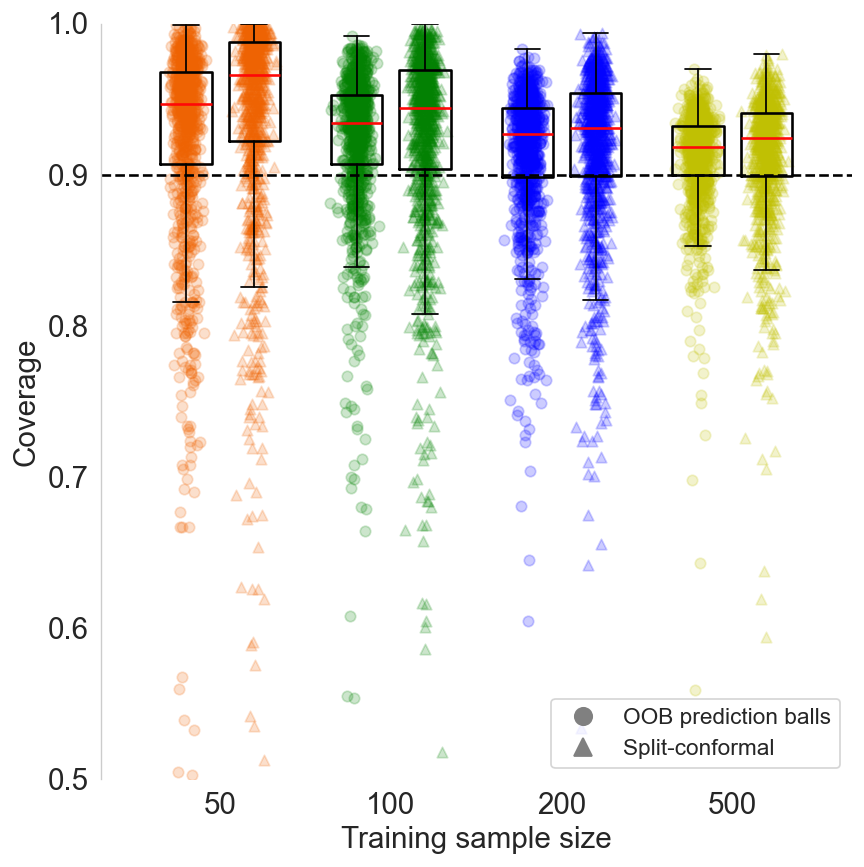}
        \subcaption{Type IV ($\bX=(X_{0.25}, X_{0.25}, X_{0.25})^\top$)}
    \end{minipage}
    \caption{\small Comparison between OOB and SC balls for the reported coverages (Types II and IV) across the simulated data sets, for $\sigma=\frac{\sqrt{3}}{2}$ and $\alpha = 0.10$. For Type IV, we considered $\bX=(X_{0.25}, X_{0.25}, X_{0.25})^\top$. Each dot in the boxplots is associated with a data set $\mathcal{L}_n^{(j)}$, $j=1,\ldots, N$, and its vertical position measures the estimated coverage probability (as in \eqref{type_IV_estimation}) conditioned to $\mathcal{L}_n^{(j)}$.}
    \label{fig:euc_coverage_sigma_09_1}
\end{figure}

\subsection{\texorpdfstring{Illustration of OOB balls on different metric spaces}{Illustration of OOB balls on different metric spaces}} \label{supp_subsec:illustration_oob_balls}

Figure \ref{fig:population_prediction_balls_S2_H2} illustrates the population prediction balls on the unit sphere and hyperboloid for different significance levels. Figure \ref{fig:population_prediction_balls_SPD} is the analog plot for the space of SPD matrices.

\begin{figure}[H] 
    \centering
    \begin{subfigure}{0.44\textwidth}
        \centering
        \includegraphics[height=5cm, align=c]{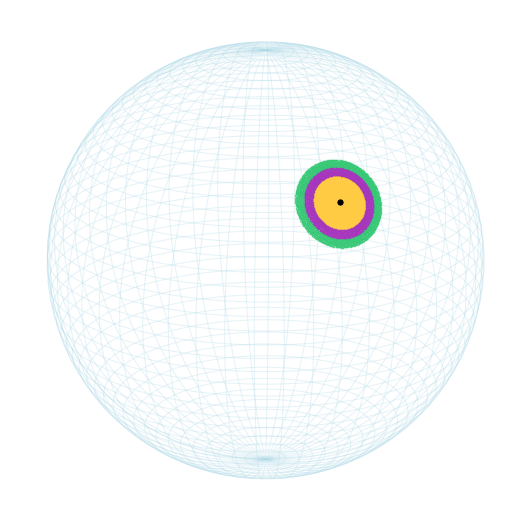}  
        \subcaption{Sphere $\mathbb{S}^2$}\label{fig:population_prediction_balls_S2_H2_a}
    \end{subfigure}\hfill
    \begin{subfigure}{0.44\textwidth}
        \centering
        \includegraphics[height=4.58139cm, align=c]{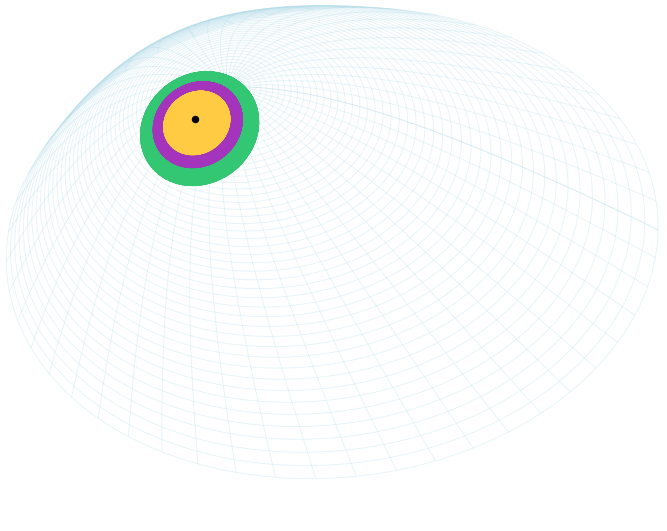}
        \subcaption{Hyperboloid $\mathbb{H}^2$}\label{fig:population_prediction_balls_S2_H2_b} 
    \end{subfigure}
    \caption{\small Illustration of OOB balls on $\mathbb{S}^2$ and $\mathbb{H}^2$ for the examples of Section \ref{subsec:sphere_hyp}, with $\kappa=200$ and significance levels $\alpha=0.25$, $0.10$, and $0.01$, corresponding to green, purple, and yellow (superimposed), respectively.
    }
    \label{fig:population_prediction_balls_S2_H2}
\end{figure}

\vspace{-0.65cm}

\begin{figure}[H]
    \centering
    \begin{subfigure}{0.3\textwidth}
        \centering
        \includegraphics[height=3.75cm]{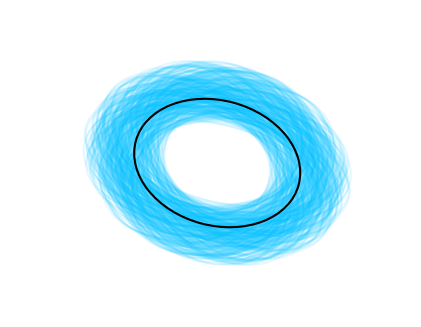}  
        \subcaption{AI distance}
    \end{subfigure}\hfill
    \begin{subfigure}{0.3\textwidth}
        \centering
        \includegraphics[height=3.75cm]{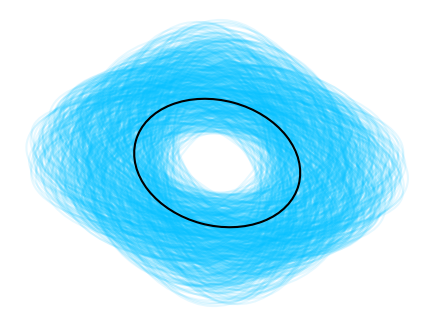}  
        \subcaption{LC distance}
    \end{subfigure}\hfill
    \begin{subfigure}{0.3\textwidth}
        \centering
        \includegraphics[height=3.75cm]{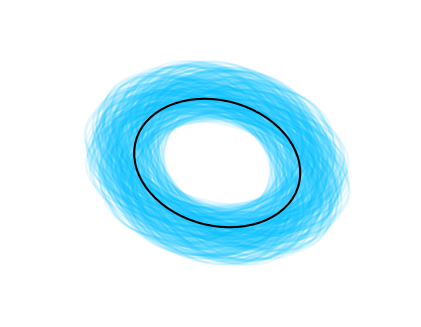}  
        \subcaption{LE distance}
    \end{subfigure}
    \caption{\small Illustration of OOB balls in $\mathcal{S}_2^+$ for the example discussed in Section \ref{subsec:SPD}, with $d=2$ and $\alpha = 0.10$.
    SPD matrices are represented via their eigen-decomposition: eigenvectors determine the orientation of the ellipse axes, and eigenvalues determine their lengths. The visualization is inspired by Figure 3.3 in \cite{Chacon2018|SM}.}
    \label{fig:population_prediction_balls_SPD}
\end{figure}

\subsection{Additional numerical results} \label{supp_subsec:github}
A comprehensive list of figures and tables from Section 5 is available at \url{https://github.com/dieseor/oob_balls}.

\begin{itemize}
    \item Types II and IV coverages of OOB and SC balls in the Euclidean space for $\sigma = \frac{\sqrt{3}}{2}$ and $\alpha=0.05$, $0.01$, and for $\sigma = \sqrt{3}$ and $\alpha=0.10$, $0.05$, $0.01$.
    \item Radius and coverage Types II and IV of OOB balls on the sphere and the hyperboloid for $\alpha=0.05$, $0.01$.
    \item Radius and coverage Types II and IV in $\mathcal{S}_2^+$ for the AI distance with $\alpha=0.05$, $0.01$.
    \item Radius and coverage Types I--IV in $\mathcal{S}_2^+$ for the LC and LE distances with $\alpha=0.10$, $0.05$, $0.01$.
\end{itemize}

\section{Connection of the HvMF and vMF distributions and normalizing constants}
\label{supp_sec:hvmf}

The Hyperboloid von--Mises Fisher (HvMF) distribution on $\Hd$, $d\geq1$, with location $\bmu\in\Hd$ and concentration $\kappa>0$ was introduced by \cite{Jensen1981|SM} as that induced by the density
\begin{align*}
    \bx\in\Hd\mapsto f_{\mathrm{HvMF}}(\bx;\bmu,\kappa)\defin c^\mathrm{HvMF}_{d}(\kappa)e^{\kappa (\bx,\bmu)}
\end{align*}
with respect to the Lebesgue measure on $\Hd$. Above, $c^\mathrm{HvMF}_{d}(\kappa)$ is a normalizing constant. The HvMF is the analog on the hyperboloid of the celebrated von Mises--Fisher (vMF) distribution on $\Sd$. The vMF distribution with location $\bmu\in\Sd$ and concentration $\kappa\geq 0$ has a density with respect to the Lebesgue measure on $\Sd$ given by
\begin{align*}
  \bx\in\Sd\mapsto f_{\mathrm{vMF}}(\bx;\bmu,\kappa)\defin c^\mathrm{vMF}_{d}(\kappa) e^{\kappa\bx^\top\bmu},\quad c_{d}^\mathrm{vMF}(\kappa)\defin\frac{\kappa^{(d-1)/2}}{(2\pi)^{(d+1)/2}\mathcal{I}_{(d-1)/2}(\kappa)},
\end{align*}
with $\mathcal{I}_\nu$ denoting the modified Bessel function of the first kind and order $\nu$. The normalizing constant for $\kappa=0$ (uniform distribution) is defined as the limit $\lim_{\kappa\to0}c_{d}^\mathrm{vMF}(\kappa)=2\pi^{(d+1)/2}/\Gamma((d+1)/2)$. The HvMF distribution does not allow $\kappa=0$, as this generates an improper uniform density on the (non-compact) hyperboloid.

In the following proposition, we reapproach the derivations by \cite[pp. 196--197]{Jensen1981|SM} to clarify the connection of the HvMF and vMF distributions, and to detail the derivation and computation of the normalizing constant. The complexity of computing the normalizing constant for the HvMF is comparable to that of the vMF for odd dimension $d$ and is slightly more direct for even $d$, in the sense of simplifying the finite series
\begin{align*}
    \mathcal{I}_{m+1/2}(z)=&\;\frac{1}{\sqrt{2\pi z}}\sum_{k=0}^m \frac{(m+k)!}{k!(m-k)!}\frac{(-1)^ke^{z}+(-1)^{m+1}e^{-z}}{(2z)^{k}}
\end{align*}
for integers $m\geq 0$ (see Equation 8.467 in \cite{Zwillinger2014|SM}) to the series
\begin{align}
    \mathcal{K}_{m+1/2}(z)=&\;\sqrt{\frac{\pi}{2z}} e^{-z} \sum_{k=0}^m \frac{(m+k)!}{k!(m-k)!}\frac{1}{(2z)^{k}} \label{eq:Km}
\end{align}
(see Equation 8.468 in \cite{Zwillinger2014|SM}).

\begin{proposition} \label{prop:hvmf}
For any $\bx\in\Hd$, $d\geq 1$, consider the parametrization $\bx=(\cosh(x_r), \allowbreak \sinh(x_r) \allowbreak \bx_s^\top)^\top$, where $(x_r,\bx_s^\top)^\top\in\R_{\geq0}\times \Sdm$. Analogously, set $\bmu=(\cosh(\mu_r),\sinh(\mu_r) \bmu_s^\top)^\top\in\Hd$. Then, the density of the HvMF distribution with location $\bmu\in\Hd$ and concentration $\kappa>0$ with respect to the Lebesgue measure on $\Hd$ is
\begin{align*}
    f_{\mathrm{HvMF}}(\bx;\bmu,\kappa)=&\;\nonumber c_d^\mathrm{HvMF}(\kappa)\frac{\sinh^{d-2}(x_r)\exp(-\kappa\cosh(\mu_r)\cosh(x_r))}{c_{d-1}^\mathrm{vMF}\left(\kappa\sinh(\mu_r)\sinh(x_r)\right)}\\
    &\;\times f_{\mathrm{vMF}}(\bx_s;\bmu_s,\kappa\sinh(\mu_r)\sinh(x_r)).
\end{align*}
The normalizing constant $c_d^\mathrm{HvMF}(\kappa)$ is given by
\begin{align*}
    c_d^\mathrm{HvMF}(\kappa)=\frac{\kappa^{(d-1)/2}}{(2\pi)^{(d-1)/2} 2\mathcal{K}_{(d-1)/2}(\kappa)},
\end{align*}
where
\begin{align*}
    \mathcal{K}_{(d-1)/2}(\kappa)\defin\frac{\pi^{1/2}(\kappa/2)^{(d-1)/2}}{\Gamma(d/2)}\int_0^\infty e^{-\kappa\cosh(t)}\sinh^{d-1}(t)\,\rd t
\end{align*}
is the modified Bessel function of the second kind and order $(d-1)/2$.
\end{proposition}

\begin{proof}[Proof of Proposition~\ref{prop:hvmf}] \!
Following \cite{Jensen1981|SM}, for $\bx \in \mathbb{H}^d$ and $u\in\R$, define the hyperbolic-spherical coordinates:
\begin{align*}
\begin{cases}
    x_1=\cosh(u)\\
    x_2=\sinh(u)
\end{cases}
\end{align*}
when $d=1$ and
$$
\begin{cases}
\begin{aligned}
    x_1      =&\; \cosh(u) \\
    x_2      =&\; \sinh(u)\cos(\theta_1) =\sinh(u)y_{1}\\
    x_3      =&\; \sinh(u)\sin(\theta_1)\cos(\theta_2) =\sinh(u)y_{2}\\
    &\;\;\vdots \\
    x_d      =&\; \sinh(u)\sin(\theta_1)\cdots\sin(\theta_{d-2})\cos(\theta_{d-1})=\sinh(u)y_{d-1} \\
    x_{d+1}  =&\; \sinh(u)\sin(\theta_1)\cdots\sin(\theta_{d-2})\sin(\theta_{d-1})=\sinh(u)y_{d}
\end{aligned}
\end{cases}
$$
when $d>1$, with $(\theta_1,\ldots,\theta_{d-1})^\top\in (0,\pi)^{d-2}\times[0,2\pi)$ being the angular hyperspherical coordinates of $\by\in\Sdm$. The Jacobian of the inverse $h^{-1}$ of the transformation $\bx=h(u,\theta_1,\ldots,\theta_{d-1})$ is $J(u,\theta_1,\ldots,\theta_{d-1})=\sinh^{d-1}(u)J(\theta_1,\ldots,\theta_{d-1})$, where $J(\theta_1,\ldots,\theta_{d-1})$ is the Jacobian of the angular hyperspherical coordinates $(\theta_1,\ldots,\theta_{d-1})\mapsto\by\in\Sdm$. Therefore, $\mu_d(\rd \bx)= \sinh^{d-1}(u)\,\rd u \,\sigma_{d-1}(\rd \by)$, with $\mu_d$ denoting the Lebesgue measure on $\Hd$ and $\sigma_{d-1}$ the Lebesgue measure on $\Sdm$.

Now, recall that $f_{\mathrm{HvMF}}(\bx;\bmu,\kappa)=c^\mathrm{HvMF}_{d}(\kappa)\exp\{\kappa (-x_1\mu_1+\bx_{-1}^\top\bmu_{-1})\}$. Using the hyperbolic-spherical coordinates $\bx=(\cosh(u),\sinh(u) \by^\top)^\top$, $\by \in \mathbb{S}^{d-1}$, it readily follows that
\begin{align}
    f_{\mathrm{HvMF}}(\bx;\bmu,\kappa)=&\;c^\mathrm{HvMF}_{d}(\kappa)\exp\{-\kappa\cosh(u)\cosh(\mu_r)+\kappa\sinh(u)\sinh(\mu_r)\bx_{s}^\top\bmu_{s}\}\sinh^{d-1}(u)\nonumber\\
    =&\;c^\mathrm{HvMF}_{d}(\kappa)\exp\{-\kappa\cosh(\mu_r)\cosh(u)\} \sinh^{d-1}(u) [c_{d-1}^\mathrm{vMF}(\kappa\sinh(\mu_r)\sinh(u))]^{-1}\nonumber\\
    &\;\times f_\mathrm{vMF}(\by; \bmu_{s}, \kappa\sinh(\mu_r)\sinh(u)). \label{eq:densprod}
\end{align}
From \eqref{eq:densprod}, it follows that the normalizing constant is
\begin{align}
    c_d^\mathrm{HvMF}(\kappa)^{-1}=\int_0^\infty \frac{\sinh^{d-1}(u)e^{-\kappa\cosh(\mu_r)\cosh(u)}}{c_{d-1}^\mathrm{vMF}\left(\kappa\sinh(\mu_r)\sinh(u)\right)}\,\rd u.\label{eq:cdhvmf}
\end{align}

Let us assume that $c_d^\mathrm{HvMF}(\kappa)^{-1}$ does not depend on $\mu_r\geq 0$. Then, for $\mu_r=0$, it easily follows~that
\begin{align*}
    c_d^\mathrm{HvMF}(\kappa)^{-1}=&\;
    \frac{2\pi^{d/2}}{\Gamma(d/2)}\int_0^\infty \sinh^{d-1}(u)e^{-\kappa\cosh(u)}\,\rd u
    =\frac{(2\pi)^{(d-1)/2} 2\mathcal{K}_{(d-1)/2}(\kappa)}{\kappa^{(d-1)/2}}.
\end{align*}

We now show that \eqref{eq:cdhvmf} does not depend on $\bmu$ (and in particular, on $\mu_r\geq 0$). Define the group of hyperbolic transformations on $\R^{d+1}$ as:
$$
\mathrm{SH}(d) \defin \{ 
\bA \in \R^{(d+1) \times (d+1)}: \bA^\top \tilde{\bI}_{d+1} \bA = \tilde{\bI}_{d+1}, \, \det(\bA)=1, \, A_{11}>0
\},
$$
where 
$$
\tilde{\bI}_{d+1} 
=
\begin{bmatrix}
-1 & \boldsymbol{0}_d \\
\boldsymbol{0}_d & \bI_d \\
\end{bmatrix}.
$$
The group of hyperbolic transformations acts transitively on $\mathbb{H}^d$ \citep[see Lemma~2(ii) of][]{Jensen1981|SM}, i.e., for any pair of mean vectors $\bmu, \tilde{\bmu} \in \mathbb{H}^d$, there exists an action $\bA\in \mathrm{SH}(d)$ such that $\tilde{\bmu} = \bA \bmu$. The equality $c_d^{\mathrm{HvMF}}(\bmu, \kappa) = c_d^{\mathrm{HvMF}}(\tilde{\bmu}, \kappa)$ follows from
\begin{align}
c_d^{\mathrm{HvMF}}(\bmu, \kappa)^{-1} =&\; \int_{\mathbb{H}^d} e^{-\kappa(\bx, \bmu)} \,\mu_d(\rd\bx) 
= \int_{\mathbb{H}^d} e^{-\kappa(\bA\bx, \tilde{\bmu})} \,\mu_d(\rd \bx) \label{invariant_constant_1} \\
=&\;\int_{\mathbb{H}^d} e^{-\kappa(\bz, \tilde{\bmu})} \,\mu_d(\rd \bz) = c_d^{\mathrm{HvMF}}(\tilde{\bmu}, \kappa)^{-1}. \label{invariant_constant_2}
\end{align}
The second equality in \eqref{invariant_constant_1} holds because the Minkowski pseudo-inner product is invariant under hyperbolic transformations, i.e., $(\bx,\by) = (\bA\bx,\bA\by)$ for every $\bx, \by \in \mathbb{H}^d$ and every $\bA \in \mathrm{SH}(d)$. The invariance follows from the identity $(\bx, \by) = \bx^\top \tilde{\bI}_{d+1} \by$ for any $\bx, \by \in \mathbb{R}^{d+1}$. For the first equality in \eqref{invariant_constant_2}, first observe that for every $\bA \in \mathrm{SH}(d)$, its inverse $\bA^{-1} \in \mathrm{SH}(d)$ is given by $\bA^{-1} = \tilde{\bI}_{d+1} \bA^\top \tilde{\bI}_{d+1}$. Now, we apply the substitution $\bx = \bA^{-1} \bz$ (note that $\det(\bA^{-1}) = 1$). Observe also that the domain of integration remains invariant since the mapping $\bx \mapsto \bA^{-1}\bx$ is a bijection from $\mathbb{H}^d$ to itself. Surjectivity holds because for any $\bz \in \mathbb{H}^d$, $\bz = \bA^{-1}(\bA\bx)$, and $\bA\bx \in \mathbb{H}^d$ because the unit hyperboloid is invariant under hyperbolic transformations \citep[Lemma~1(ii) of][]{Jensen1981|SM}.
\end{proof}


\fi

\end{document}